\documentclass[a4paper,12pt]{article}
\pdfoutput=1 

\usepackage{jinstpub} 

\title{Hadronic vs Electromagnetic Pulse Shape Discrimination in CsI(Tl) for High Energy Physics Experiments}

 \author{S. Longo}
 \author{and J. M. Roney}
 \affiliation{Department of Physics and Astronomy, University of Victoria,\\3800 Finnerty Rd, Victoria, BC, V8P 5C2, Canada}

\emailAdd{longos@uvic.ca}

\abstract{Pulse shape discrimination using \csi{} scintillators to perform neutral hadron particle identification is explored with emphasis towards application at high energy electron-positron collider experiments.  Through the analysis of the pulse shape differences between scintillation pulses from photon and hadronic energy deposits using neutron and proton data collected at TRIUMF, it is shown that the pulse shape variations observed for hadrons can be modelled using a third scintillation component for \csi{}, in addition to the standard fast and slow components.  Techniques for computing the hadronic pulse amplitudes and shape variations are developed and it is shown that the intensity of the additional scintillation component can be computed from the ionization energy loss of the interacting particles.  These pulse modelling and simulation methods are integrated with GEANT4 simulation libraries and the predicted pulse shape for \csi{} crystals in a $5 \times 5$ array of $5 \times 5 \times 30 \text{ cm}^3$ crystals is studied for hadronic showers from 0.5 and 1 GeV/c $K_L^0$ and neutron particles.   Using a crystal level and cluster level approach for photon vs hadron cluster separation we demonstrate proof-of-concept for neutral hadron detection using \csi{} pulse shape discrimination in high energy electron-positron collider experiments.}

\newcommand{\csi}{CsI(Tl)}

\newcommand{\mus}{ \mu \text{s}}
\newcommand{\dedx}{$\frac{dE}{dx}$}
\newcommand{\tauh}{$\tau_\text{Hadron}$}
\newcommand{\kl}{$K^0_L$}
\newcommand{\fdedx}{$f(\frac{dE}{dx})$}

\usepackage{float}
\usepackage{algorithmic}
\usepackage[caption = false]{subfig}
\usepackage{enumitem}
\setlist{nosep} 
\def\babar{\mbox{\sl B\hspace{-0.4em} {\small\sl A}\hspace{-0.37em} \sl B\hspace{-0.4em} {\small\sl A\hspace{-0.02em}R}}}
\begin{document}
\maketitle
\flushbottom
\section{Introduction}

We study the scintillation response of Thallium doped Cesium Iodide (\csi{}) to charged and neutral hadrons in order to evaluate the potential for the application of inorganic scintillator pulse shape discrimination (PSD) to improve electromagnetic vs hadronic calorimeter cluster identification at high energy $e^+ e^-$ collider experiments.  In the low energy regime ($E_k<20$ MeV) it has been established that charged particle identification for electron, proton and alpha particles can be achieved using \csi{} scintillation pulse shape discrimination \cite{Storey,Benrachi,Skulski}. As a result this technique has been used for charged particle identification in nuclear physics detectors for example in heavy-ion detectors such as AMPHORA \cite{AMPHORA} and CHIMERA \cite{CHIMERA}.  Pulse shape discrimination has not yet been applied at high energy physics experiments using \csi{} detectors.  For example the past \babar{} \cite{BaBarNIM1,BaBarNIM2} and Belle \cite{BelleNIM} experiments, which made use of \csi{} electromagnetic calorimeters, only extracted pulse amplitude and timing information \cite{BaBarNIM1,BaBarNIM2,BelleNIM}.   With upcoming/present experiments such as Belle II \cite{Belle2TDR} and BESIII \cite{BESIIITDR} applying new detector technologies such as online waveform processing with FPGAs in the front-end electronics for the \csi{} calorimeter crystals \cite{BelleIIECLreconstruction}, online pulse shape characterization is now feasible to extract pulse shape information in addition to the standard crystal energy and timing variables.  In this study we focus on the objective of applying pulse shape discrimination to improve electromagnetic vs hadron calorimeter cluster identification.  Improvements in cluster identification would result in the reduction of systematic uncertainties related to particle identification of photons vs long lived neutron hadrons and low momentum pion vs muon separation.  These improvements would complement the large data samples planned to be collected by intensity frontier experiments to perform precision tests of the Standard Model.  

The reconstruction of long lived neutral hadrons are an important but challenging task at high energy particle detectors. For example, $K^0_L$ detection is critical for important physics analyses at $e^+ e^-$ B-Factories
such as the measurement of the Charge-Parity violation parameter, $\sin{2\beta}$,  using the decay of $B \rightarrow J/\psi  K^0_L$ \cite{BaBarsine2beta,Bellesine2beta}. The past B-factories \babar{} and Belle relied on event topologies and calorimeter energy spatial distributions to separate neutral hadrons from photons \cite{BaBarsine2beta,Bellesine2beta}.  In these cases, the neutral hadron experimental signature is characterized by the lack of associated charged particles in the tracking detectors, the transverse spatial distribution of energy deposits in the CsI(Tl) calorimeter and/or the characteristics of energy deposited in detectors behind the CsI(Tl) calorimeter. 
These methods however have shown to lead to low purities for $ J/\psi + K^0_L$ samples (51\%)  compared to sample purities achieved for $J/\psi + K^0_S$ (96\%) as the $K^0_S$ can be reconstructed using tracking information \cite{BaBarsine2beta}.  

The question we address in this paper is: Do the hadronic showers initiated by higher momentum ($|\vec{P}| = 0.1-1$ GeV/c) $K_L^0$ or neutrons produce significant enough energy deposits from secondary charged hadrons such that PSD can be applied to substantially improve the discrimination between electromagnetic and hadronic showers in a \csi{} calorimeter?  We focus on separating showers from $K_L^0$ or neutrons and photons in particular as these particles cannot be identified with tracking detectors however we note that electromagnetic vs hadron shower identification would also have application in improving charged particle identification in high energy experiments as well.   This technique has been explored for fast neutron detection using small $2.54 \text{ cm}$ diameter \csi{} crystals in references \cite{Bartle,McLean} where fast neutron identification was demonstrated using this principle.  These studies show that the secondary charged hadrons created from the inelastic neutron interactions in the crystal can transfer a significant amount of energy to the \csi{} \cite{Bartle,McLean}.   By applying \csi{} PSD, the different inelastic neutron interactions in the crystal then can be identified \cite{Bartle,McLean}.  From these studies it is expected that the crystals in $K^0_L$ initiated showers will have hadronic \csi{} pulse shapes.  

We begin in Section \ref{ExperimentalData} with a description of the fast neutron data and proton testbeam data,  collected at the TRIUMF Proton Irradiation Facility (PIF) \cite{Ewart}, and used in this paper. In addition, the experimental setup is also described in Section \ref{ExperimentalData}.  In Section \ref{DevelopmentofModel} we develop a hadron scintillation component model to characterize the \csi{}  pulse shape variations observed in the fast neutron data.   This model is validated in Section \ref{protondata} by applying it to the proton testbeam data.  In Section \ref{simtech} simulation methods for the hadron scintillation component model are developed and integrated with GEANT4\footnote{Simulation results are computed using GEANT4 version 10.2.2 and the \texttt{FTFP\_BERT\_HP} physics lists with range cuts for all particles set to 0.07 mm. } particle interactions in matter simulation libraries \cite{geant}. Simulated results for neutron and proton interactions in a \csi{} crystal are then computed and quantitatively compared with the fast neutron and proton testbeam data collected.  Finally in Section \ref{ClusterStudy} the pulse shape characterization and simulation techniques are applied to compute the pulse shape response for a \csi{} crystal cluster to hadronic showers from a sample of simulated $K^0_L$ and neutron events.   By comparing the predicted cluster response for the hadronic and electromagnetic showers, neutral hadron vs photon discrimination using \csi{} PSD is demonstrated.

\section{Experimental Data}
\label{ExperimentalData}

\subsection{Neutron and Proton Data}

Proton data from proton testbeams was collected at the TRIUMF Proton Irradiation Facility (PIF) \cite{Ewart}.  Scintillation pulses from protons with primary kinetic energies of 67.0, 57.7, 40.1 and 20.0 MeV were recorded.  The proton beam in the 57.7, 40.1 and 20.0 MeV runs was partially degraded prior to the \csi{} detector to allow for multiple proton energies in the same beam.  The \csi{} detector was self-triggered at rate of approximately 2.1 kHz while recording the proton pulses. 

A sample of fast neutron events was also collected in a neutron run.  During this run the TRIUMF cyclotron was in operation performing isotope production which involved impinging up to 500 MeV protons on a target located on the other side of a wall of concrete shielding separating the PIF area from target area.  From this process a sample of high energy neutrons approximately following a 1/E energy distribution up to a maximum energy of 500 MeV \cite{Ewart} interacted with the detector in the PIF area during a 10 hour run and as a result this run contained a sample of fast neutron events.  This dataset also included energy deposits from cosmic muons.  Energy calibration was completed using low energy photon peaks from $^{137}$Cs (0.662 MeV), $^{40}$K (1.46 MeV) and $^{208}$Tl (2.61 MeV) backgrounds.  As a result, light output yield values throughout this paper are expressed in photon equivalent energy units.

\subsection{\csi{} Detector}

\csi{} scintillation pulses were digitized using a spare crystal from the past B-factory experiment \babar{} \cite{BaBarNIM1,BaBarNIM2}. The crystal was manufactured by Shanghai Institute of Ceramics and had a length of 30 cm and a trapezoidal geometry with front face size of approximately $4 \times 4 \text{ cm}^2$ and light readout face size of approximately $5 \times 5 \text{ cm}^2$.  The crystal had thin wrappings of Teflon and Mylar in order to improve the light collection efficiency.  A Hamamatsu R580 photomultiplier tube (PMT) with diameter of 38 mm was used for scintillation light detection \cite{PMT}.  Using a spring assembly, the PMT was pressed against the crystal face and an air optical coupling was used to interface the PMT and crystal.  The PMT output was connected to a CAEN V1724 digitizer that was triggered on a voltage threshold.  Once triggered the scintillation pulses were digitized with sampling time of 10 ns and saved for offline analysis.   

\section{Hadronic Scintillation Component Model}
\label{DevelopmentofModel}
\subsection{Charge Ratio Characterization}

We begin by characterizing the scintillation pulse shapes in the PIF neutron measurement using the short-over-long charge ratio PSD technique that has been applied in past \csi{} PSD studies \cite{Skulski,McLean}.  This technique takes advantage of the observation that proton and alpha energy deposits in \csi{} result in faster scintillation emission compared to photon energy deposits \cite{Storey}.  This results in a greater percentage of the scintillation emission to occur earlier in the scintillation pulse for hadron energy deposits leading to higher values of the charge ratio.  Specifically we use the charge ratio, $R_\text{PSD}$, defined in equation \ref{ChargeRatio} where $Q(t)$ is the charge output of the PMT as a function of time. 

\begin{equation}
\label{ChargeRatio}
R_\text{PSD} = \frac{Q( 1.2 \mus )}{Q( 7.4 \mus)}
\end{equation}

\begin{figure}[H]
\centering
\includegraphics[width=0.6\textwidth]{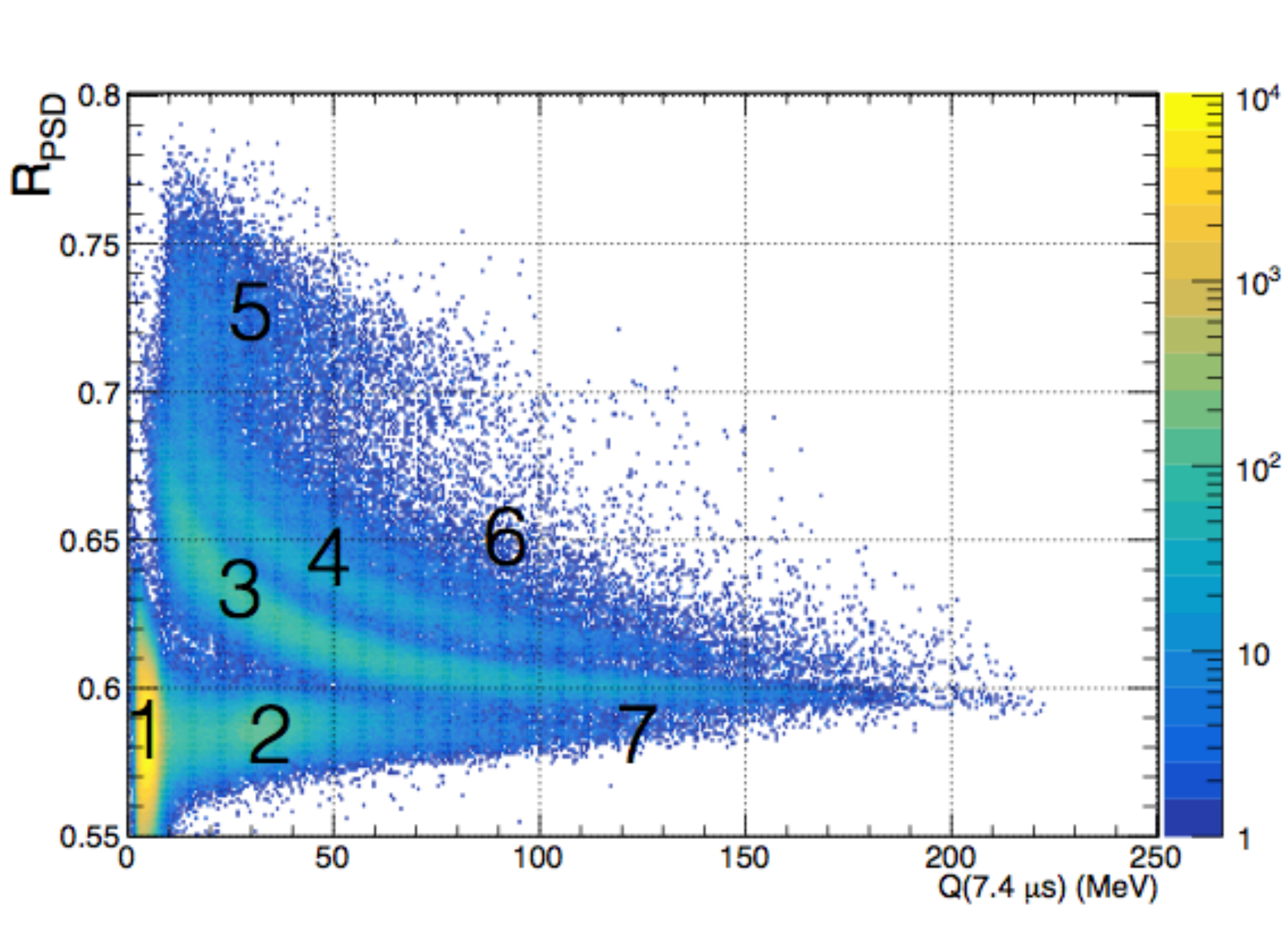}
\caption{$R_\text{PSD}$ vs $Q( 7.4 \mus)$ pulse shape characterization of pulses recorded in the 10 hour PIF neutron measurement.}
\label{ChargeRatioPMTrun0}
\end{figure}

A short charge time of $1.2 \mus$ is used so that comparison can be made with reference \cite{McLean} where the same short gate time was applied and stated to be the optimal time for \csi{} PSD.  Our long gate of $7.4 \mus$ was chosen as this is the typical length of time available before pile-up effects in high radiation environments such as at $e^+ e^-$ collider experiments.  Figure \ref{ChargeRatioPMTrun0} displays the two dimensional histogram of $R_\text{PSD}$ vs $Q( 7.4 \mus)$. 

From Figure \ref{ChargeRatioPMTrun0}, it is observed that a diverse spectra of pulse shapes are present in the 10 hour PIF neutron run.  In order to discuss the material interactions causing the band structures observed we label seven regions in Figure \ref{ChargeRatioPMTrun0} for reference.  Region 2 is identified as the cosmic muon peak centred around a charge ratio of approximately 0.585. At higher values of $R_\text{PSD}$,  where hadron pulses are expected, systematic band structures are present. These band structures were reported in two past studies of the response of \csi{} to fast neutrons \cite{Bartle,McLean}.  In these references the bands are attributed to secondary proton, deuteron and triton particles generated by neutron interactions in the crystal \cite{Bartle,McLean}.  Using the simulation methods discussed in Section \ref{simtech}, we use Monte Carlo (MC) truth to confirm that the Region 3 band arises from secondary protons from neutron scatters.  We find however that the Region 4 band is predominately from neutron interactions where two secondary protons are created.  Specifically for all labelled regions we find that the corresponding interaction is given by:

\begin{itemize}
\item[1 - ] Secondary photons from low energy ($E_k<10$ MeV) neutrons and natural radioactive background.
\item[2 - ]  Cosmic muon peak.
\item[3 - ]  Neutron scatter where single secondary proton was produced.  The value of the deposited energy
is determined from the secondary proton kinetic energy as it will stop in the crystal
volume.
\item[4 - ]  Neutron scatter where two secondary protons where produced.  
\item[5 - ]  Neutron scatter where secondary alpha was produced.  
\item[6 - ]  Neutron scatter where secondary proton/deuteron and alpha were produced.
\item[7 - ]  Neutron scatter where high momentum proton was produced and escapes crystal volume.
\end{itemize}

\subsection{Pulse Shape Variations from Photon Pulse}

For greater than 1 MeV photon energy deposits in \csi{} it is well established that the scintillation pulse shape can be described using a two component scintillation model consisting of a fast and slow component emission as shown in equation \ref{TwoCompModel} in charge form and equation \ref{TwoCompModel_Current} in current-form \cite{Valentine}.  

\begin{equation}
\label{TwoCompModel}
 Q_\gamma(t) =L_\gamma (N_\text{fast} (1-e^\frac{-t}{\tau_\text{fast}}) + N_\text{slow} (1-e^\frac{-t}{\tau_\text{slow}}))
\end{equation} 

\begin{equation}
\label{TwoCompModel_Current}
 I_\gamma(t) =L_\gamma \Big(  \frac{N_\text{fast}}{\tau_\text{fast}}  e^\frac{-t}{\tau_\text{fast}} + \frac{N_\text{slow}}{\tau_\text{slow}} e^\frac{-t}{\tau_\text{slow}}\Big)
\end{equation} 

\noindent Where $L_\gamma$ is the total light output from the photon energy deposit, $\tau_\text{fast}$ is the time constant of the fast scintillation component, $\tau_\text{slow}$ is the time constant of the slow scintillation component, $N_\text{fast}$ is the relative intensity of the fast scintillation component yield and $N_\text{slow}$ is the relative intensity of the slow scintillation component yield defined such that $ N_\text{fast}  +  N_\text{slow} = 1$.

In order to study the origin of the pulse shape difference between photon pulses and pulses from hadron energy deposits, we construct a template photon pulse by individually fitting equation \ref{TwoCompModel} to the pulses in Region 1 (low energy photons) of Figure \ref{ChargeRatioPMTrun0} defined by, $3 <  Q( 7.4 \mus) < $ 6 MeV and $0.569 < R_\text{PSD} < 0.593$.  From this large sample of fits we extract the mean values of the four scintillation parameters to define a template photon pulse.  The results for the template photon pulse parameters are shown in Table \ref{GammaParameters}.  We also include in Table \ref{GammaParameters} the photon pulse shape parameters found in a previous investigation of the photon \csi{} pulse shape near room temperature by reference \cite{Valentine}.  We note that the template photon pulse parameters we measured are in agreement with reference \cite{Valentine}.     

\begin{table}[h]
\centering
\caption{Mean values for photon pulse shape parameters determined by fitting shape parameters to a large sample of 3-6 MeV photon pulses.  The systematic uncertainty is $\pm 1$\% and statistical errors are negligible.  Our values are compared to those from Valentine et al \cite{Valentine}. Note that $N_\text{slow}/N_\text{fast}$ corresponds to $Q_2$/$Q_1$ in reference \cite{Valentine} and $N_\text{slow} = 1- N_\text{fast}$.}

\label{GammaParameters}
\begin{tabular}{|c|c|c|c|c|}
\hline
Parameter & $\tau_\text{fast}$ & $\tau_\text{slow}$ & $N_\text{fast}$ & $N_\text{slow}/N_\text{fast}$  \\ \hline
This Study & $851 \pm 9 \text{ ns}$ &  $5802 \pm \text{58 ns}$ & $0.569 \pm 0.006$ & $0.756 \pm 0.007$ \\ \hline
From Ref. \cite{Valentine} & $832 \pm 42 \text{ ns}$ &  $5500 \pm 275 \text{ ns}$ & $0.568 \pm 0.028$ & $0.760 \pm 0.038$  \\ \hline
\end{tabular}
\end{table}

 We calculate for pulses in regions 2-7 the energy normalized charge difference, $\text{D}_{Q}(t)$, defined by equation \ref{PulseDifference} where we normalize to the charge at $7.4 \mus$.
 
\begin{equation}
\label{PulseDifference}
\text{D}_{Q}(t)  = \frac{Q(t)}{Q(7.4 \mus)} -\frac{Q_\gamma(t)}{Q_\gamma(7.4 \mus)}
\end{equation} 

\begin{figure}[h]
\centering
\subfloat[]{\includegraphics[width=0.5\textwidth]{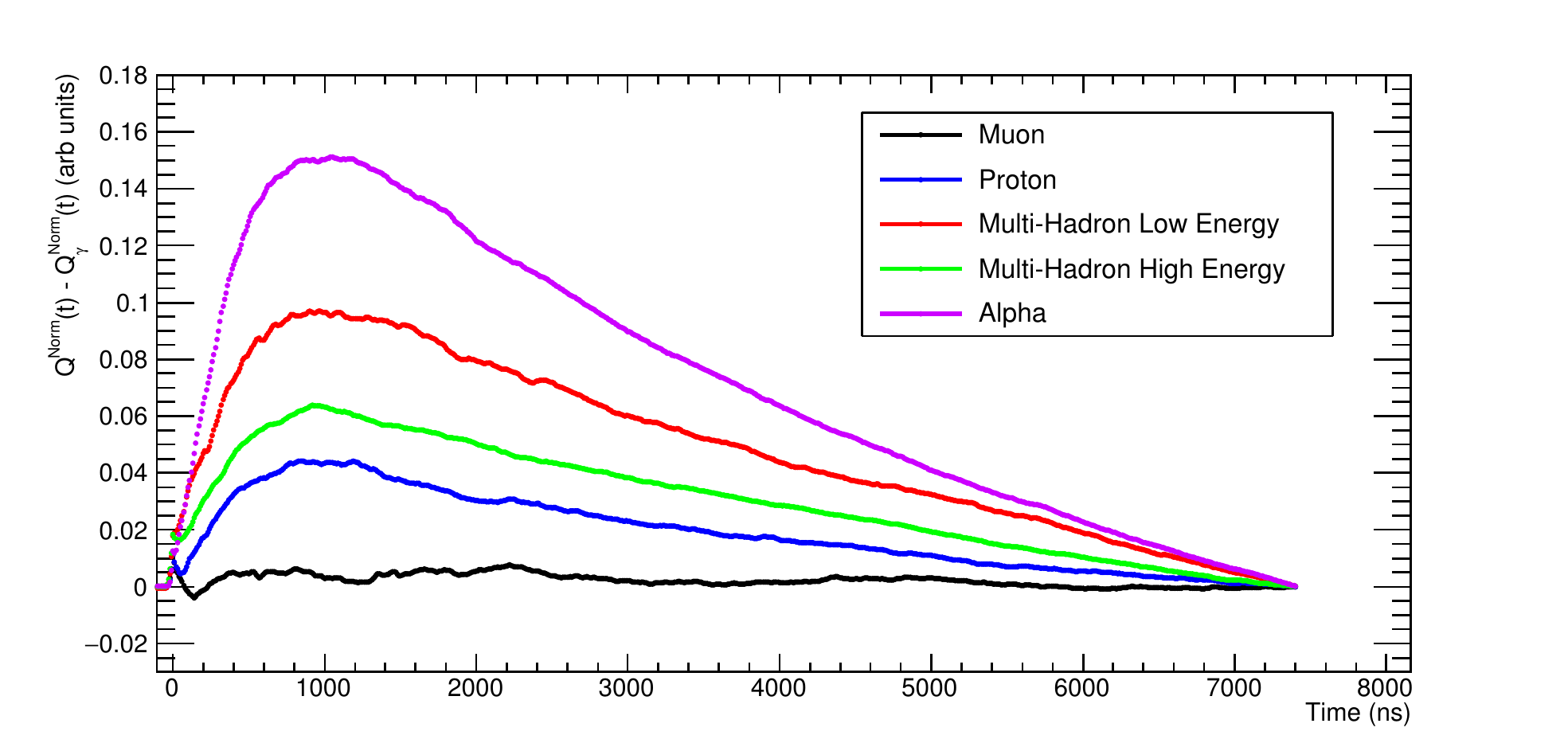}
} 
\subfloat[]{\includegraphics[width=0.5\textwidth]{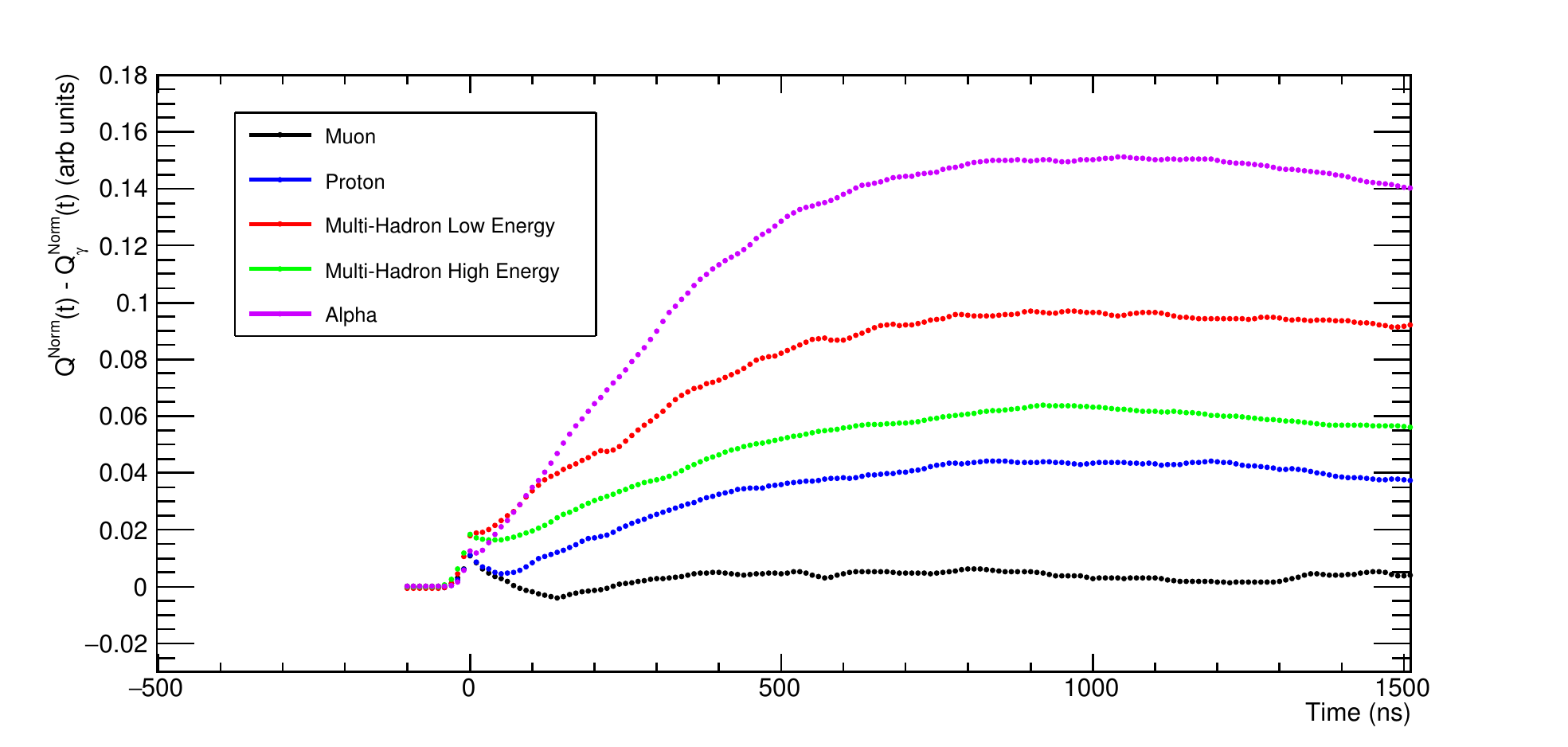}\label{DataMinusGammab}}
\caption{a) Plots of $\text{D}_{Q}(t)$ for sample pulses in the various regions defined from the pulse shape spectrum in Figure \ref{ChargeRatioPMTrun0}. "Muon" is from a sample pulse in from region 2; "Proton", region 3; "Multi-hadron low energy". region 4; "Multi-hadron high energy", region 6; and  "Alpha", region 5.  b) Zoom in on first microseconds of pulse differences showing approximate form of an integrated exponential.}
\label{DataMinusGammaa}
\end{figure}

From $\text{D}_{Q}(t)$ we gain insight to the origin of the pulse shape variations observed in non-photon pulses.  Typical $\text{D}_{Q}(t)$ for the pulse regions 2-6 are shown in Figure \ref{DataMinusGammaa}.   From these charge difference plots it is observed that only the amplitude of $\text{D}_{Q}(t)$ is dependent on $R_\text{PSD}$ and the shape of $\text{D}_{Q}(t)$ remains constant, independent of the charge ratio or energy deposited.  This result demonstrates the wide spectrum of the pulse shapes observed in Figure \ref{ChargeRatioPMTrun0} all deviate from the photon pulse in the same way and thus likely occur from the same origin.  We note that $\text{D}_{Q}(t)$ was also studied by reference \cite{Skulski} for low energy alpha particles where the same shape was observed.  In addition the peak value of $\text{D}_{Q}(t)$ occurs at approximately $1.2 \mus$ after the trigger point accounting for why this was found to be the optimal short gate by reference \cite{McLean}.   

Studying the shape of $\text{D}_{Q}(t)$, the decay to zero charge difference at $7.4 \mus$ is a result of the charge normalization in the definition of $\text{D}_{Q}(t)$.  In the initial microseconds shown in the zoom in Figure \ref{DataMinusGammab} however it is observed that $\text{D}_{Q}(t)$ has the approximate behaviour of an integrated exponential suggesting that the pulse shape difference arises from an additional scintillation component.  To explicitly demonstrate this we calculate the current difference $ I^\text{Diff}(t)$ defined in equation \ref{CurrentDifference}.

\begin{equation}
\label{CurrentDifference}
I^\text{Diff}(t) =  I(t) -  \frac{A^\text{Tail}}{A_\gamma^\text{Tail}} I_{\gamma}(t)
\end{equation} 

\noindent Where $A^\text{Tail}$ is the integrated current in the tail region of the pulse we define by t=10-14$\mu$s.  As we find the pulse shape in the tail region is independent of particle type we use the charge in this region to scale the current-form of the gamma template when computing $I^\text{Diff}(t)$.   By computing $I^\text{Diff}(t)$ we can observe the shape of the light emission difference which occurs for hadron energy deposits compared to photons.  Plots of $I^\text{Diff}(t)$ for typical pulses in shape regions 2-7 are shown Figure \ref{pulse_differences} with a fit overlaid to a exponential with fixed decay time of 630 ns, only fitting for the amplitude, A.  From these plots it can be seen that the additional emission present for hadron energy deposits has the identical exponential form for all pulse shape regions.   For the muon pulse is it also observed that the difference is zero as the pulse shape is the same as the low energy photons. This particle-independent exponential shape for the hadron energy deposits indicates that the pulse variations for these particles originate from a third scintillation component which is not present for electromagnetic energy deposits.   We find the time constant for this hadronic scintillation component to be \tauh{}=$630 \pm 10$ ns. 

\begin{figure}[H]
\centering
\subfloat[Cosmic Muon (R2)]{\includegraphics[width=0.5\textwidth]{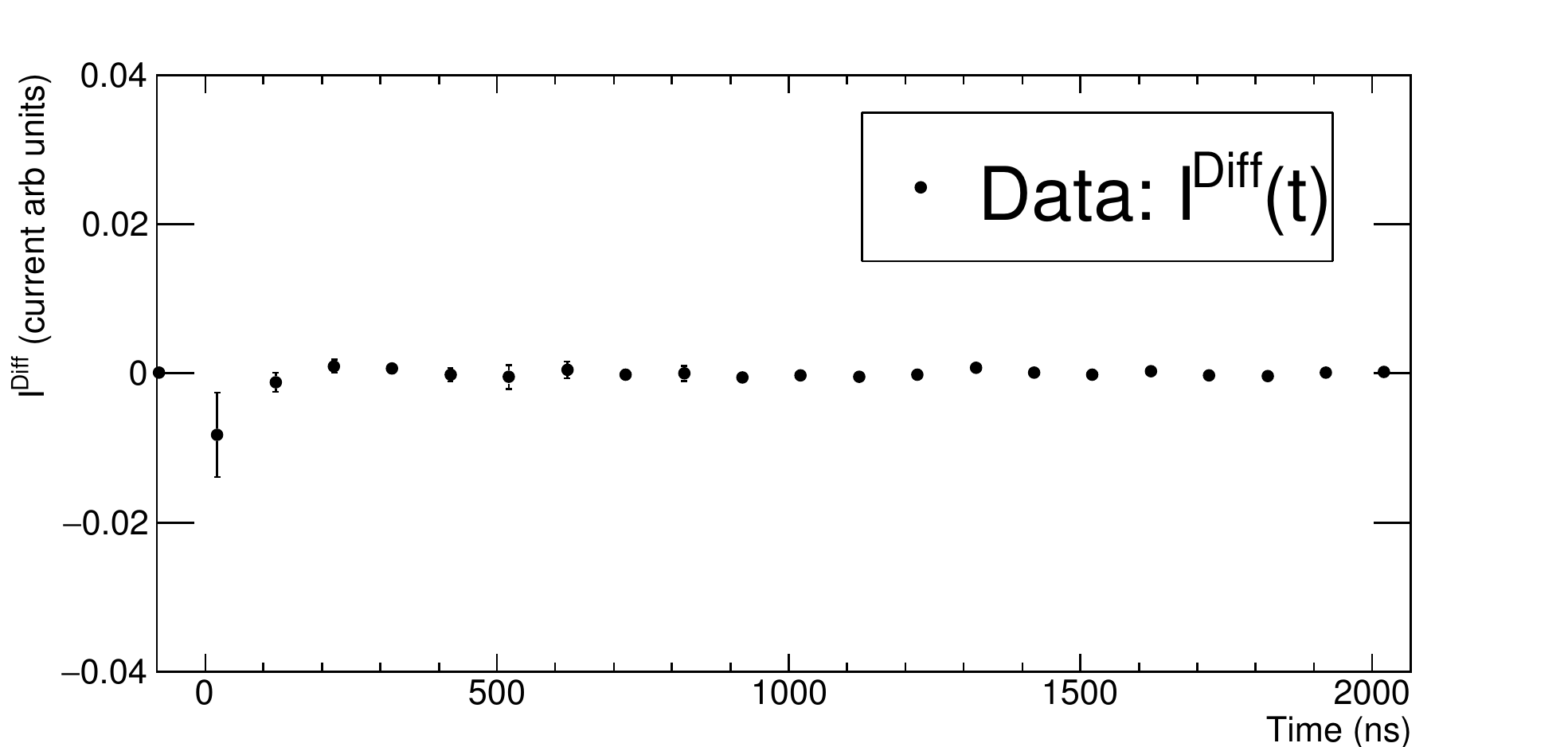}} 
\subfloat[Low Energy Proton (R3)]{\includegraphics[width=0.5\textwidth]{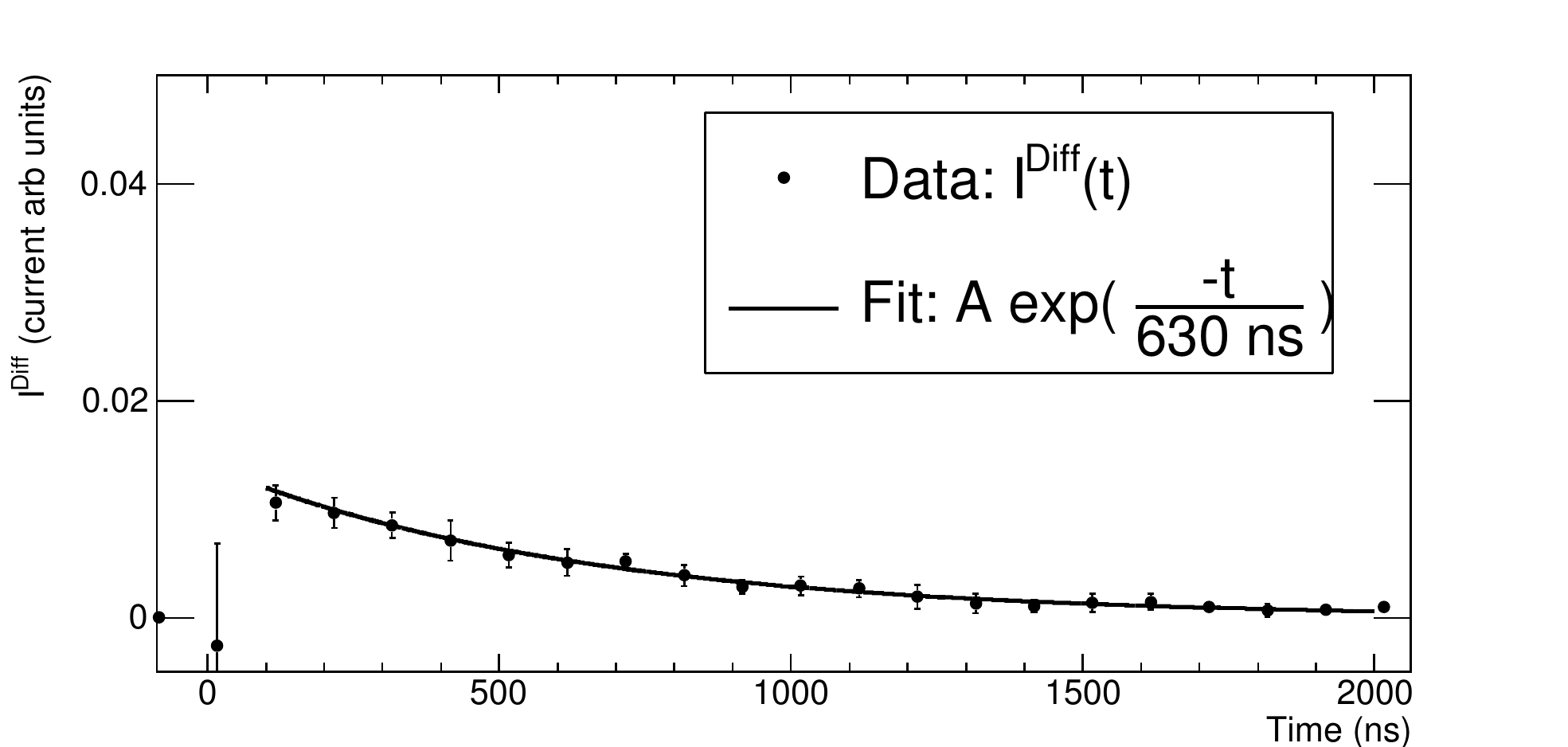}} 

\label{pulse_differences}
\end{figure}
\begin{figure}[H]
\subfloat[Low Energy Multi-Hadron (R4)]{\includegraphics[width=0.5\textwidth]{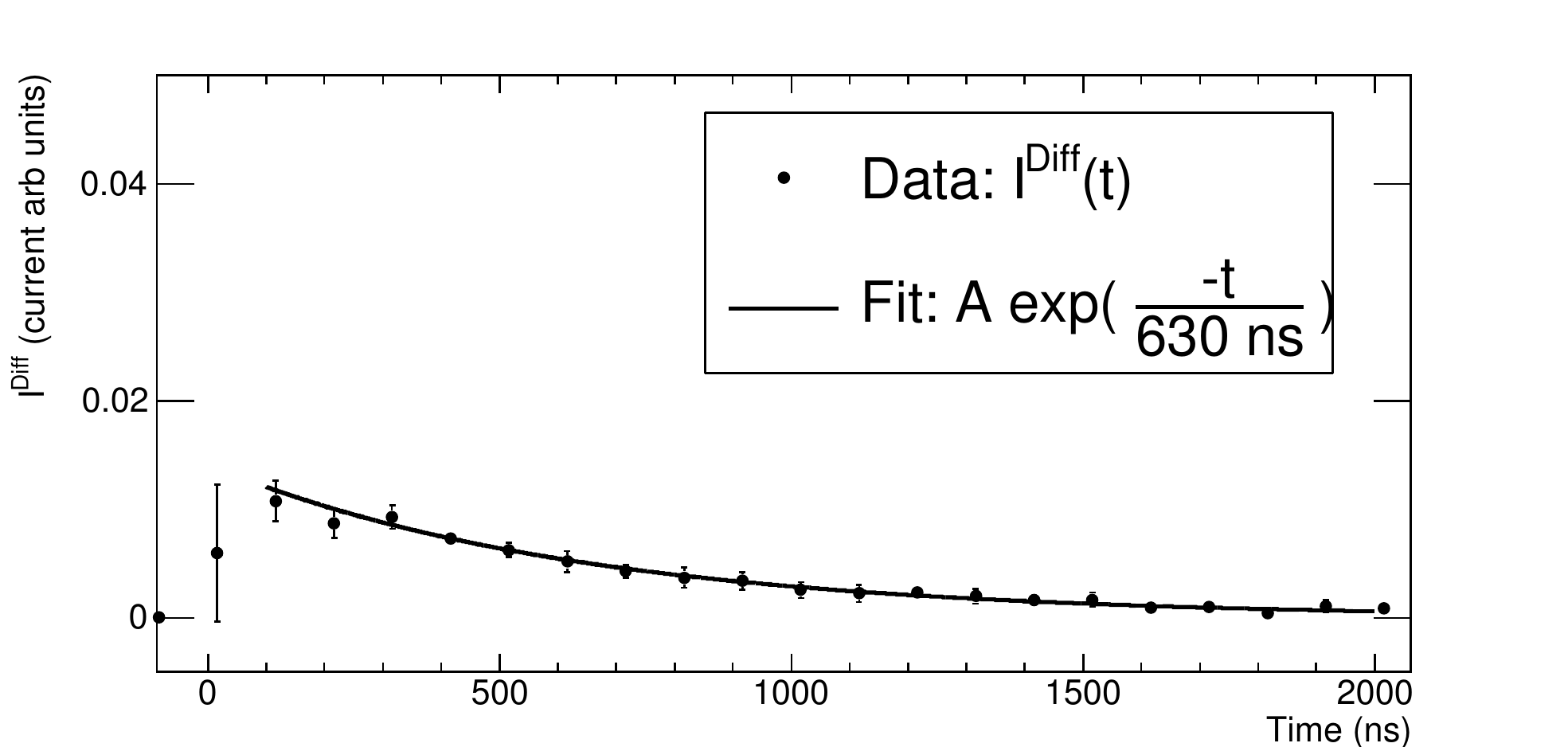}} 
\subfloat[Alpha (R5)]{\includegraphics[width=0.5\textwidth]{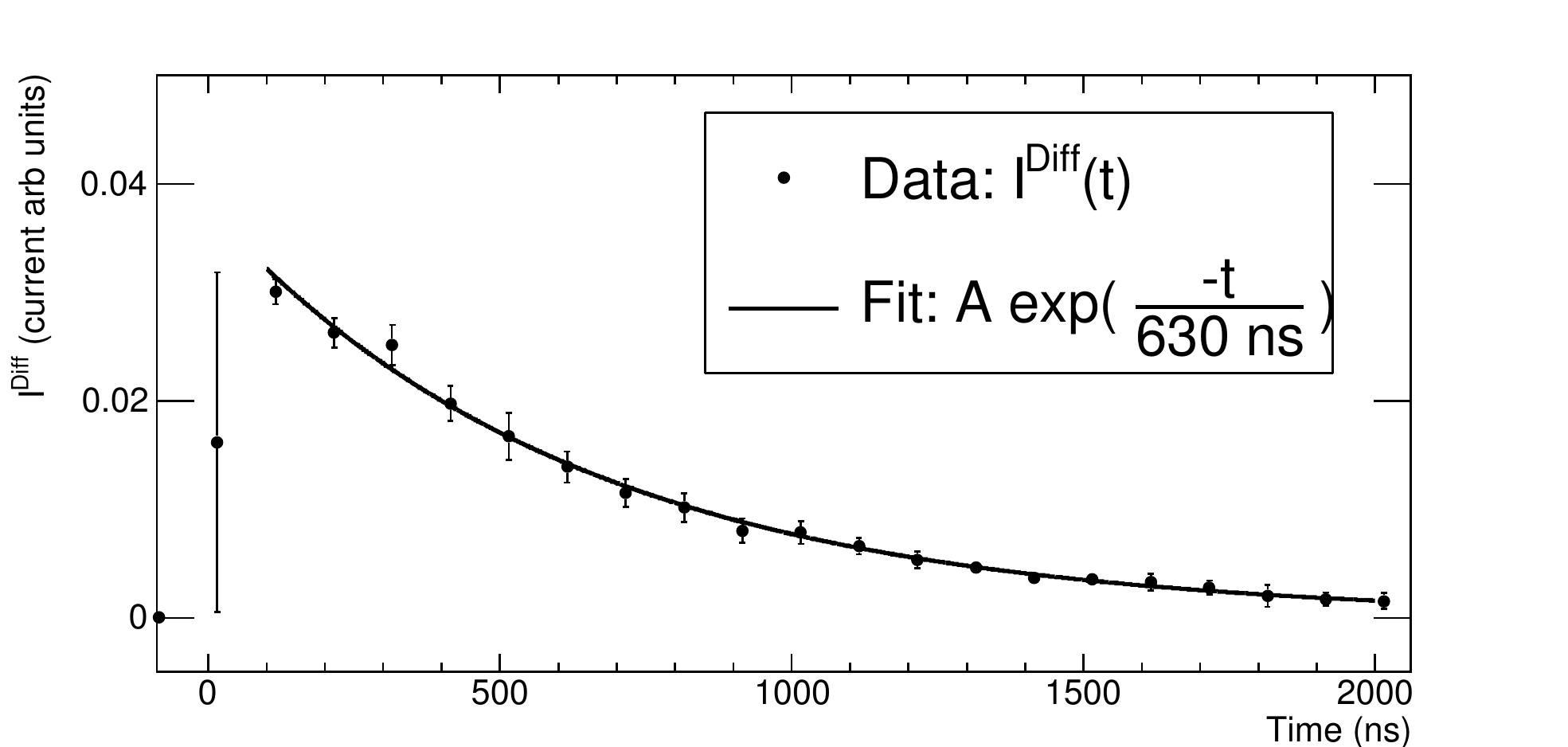}} 
\label{pulse_differences}
\end{figure}
\begin{figure}[H]
\subfloat[High Energy Multi-Hadron (R6)]{\includegraphics[width=0.5\textwidth]{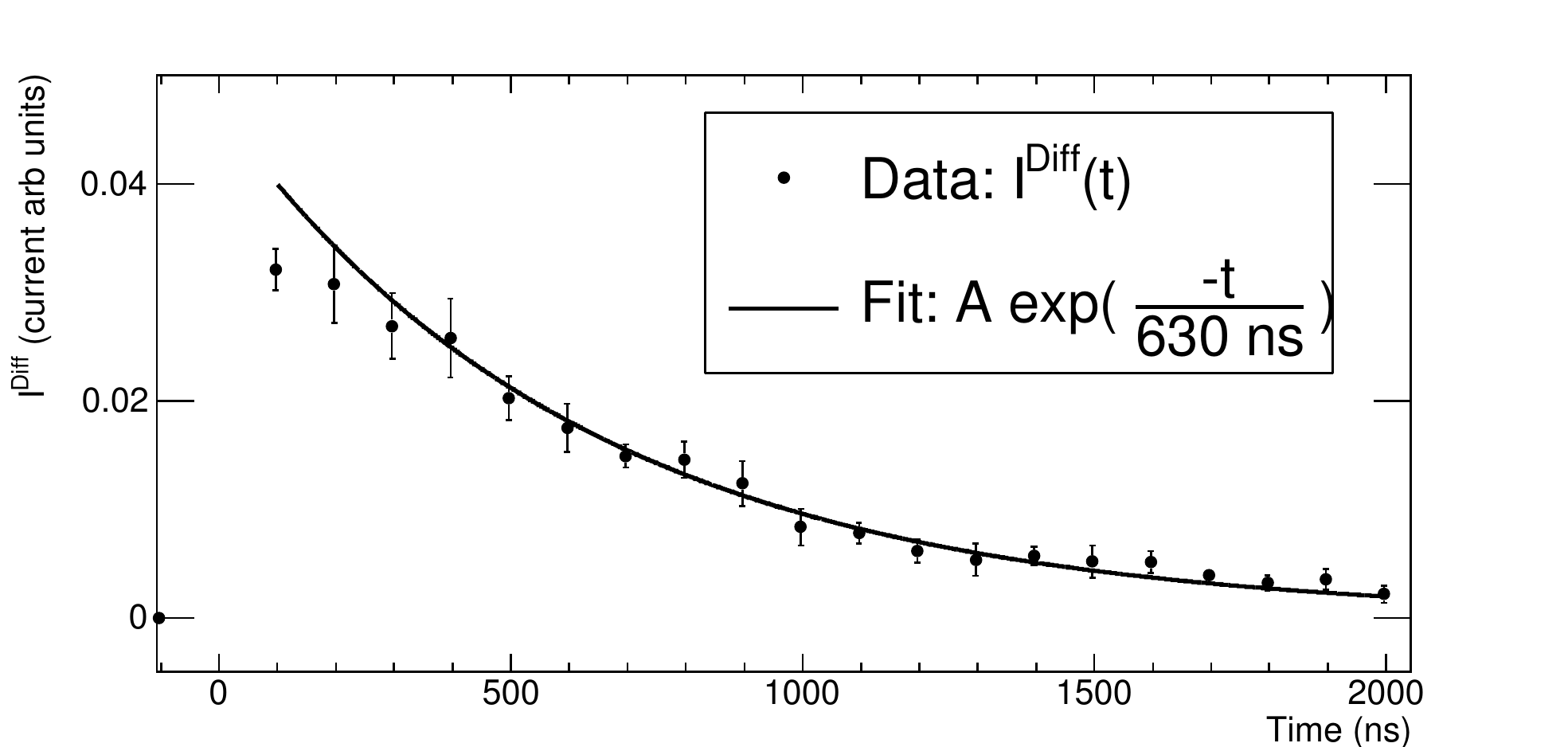}}
\subfloat[High Energy Proton (R7)]{\includegraphics[width=0.5\textwidth]{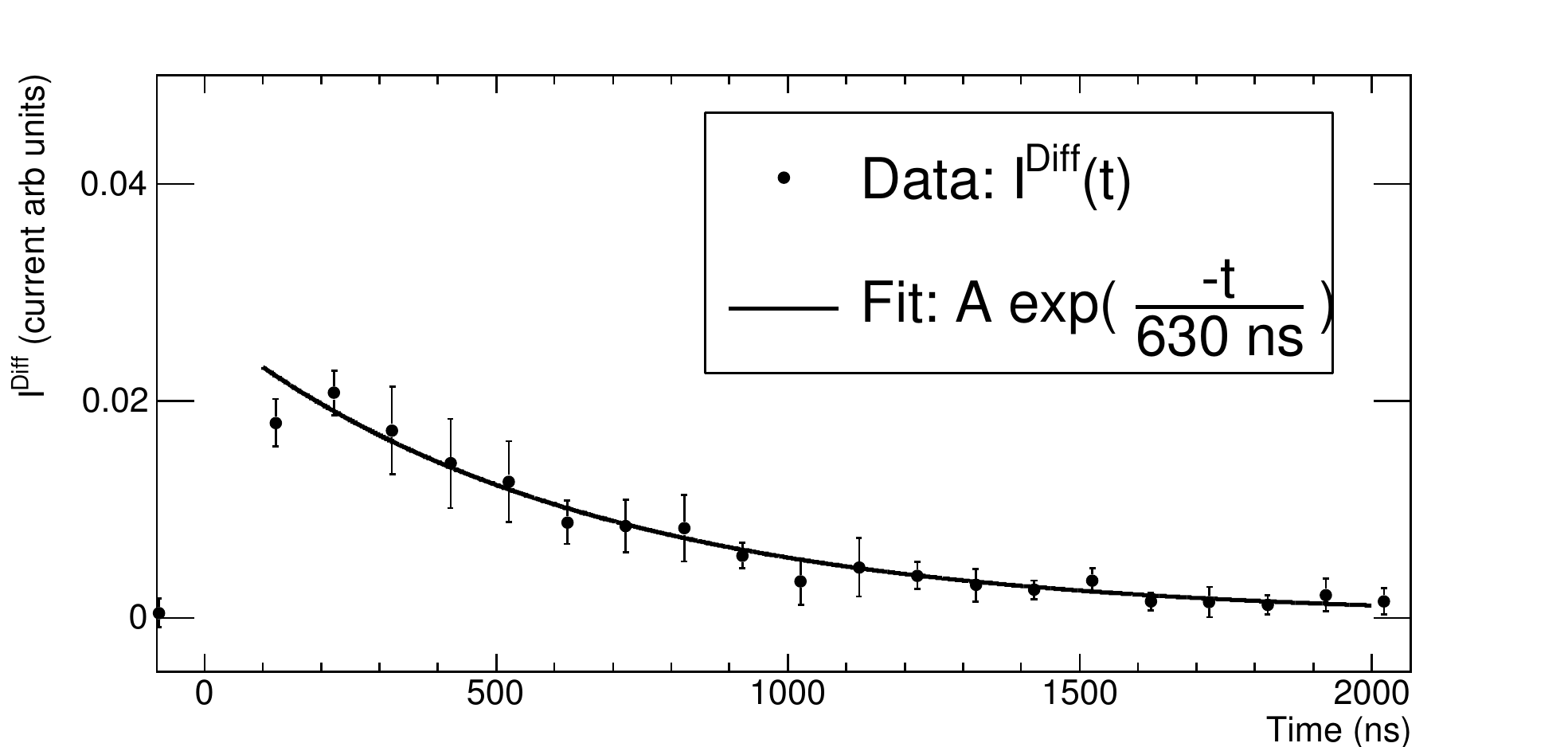}} 
\caption{Plots of $I^\text{Diff}(t)$ for the region's R2-R7 indicated Figure \ref{ChargeRatioPMTrun0}.}
\label{pulse_differences}
\end{figure}

\subsection{Hadron Scintillation Component Model}
\label{HadronComponentModel}

We incorporate this hadron scintillation component into a new three component model for \csi{} scintillation emission defined by equation \ref{ThreeCompModel}.  We note that although this model extends the number of scintillation components for \csi{}, the only free parameter of the model is the intensity of the hadron scintillation component defined as $\text{N}_\text{Hadron}$ in equation \ref{nhadrons} which, in Section \ref{simtech}, we will show depends on the ionization \dedx{} and therefore the type of particle and energy it deposits via ionization loss.  All of the remaining pulse shape parameters are fixed to the template photon parameters in Table \ref{GammaParameters} and \tauh{}=630 ns. 

\begin{equation}
\label{ThreeCompModel}
Q^\text{Hadron Component Model}(t) =  Q_\gamma(t) + \text{L}_\text{Hadron} (1- e^\frac{-t}{\tau_\text{Hadron}})
\end{equation} 

\begin{equation}
\label{nhadrons}
\text{N}_\text{Hadron} = \frac{\text{L}_\text{Hadron}}{\text{L}_{\gamma} +  \text{L}_\text{Hadron}} = \frac{\text{L}_\text{Hadron}}{ \text{L}_\text{Total}}
\end{equation} 

\noindent Where $\text{L}_\text{Hadron}$ is the total scintillation emission of the hadron scintillation component in units of photon equivalent energy.

\begin{figure}[h]
\centering
\subfloat[Alpha particle]{\includegraphics[width=0.5\textwidth]{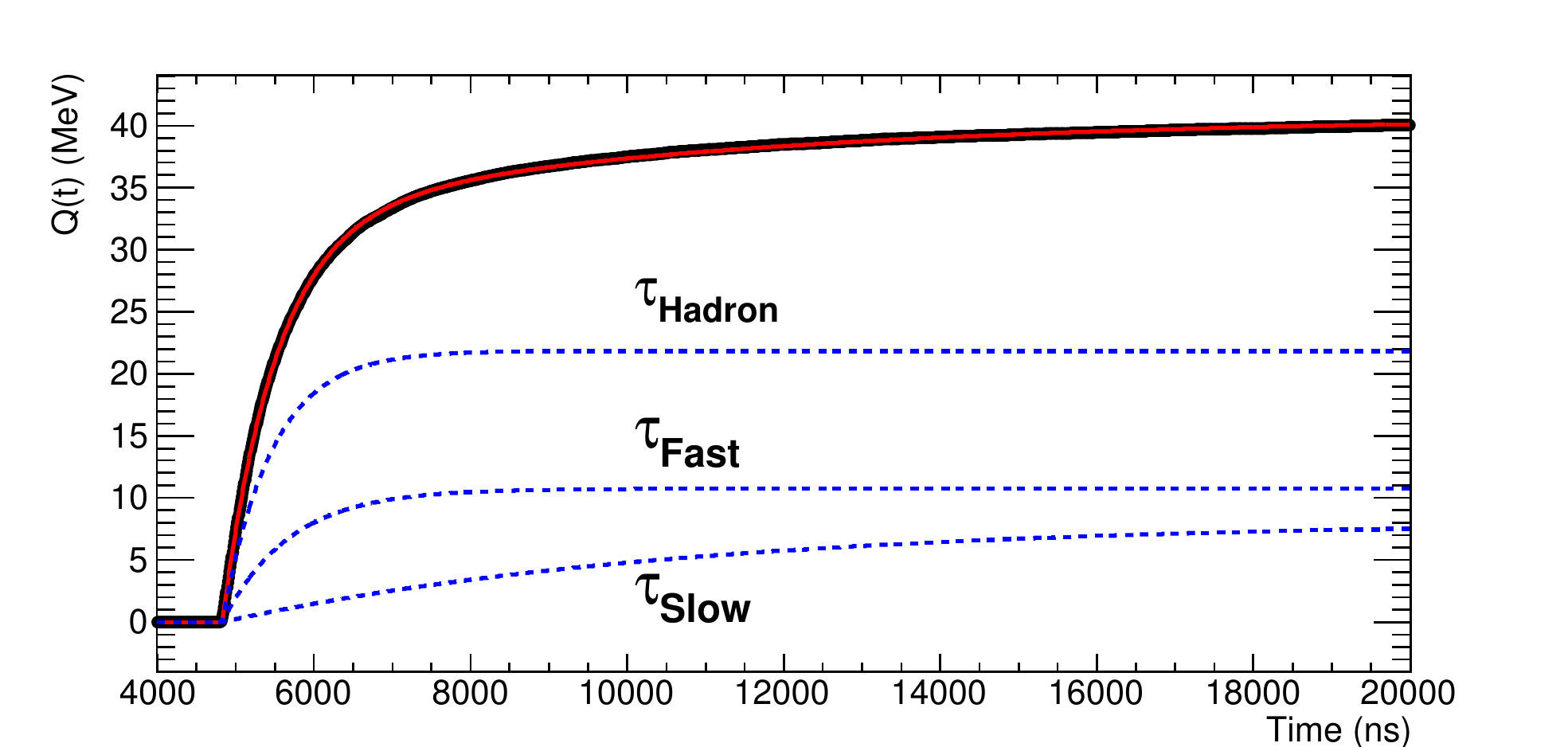}} 
\subfloat[Muon]{\includegraphics[width=0.5\textwidth]{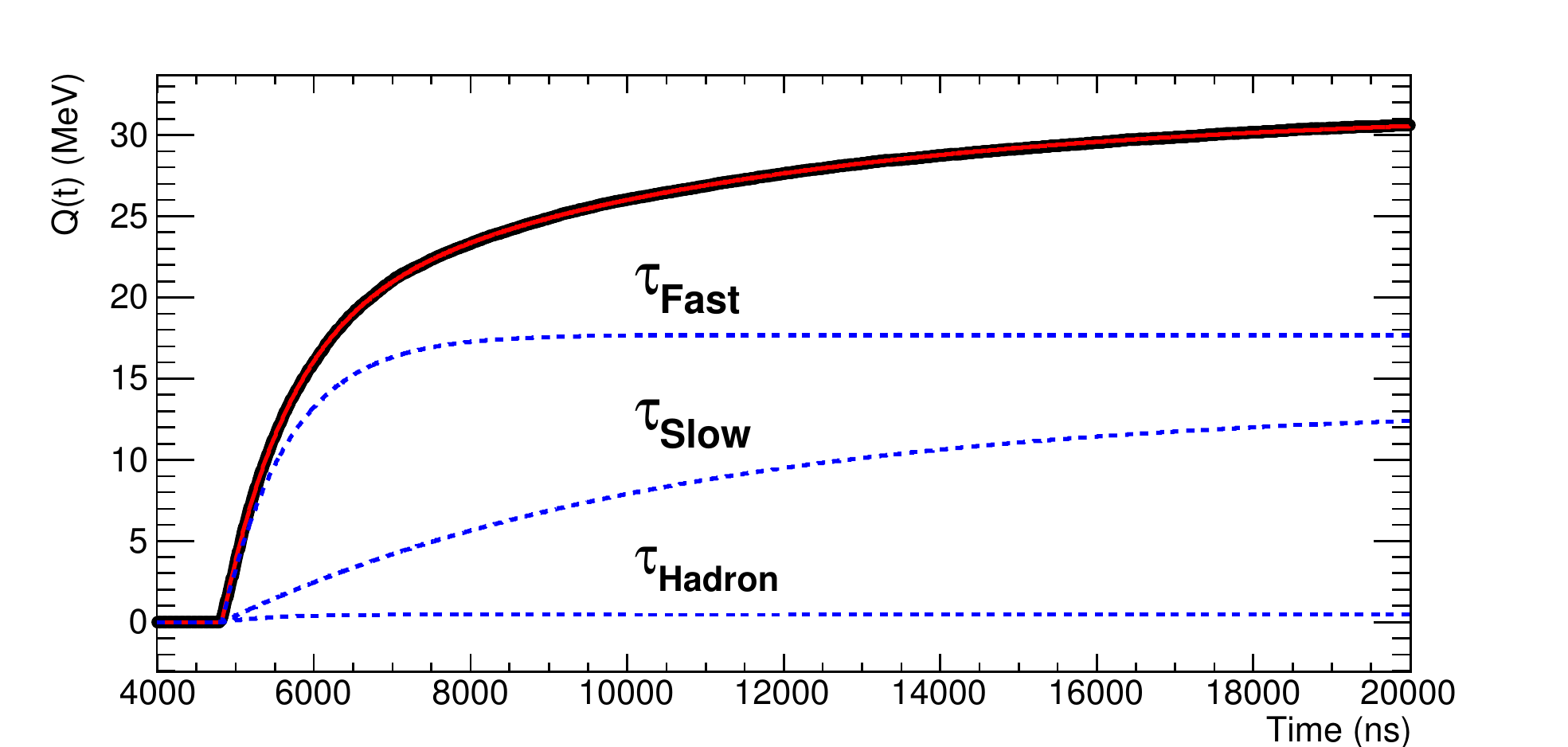}}  
\caption{Sample fits using hadron component model equation \ref{ThreeCompModel}, for typical alpha and muon pulses with pulse amplitudes of 30-40 MeV.  The contributions from the different scintillation components is shown.}
\label{chargeFits}
\end{figure}

To test this model, the PIF neutron data is re-analysed by fitting the charge pulses to equation \ref{ThreeCompModel} and extracting $\text{N}_\text{Hadron}$ for each pulse.  Typical fit results for an alpha and muon pulse shape are shown in Figure \ref{chargeFits} with the three scintillation components overlaid.

\begin{figure}[h]
\centering
\subfloat{\includegraphics[width=0.5\textwidth]{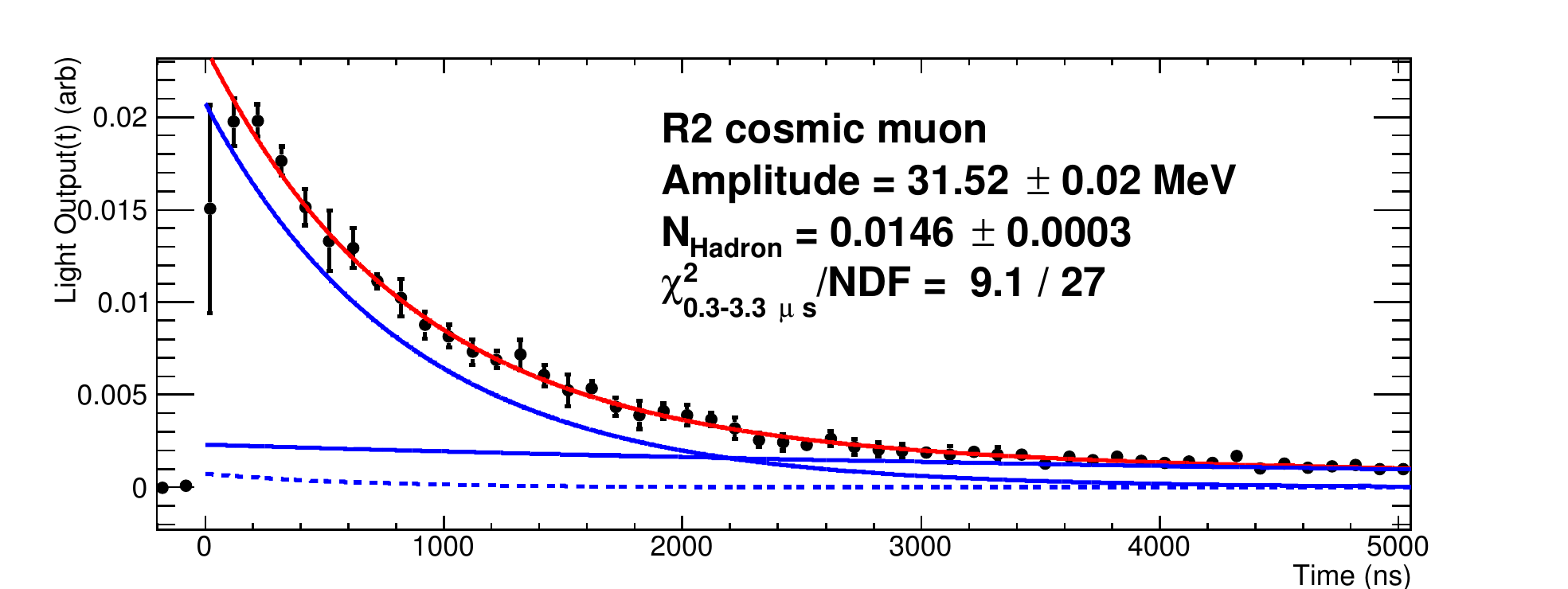}} 
\subfloat{\includegraphics[width=0.5\textwidth]{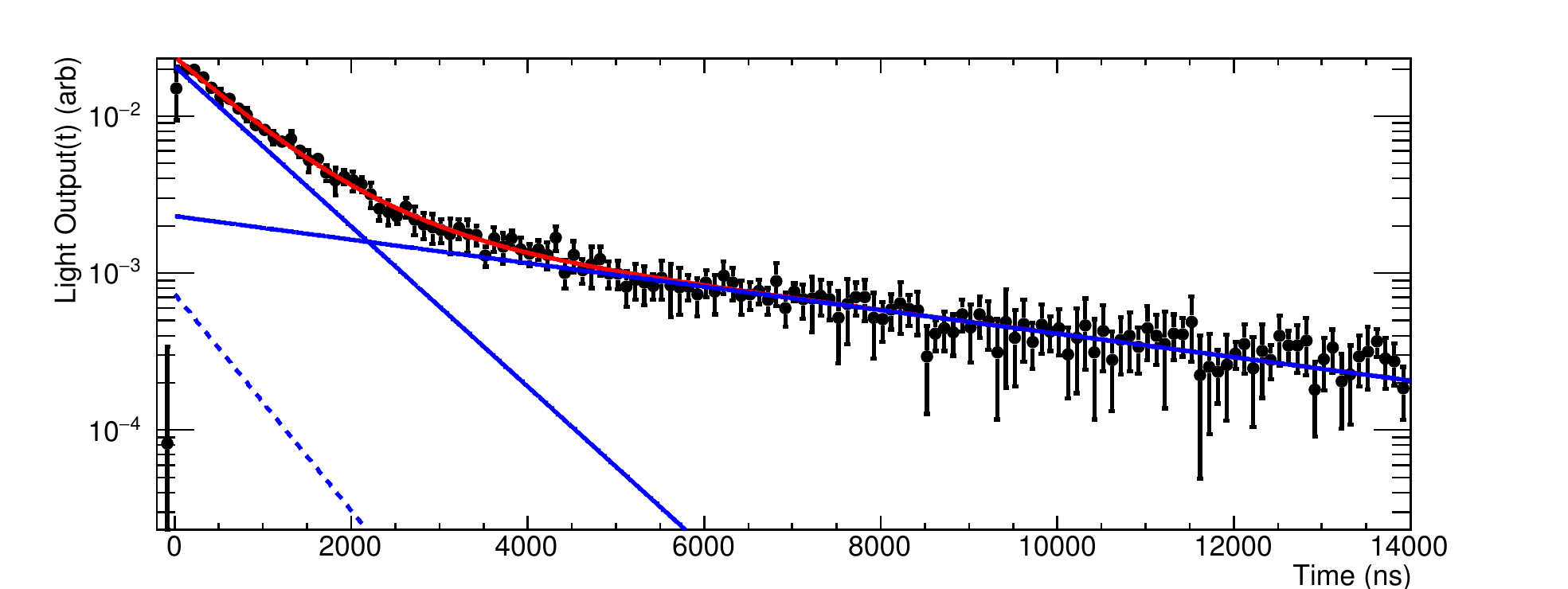}} 
\vspace{0pt}
\subfloat{\includegraphics[width=0.5\textwidth]{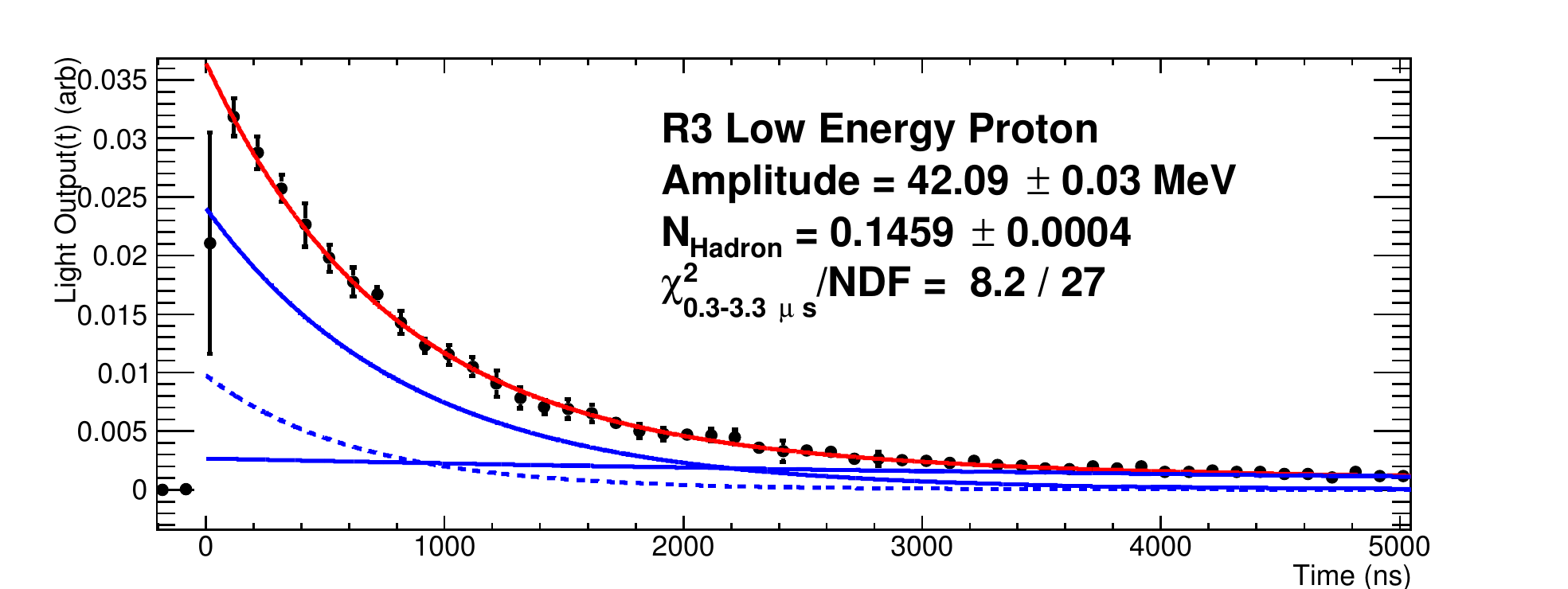}} 
\subfloat{\includegraphics[width=0.5\textwidth]{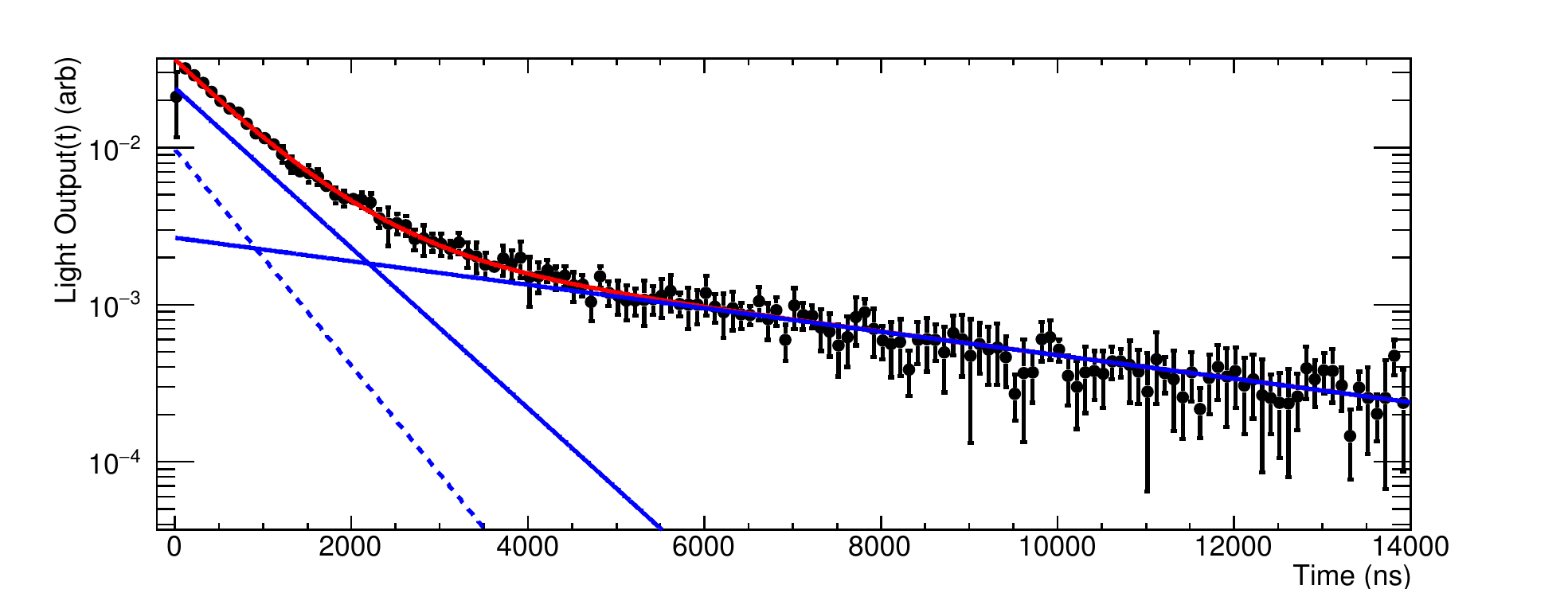}} 
\vspace{0pt}
\subfloat{\includegraphics[width=0.5\textwidth]{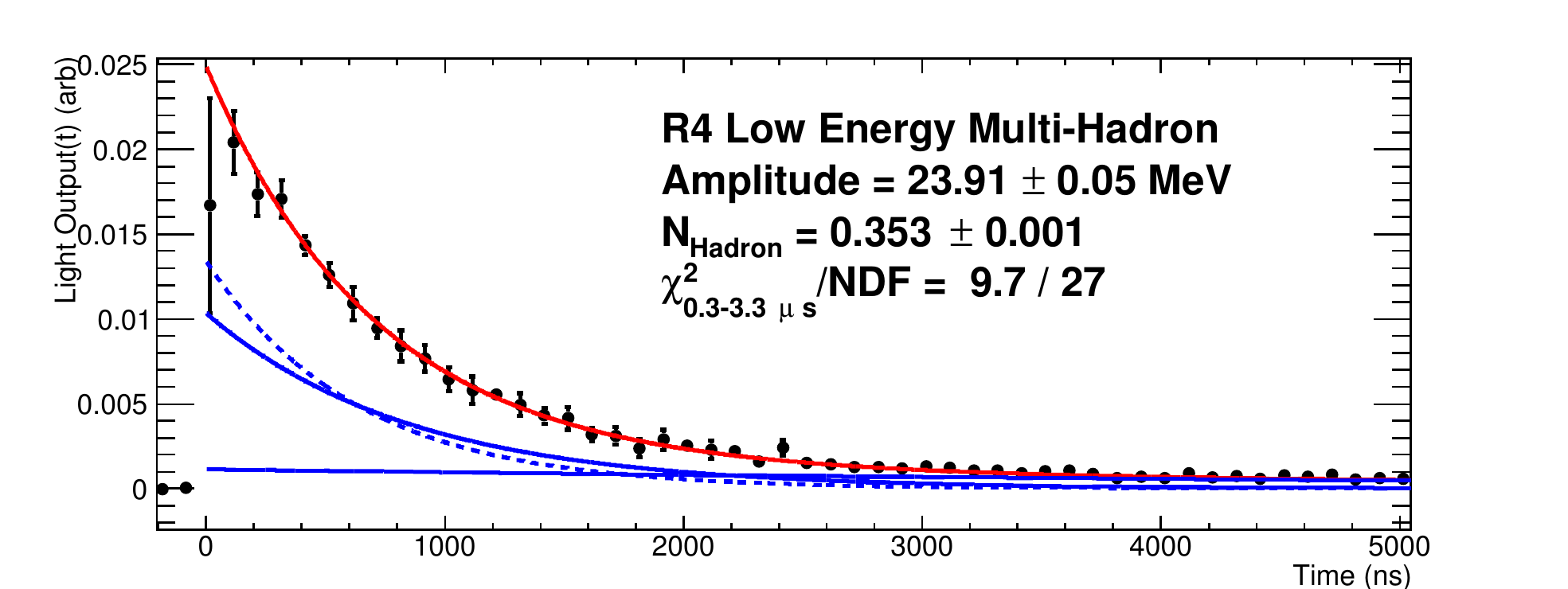}} 
\subfloat{\includegraphics[width=0.5\textwidth]{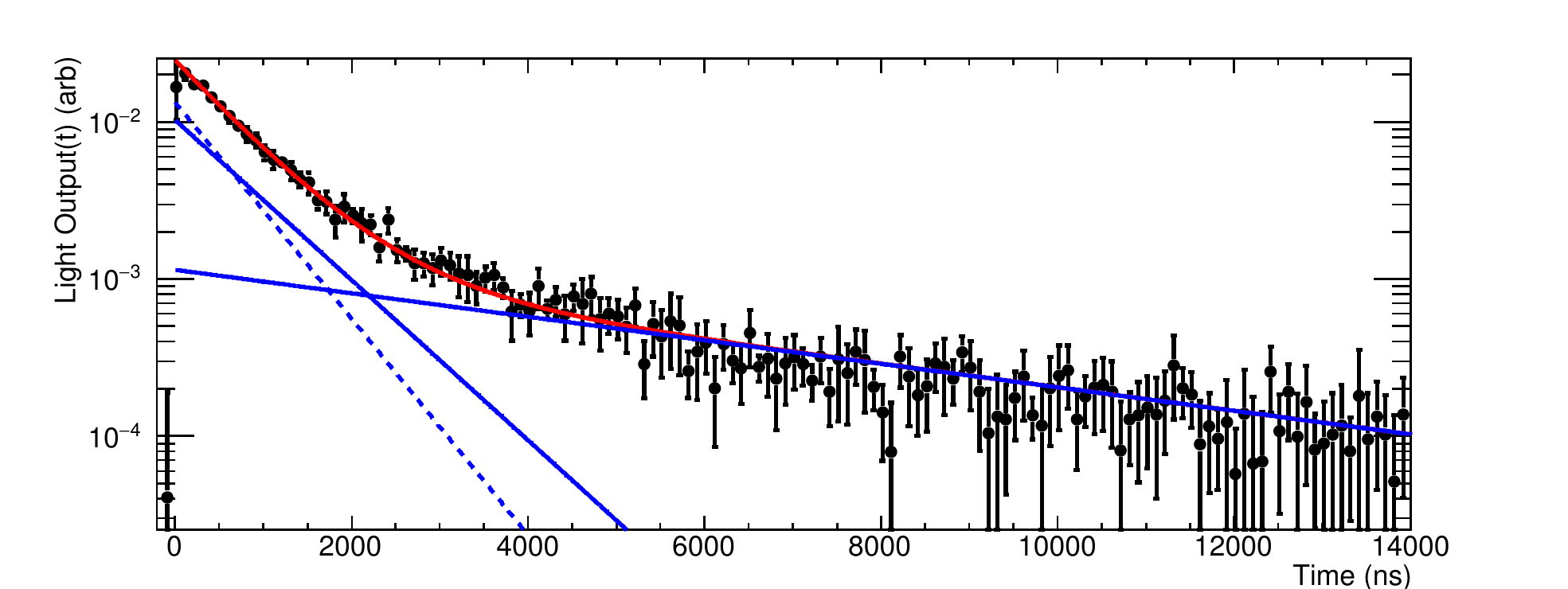}} 
\vspace{0pt}
\subfloat{\includegraphics[width=0.5\textwidth]{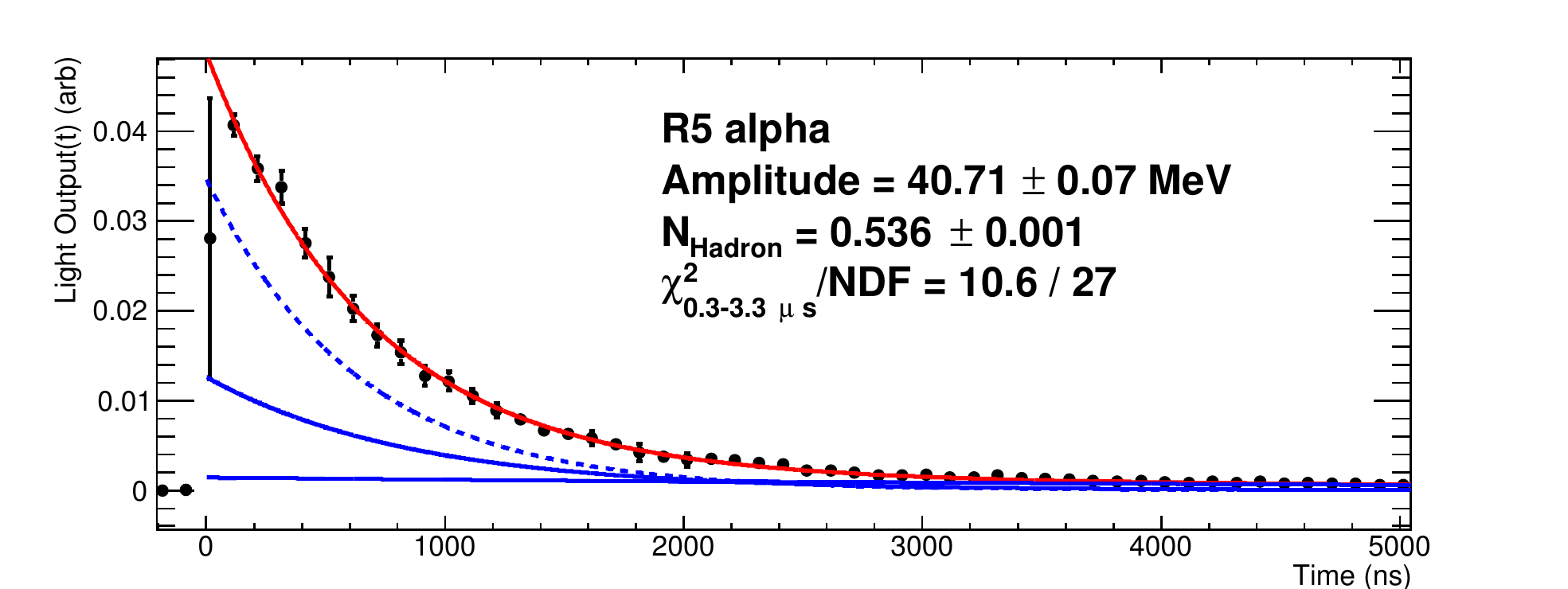}} 
\subfloat{\includegraphics[width=0.5\textwidth]{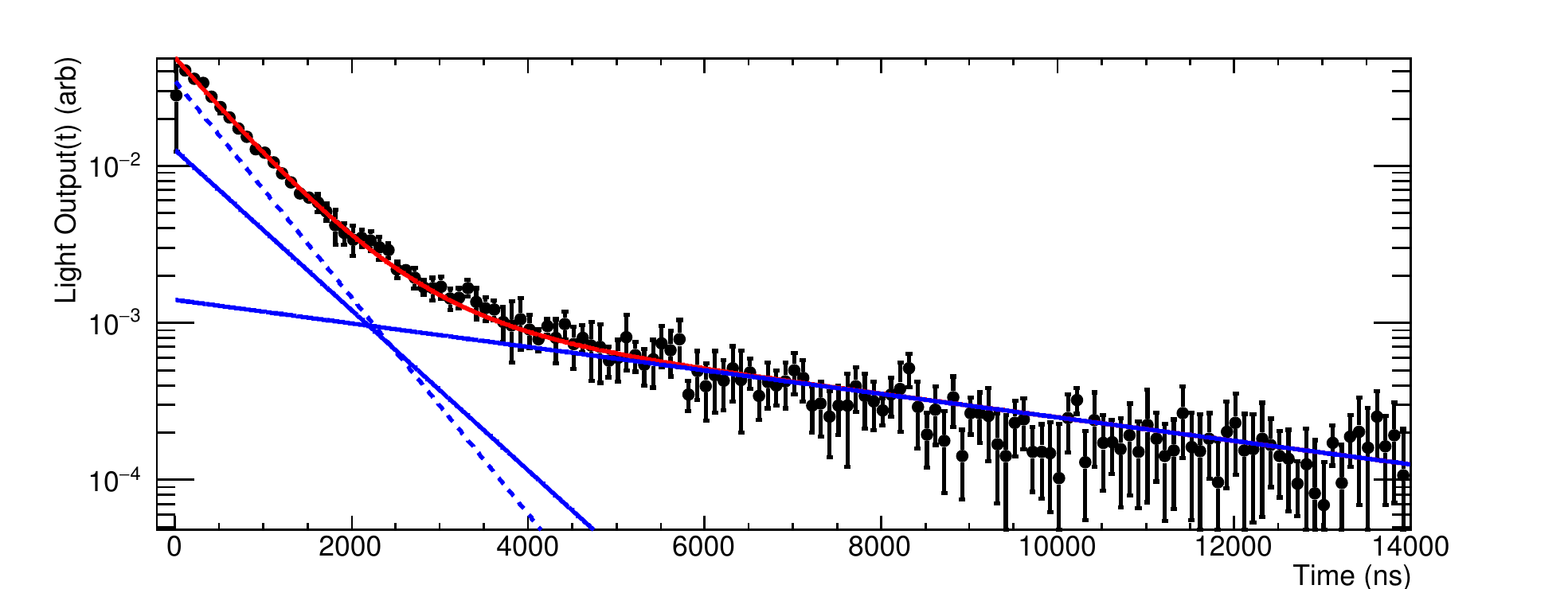}} 
\vspace{0pt}
\subfloat{\includegraphics[width=0.5\textwidth]{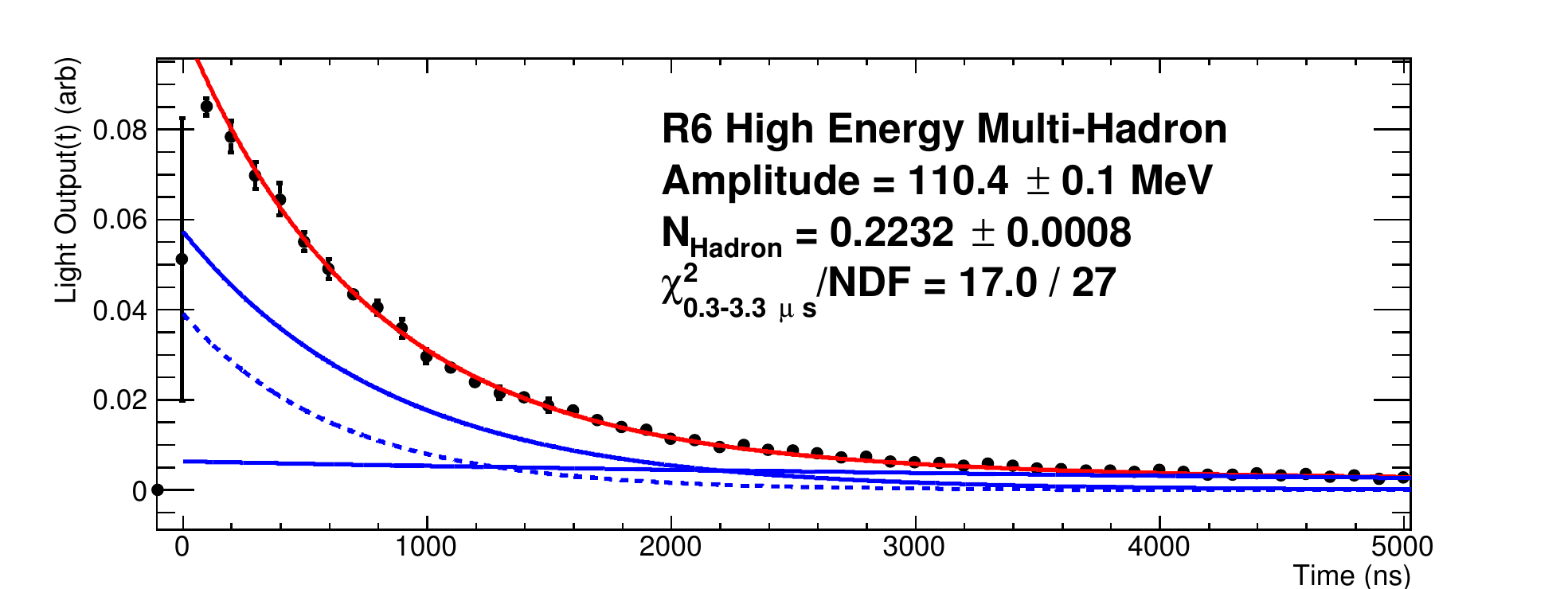}} 
\subfloat{\includegraphics[width=0.5\textwidth]{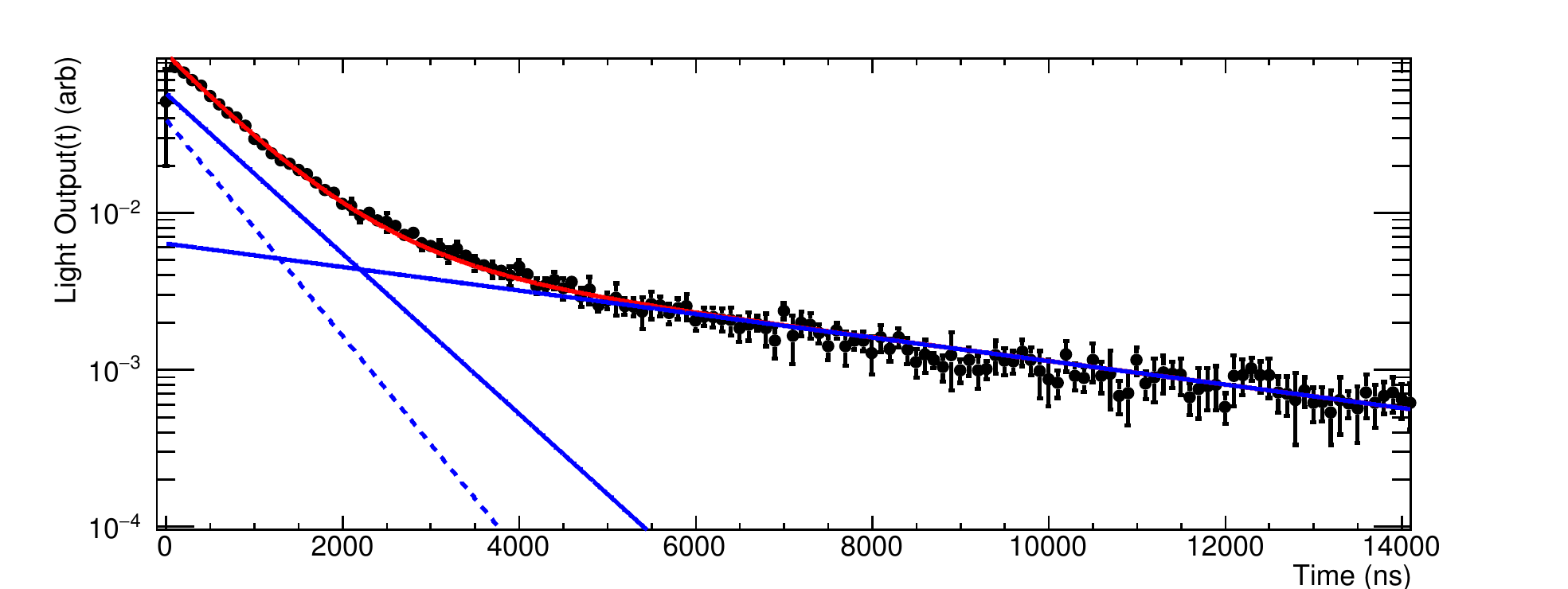}} 
\vspace{0pt}
\subfloat{\includegraphics[width=0.5\textwidth]{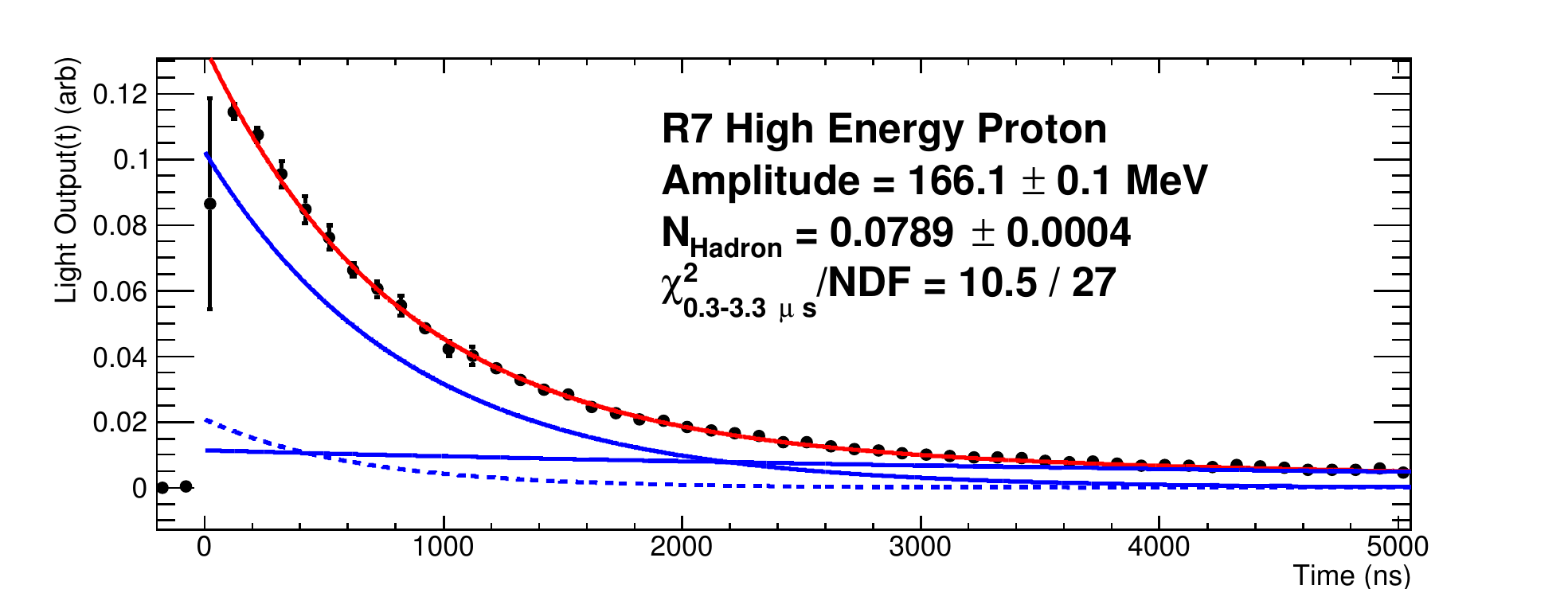}} 
\subfloat{\includegraphics[width=0.5\textwidth]{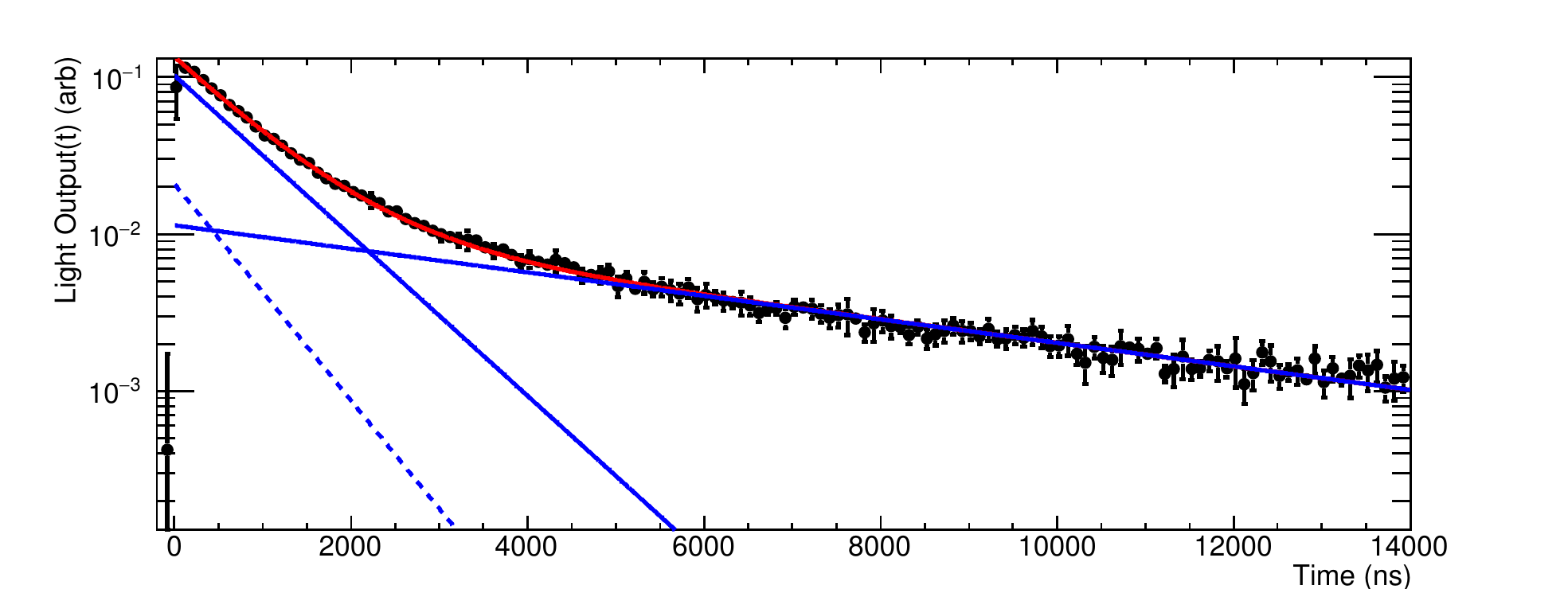}} 
\caption{The current-form of typical fit results for pulse region's R2-R7 indicated Figure \ref{ChargeRatioPMTrun0}.  Fast, slow and hadron components are overlaid.  Hadron component indicated by dashed line.  The plots on the right use a vertical log scale and expand the time from 5 $\mu$s (used for the plots on the left) to 14 $\mu$s to show the tail region of pulses.}
\label{SampleFits}
\end{figure}

As shown in Figure \ref{chargeFits} we use the charge form of the pulse to fit for the hadron intensity as by integrating the current-form we can reduce noise present in the raw current pulse. In order to evaluate the fit results, however, we use the current-form to avoid the correlations introduced when computing the charge pulse form.   The current-form of pulses with the fit result overlaid are shown in Figure \ref{SampleFits} for a typical pulse for each of the pulse shape regions 2-7.   From these plots we show visually the pulse decay shape for all pulse types is fully described up to $14 \mu $s by the addition of a 630 ns exponential at various intensities.

\begin{figure}[h]
\centering
\includegraphics[width=0.65\textwidth]{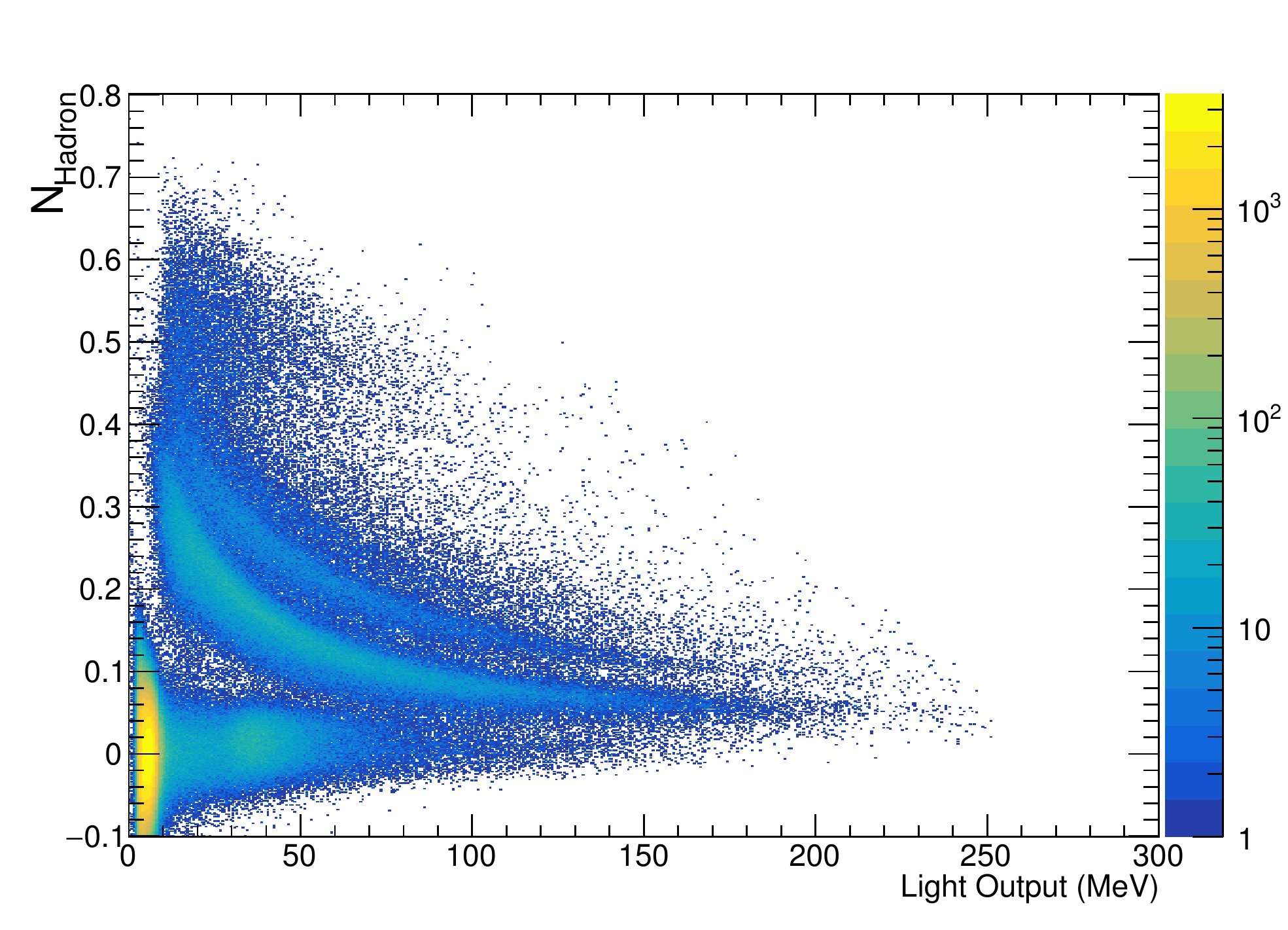}
\caption{Hadron scintillation component intensity vs pulse amplitude spectrum of PIF neutron data.  Light output is in units of equivalent photon energy.}
\label{PMT_PSDINTENSITYvsENE}
\end{figure}

In Figure \ref{PMT_PSDINTENSITYvsENE} we plot the $\text{N}_\text{Hadron}$ vs pulse amplitude and observe that the same band structures arises as the $R_\text{PSD}$ vs $Q(7.4 \mus)$ plot in Figure \ref{ChargeRatioPMTrun0}.  This demonstrates the pulse shape variations can be described by the parameter $\text{N}_\text{Hadron}$ and can be interpreted as originating from the hadron scintillation component.   In agreement with this hypothesis, the intensity of the hadron scintillation component is zero for the cosmic muon energy deposits, as well as those of photons, and increases for the hadron deposits with alpha pulses having up to 70\% of the scintillation emission in the hadron component. 

\begin{figure}[h]
\centering
\includegraphics[width=0.5\textwidth]{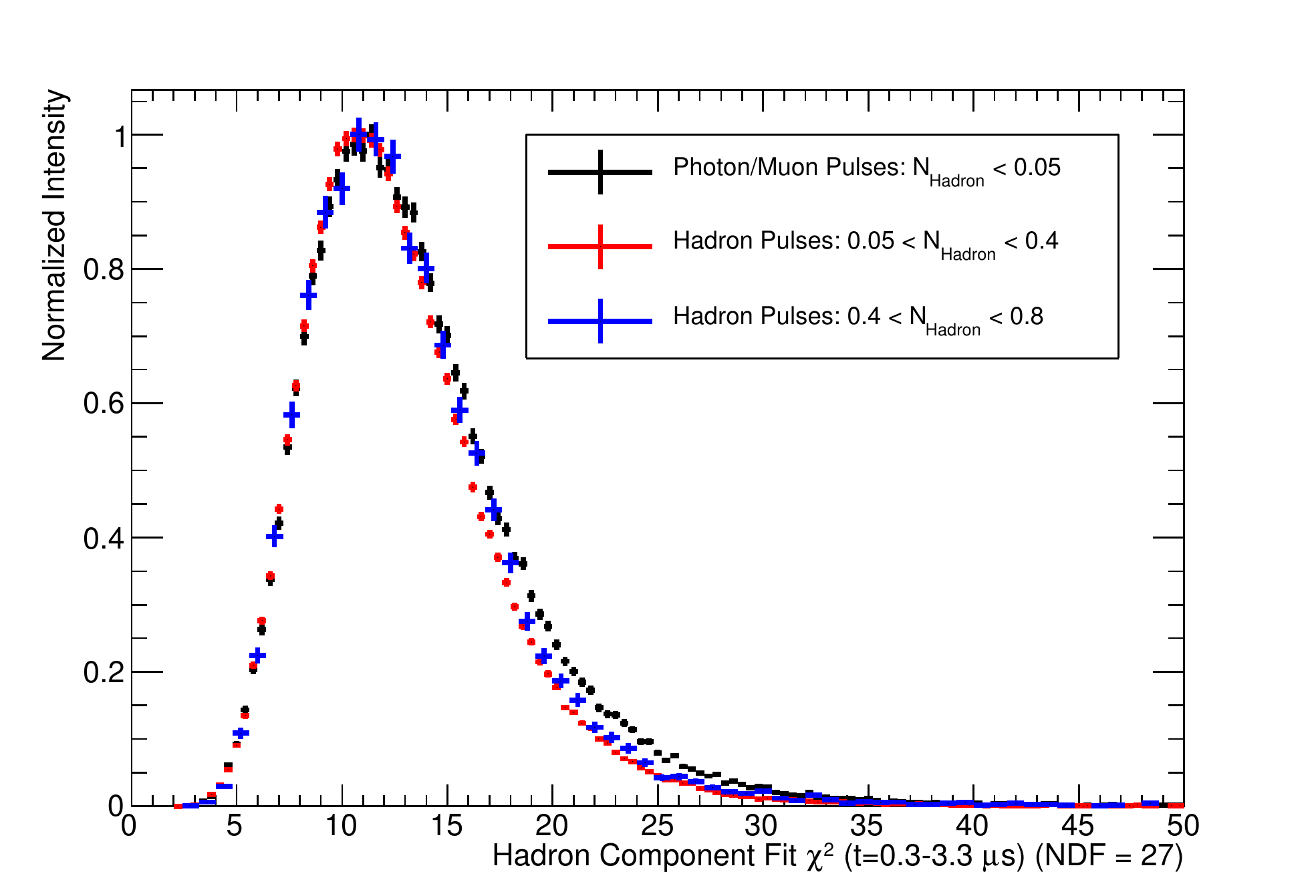}
\caption{Fit $\chi^2$ histograms for electromagnetic and hadronic pulse shapes for pulses with amplitude greater than 10 MeV.  The number of degrees of freedom (NDF) for the fit is 27.  Note as discussed in Section \ref{HadronComponentModel} the errors used for computing the $\chi^2$ are conservatively estimated as correlations within the time bin have not been taken into account, which produces a lower than expected $\chi^2$.}

\label{FitQualitya}
\end{figure}

To further demonstrate that all pulse shapes in the PIF neutron sample are well described
by the hadron component model we compute the $\chi^2$ in the pulse regions of 300 -
3300 ns where the hadron component has the highest impact in the pulse shape. We restrict the calculation of the fit quality to begin after the initial 300 ns of the pulse in order to avoid systematic effects such as the ultra-fast component and the pulse rise time, which are not included in the model as they contribute a small percentage of total charge. In order to evaluate the goodness-of-fit of the hadron component model we use the current-form of the scintillation pulse binned in 100 ns wide bins. The error of the amplitude in each bin is computed from the standard deviation of the points in the bin, which characterizes the electronic and statistical noise present in the pulse, recognizing that the points within the bin are not independent due to correlations. This conservative approach leads to an overestimate of the errors used in the $\chi^2$ calculation. As a result we focus on the relative comparison of the $\chi^2$ distributions for the photon and hadron pulses shapes as we treat all shapes consistently. In addition sample fit results for the typical waveforms are presented in Figure \ref{SampleFits} to visually show how the fit results compare with the data. In Figure \ref{FitQualitya} we divide the pulses with amplitude greater than 10 MeV into different pulse shape regions and plot the $\chi^2$ for these pulses where the number of degrees of freedom is 27 (30 bins and 3 parameters). From this we find that the model defined in equation \ref{ThreeCompModel} describes both the hadron and photon pulse shapes equally well. We note that this model does not include delayed hadronic interactions which would result in out of time pile-up pulses and yield a high $\chi^2$. From Figure \ref{FitQualitya} it can be seen that such effects are small and justifiably neglected.

\subsection{Discussion of Hadron Component Model}

By considering the pulse shape differences between photon and hadron energy deposits we have demonstrated that the pulse shape variations for \csi{} can be characterized using a third scintillation component with decay time of $630\pm10$ ns.  Using this model, the pulse shapes for \csi{} are characterized by the parameter $\text{N}_\text{Hadron}$ defined as the scintillation emission intensity of the hadron scintillation component.  This single parameter pulse shape description is advantageous compared to present approaches for pulse shape characterizing techniques for \csi{} where the four shape parameters ($\tau_\text{fast}$,$\tau_\text{slow}$, $N_\text{fast}$ and $N_\text{slow}$) of the two component scintillation model are varied to describe the \csi{} pulse shape spectrum as done in references \cite{Benrachi} and \cite{Amorini}.  In addition these studies have shown that low energy hadron energy deposits, which result in the largest pulse shape difference from photons, will result in fast time constants in the range of approximately 600-650 ns  when fit to the two component model \cite{Amorini}. This is in agreement with our model where it is expected that the $\tau_\text{Hadron}$ component would dominate the scintillation emission for these pulses and thus a pulse shape description using a two component scintillation model is expected to produce a fast time constant consistent with our hadron time constant. 

In Section \ref{simtech} we show that the instantaneous hadron scintillation component emission intensity can be computed from the ionization energy loss of the interaction particle.  A detailed analysis of the mechanism resulting in the $630\pm10$ ns scintillation component for only high \dedx{} energy deposits is beyond the scope of this paper however we note that the magnitude of this decay time is close to the $575 \pm 5$ ns thallium centre lifetime measured by reference \cite{Hamada} by observing the single component scintillation emission of \csi{} when exposed pulsed UV light.  Considering possible systematic effects such as temperature variations between experimental setups, it is possible that this is the same decay component we observe in the hadronic pulse shapes.  

\section{Hadron Component Model Applied to Proton Data}
\label{protondata}

In this section we apply the hadron component model to the proton testbeam data collected in addition to the PIF neutron run studied in the previous section.  The proton beam data was collected at kinetic energies of 20.0, 40.1, 57.7 and 67.0 MeV corresponding to proton momenta of 0.194, 0.277, 0.344 and 0.360 GeV/c, respectively.    For the following proton data runs discussed in this section fit quality cuts were applied to remove out of time pile-up pulses.

\subsection{67.0 MeV Proton Data}

\begin{figure}[h]
\centering
\includegraphics[width=0.5\textwidth]{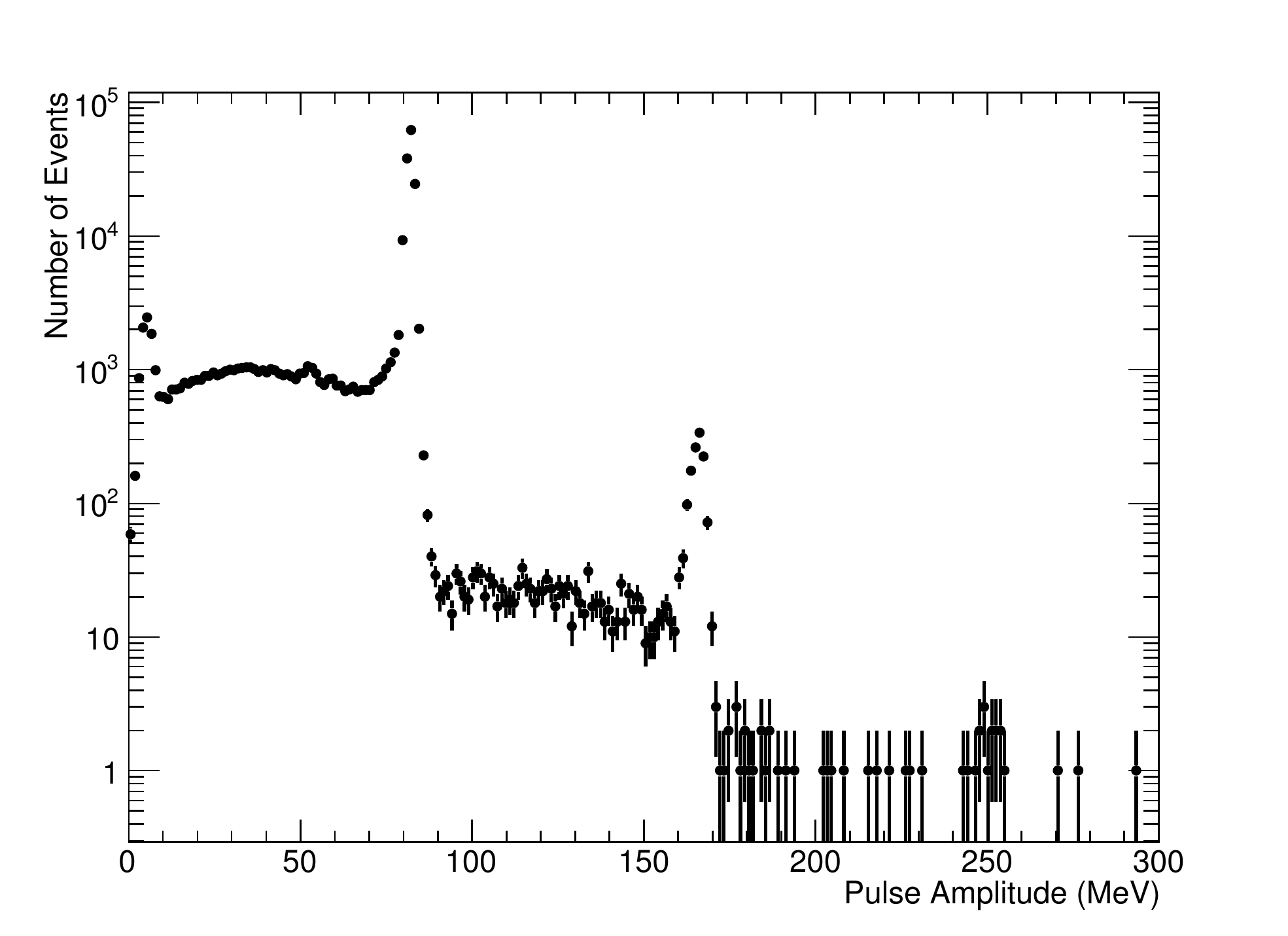}
\caption{Energy deposited spectrum for recorded pulses in 67.0 MeV proton run data.  Note that the peak of the pulse amplitude spectrum for the protons, presented in electron equivalent energy units, is higher than the proton kinetic energy due to the Birk's scintillation efficiency of CsI(Tl) for high $\frac{dE}{dx}$ energy deposits \cite{Gwin,Koba,Zeitlin,Pietras}.  The Birk's scintillation efficiency for the protons is discussed further in Section \ref{sectionBdedx}.}
\label{Mono67edep}
\end{figure}

The pulse amplitude spectrum in units of equivalent photon energy for the 67.0 MeV proton run is shown in Figure \ref{Mono67edep}.  Peaks are observed at quantized values in units of total scintillator light output for the full energy deposit of the primary proton kinetic energy.  These peaks correspond to events where single, double and triple coincident protons from the beam simultaneously entered the crystal.  Similar quantization of peaks were also observed in the other proton runs.

\begin{figure}[h]
\centering
\subfloat[67.0 MeV Data.]{\includegraphics[width=0.5\textwidth]{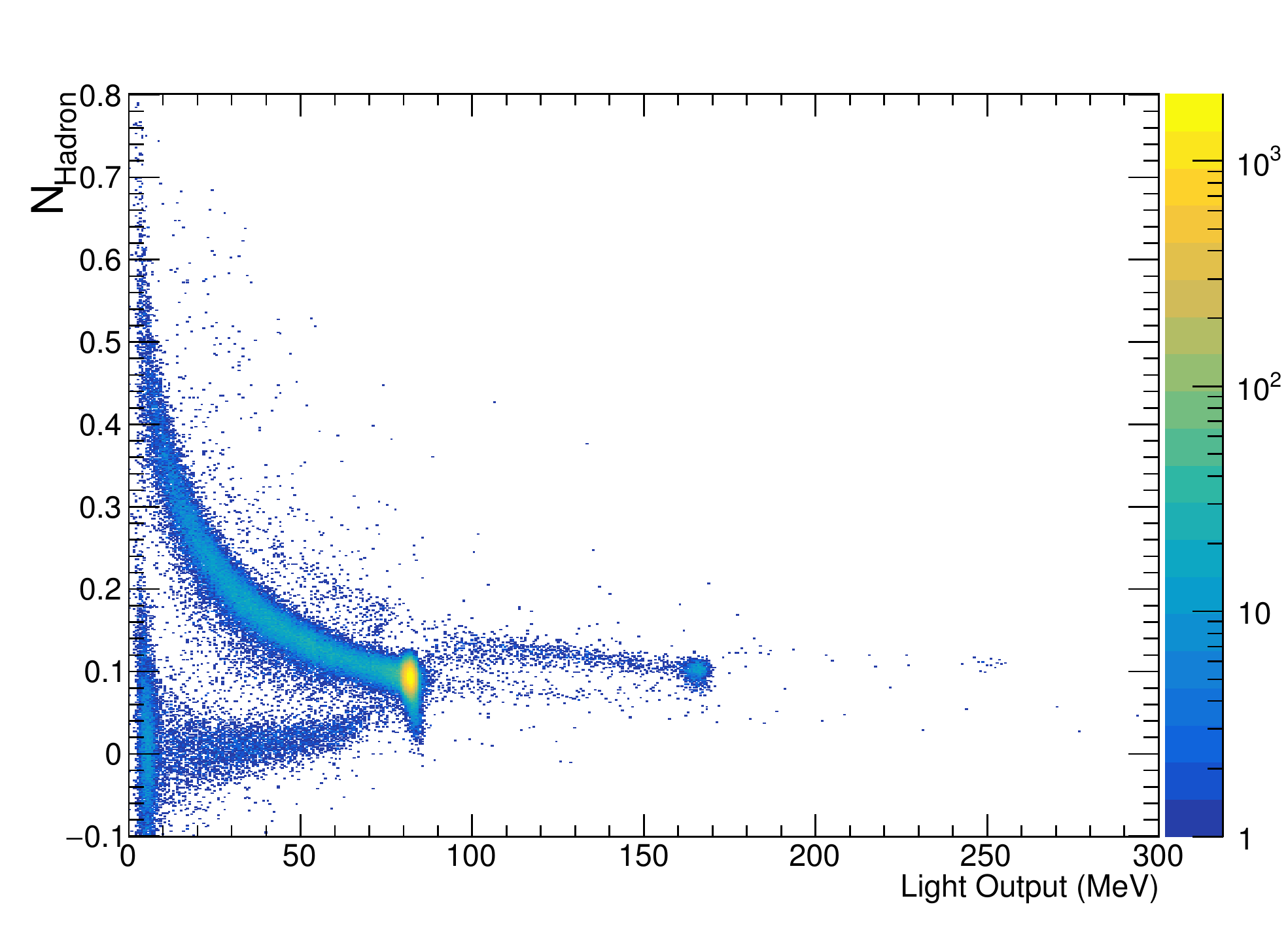}\label{Mono67a}} 
\subfloat[Simulation Truth.]{\includegraphics[width=0.5\textwidth]{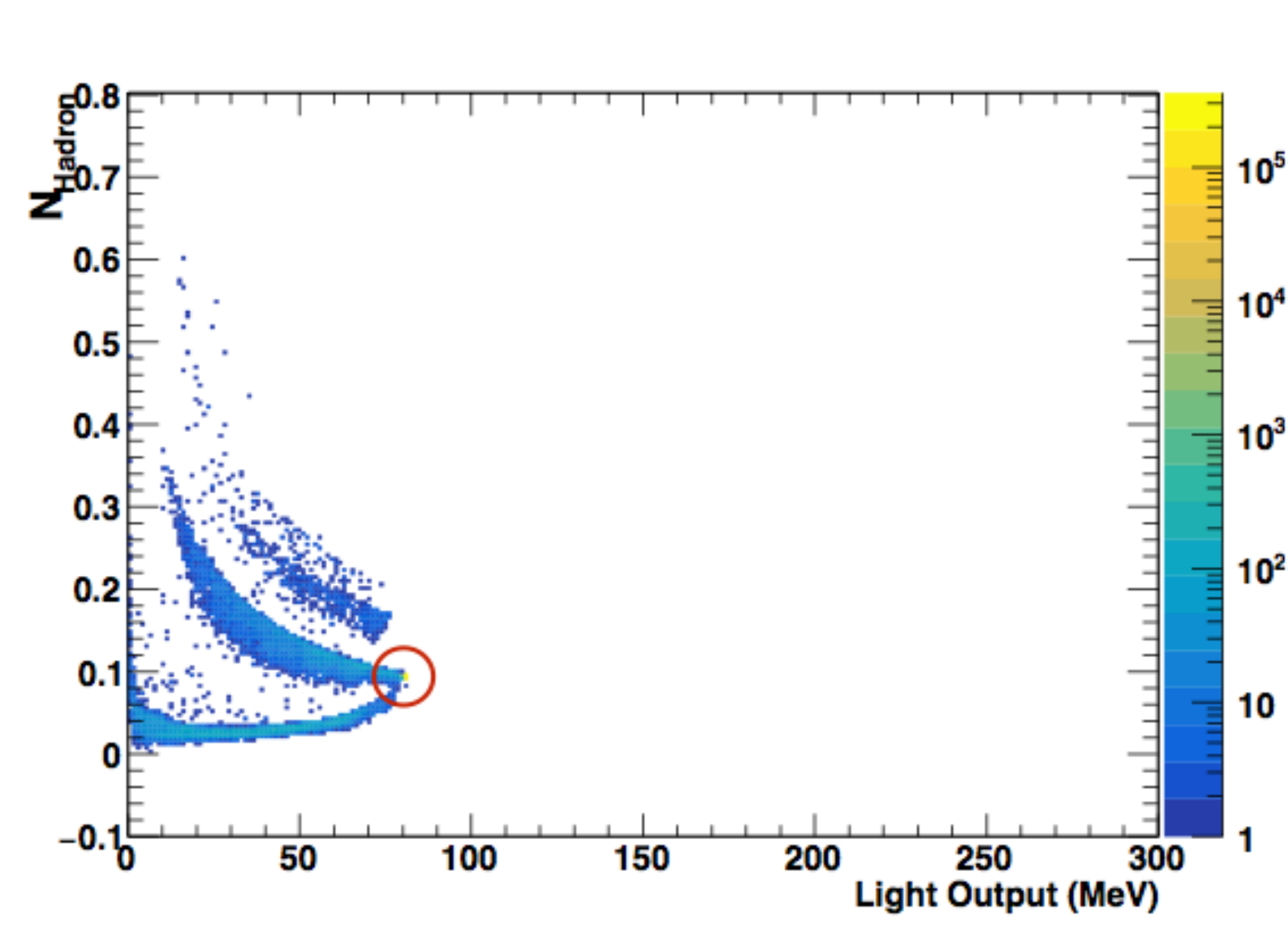}\label{Mono67b}} 
\caption{a) Data and b) Simulation truth plots of $\text{N}_\text{Hadron}$ vs pulse amplitude pulse shape spectrum for 67.0 MeV proton run.  Note in the simulation truth (b) results, a large number of events from the proton ionizing and stopping in the detector are contained in one histogram bin (circled in red) centred at (80.5 MeV, 0.0935) consistent with the position of the peak in the data plot, (a).  Pulse shape and amplitude simulation techniques are described in Section \ref{simtech}.}
\label{Mono67}
\end{figure}

The $\text{N}_\text{Hadron}$ vs pulse amplitude pulse shape spectrum for the 67.0 MeV proton run data is shown in Figure \ref{Mono67a}.  The pulses corresponding to full energy deposition from the primary 67.0 MeV protons result in pulse shapes with approximately 10\% contribution from the hadron scintillation component. Other features in the spectrum are observed in the continuum events below the main proton peak.  As was done with the neutron pulse shape spectrum, these features were understood using truth information from Monte Carlo (MC) simulations described in the Section \ref{simtech}.  Simulation truth results are shown in Figure \ref{Mono67b} for 67.0 MeV incident protons  and as expected the multi-proton peaks are not present in the simulation results as each simulated event began with a single primary proton.  In the simulation results we observe that the proton peak is present in a small number of bins (circled in red) demonstrating the consistency of the proton ionization process.  From the simulation truth we identify the pulse shape band originating at the main proton peak and trending towards 0 hadron intensity at 0 MeV as originating from events where the deposited energy by the primary proton was approximately equal to the pulse amplitude followed by the primary proton undergoing an inelastic interaction with a Cs or I atom that resulted in no secondary protons being created. The band beginning at the main proton peak and trending upward occurs when one of the secondary particles created was a proton.  The intrinsic broadening of this band in the simulation is due to events containing secondary protons with different kinetic energies.  Finally the additional bands trending upward above the main band are identified as proton inelastic interactions which result in two and three secondary protons.  

\subsection{20.0, 40.1 and 57.7 MeV Proton Data}

\begin{figure}[h]
\centering
\subfloat[20.0 MeV Protons]{\includegraphics[width=0.5\textwidth]{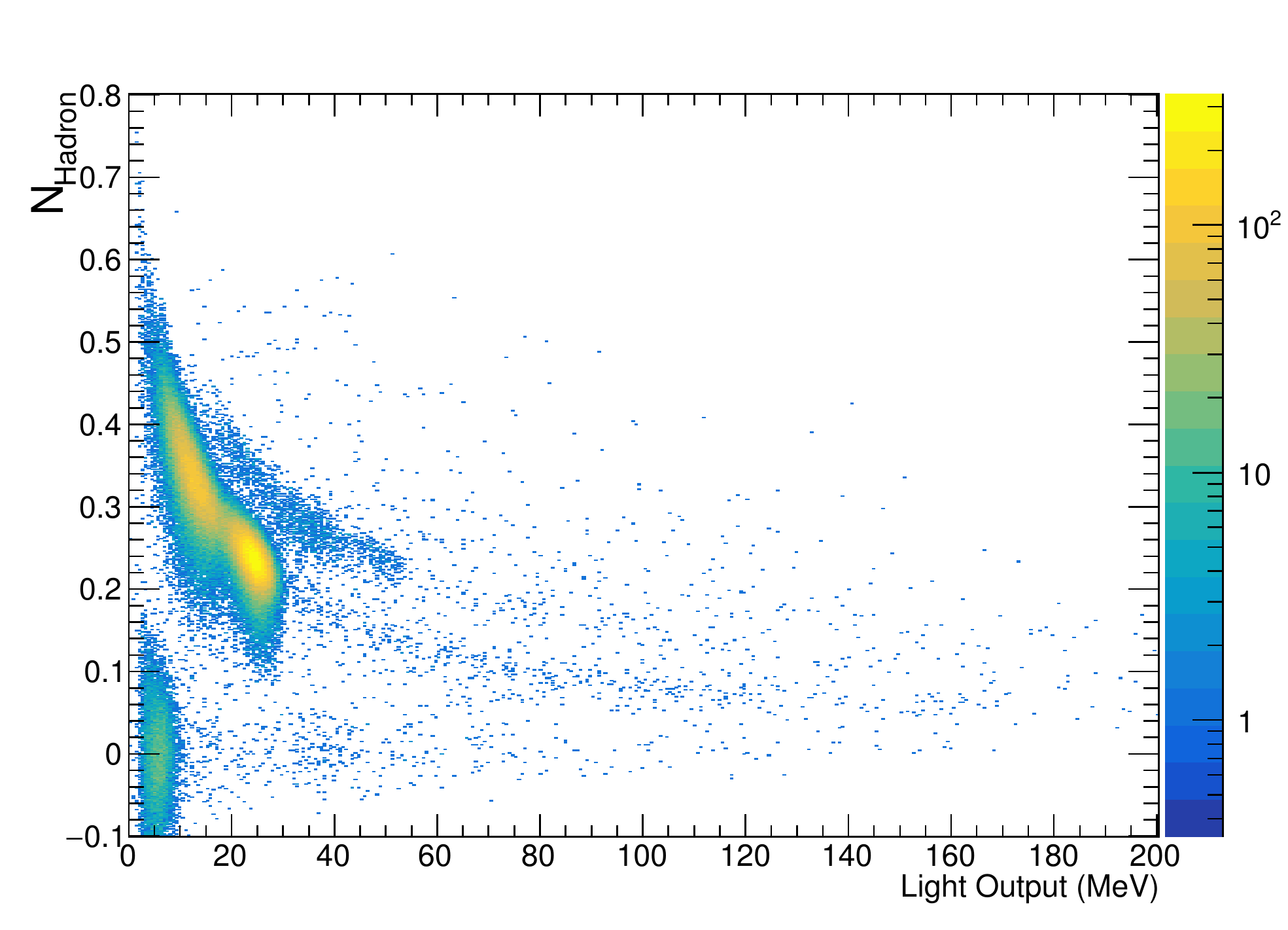}\label{Monob}} 
\subfloat[40.1 MeV Protons]{\includegraphics[width=0.5\textwidth]{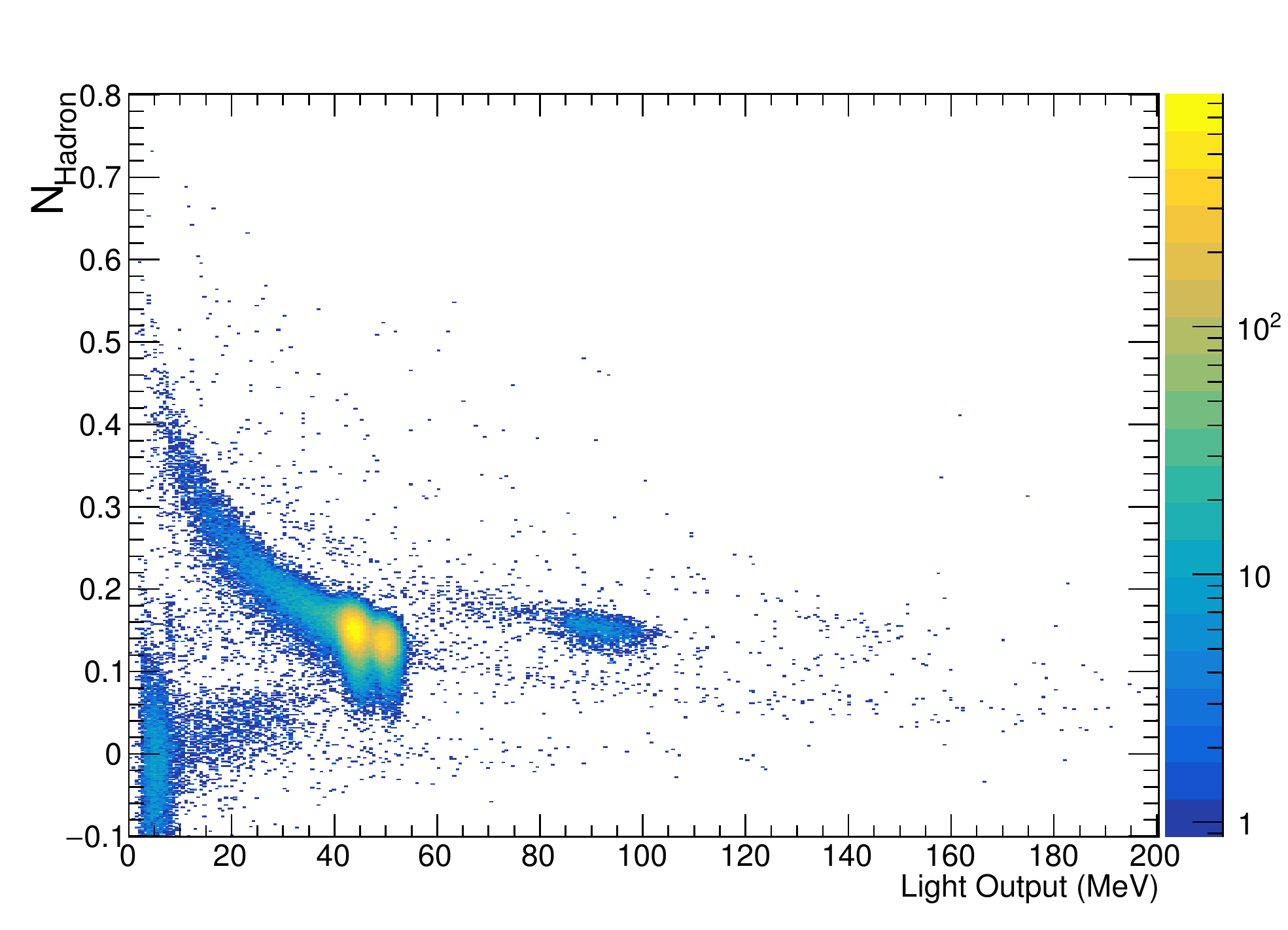}\label{Monoc}} \\
\subfloat[57.7 MeV Protons]{\includegraphics[width=0.5\textwidth]{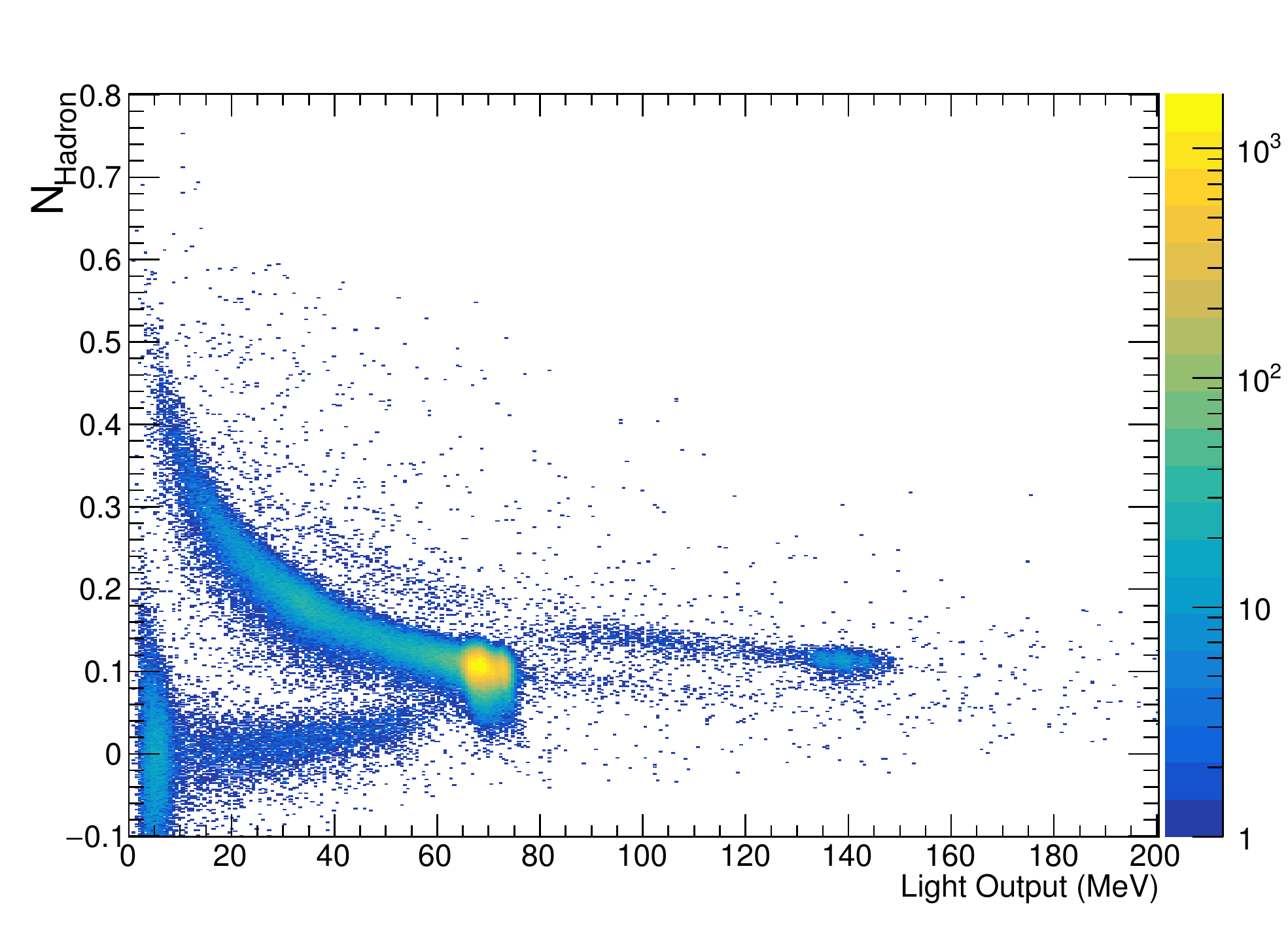}\label{Monod}} 
\subfloat[]{\includegraphics[width=0.5\textwidth]{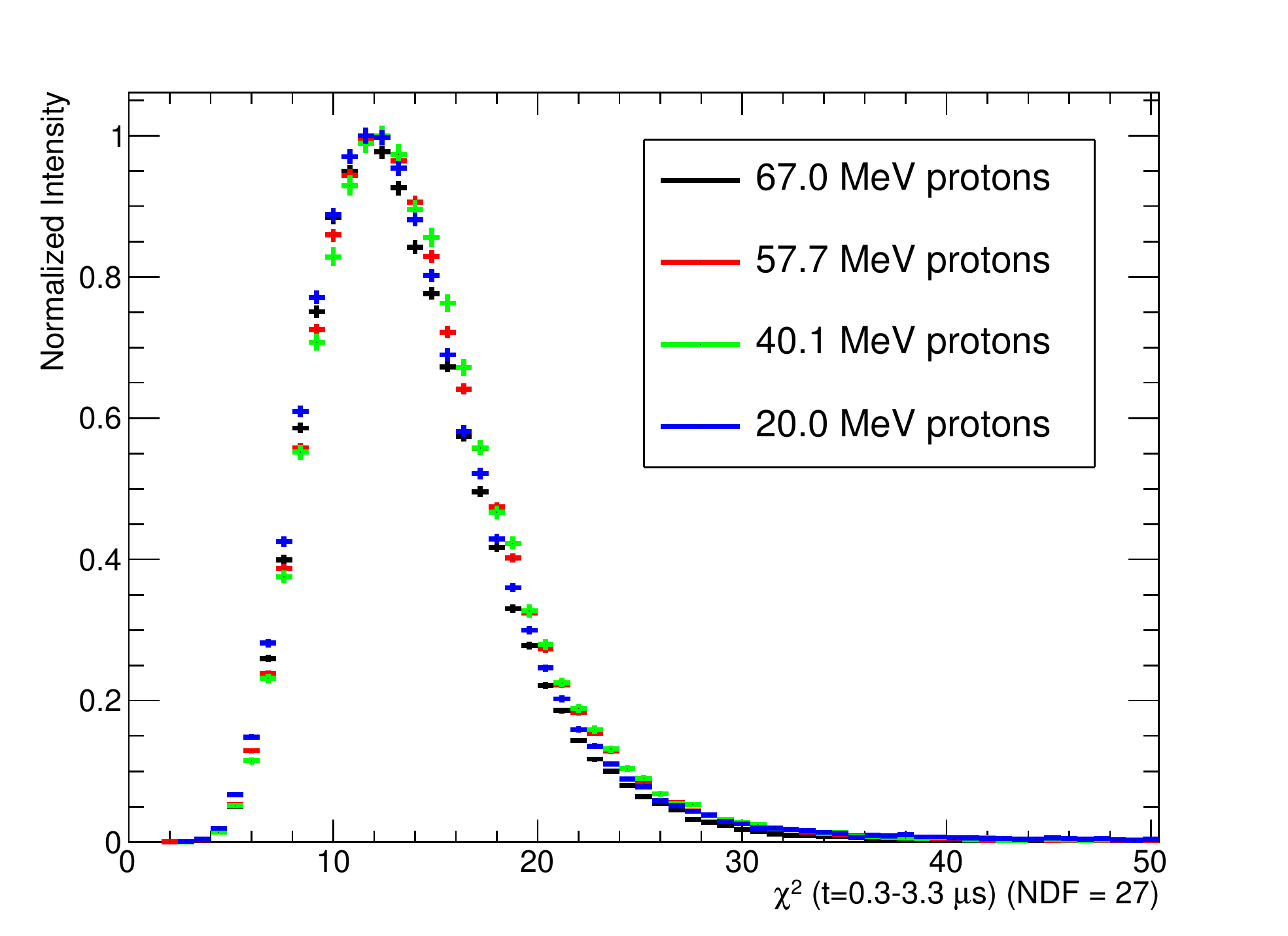}\label{MonoprotonChi}} 
\caption{Data pulse shape spectra for (a) 20.0, (b) 40.1 and (c) 57.7 MeV proton runs.  (d) Fit $\chi^2$ for pulses in main proton peaks.  The number of degrees of freedom is 27.  Note as discussed in Section \ref{HadronComponentModel} the errors used for computing the $\chi^2$ are conservatively estimated as correlations within the time bin have not been taken into account, which produces a lower than expected $\chi^2$.}
\end{figure}

During the 20.0, 40.1 and 57.7 MeV runs part of the beam had its energy degraded before reaching the crystal.  This provided one additional sample of protons with a lower energy in each of the runs.  For these runs we plot the hadron component intensity vs pulse amplitude in Figures \ref{Monob}, \ref{Monoc} and  \ref{Monod}.  In these spectra, similar features as those seen in the 67.0 MeV run are present, such that there is an intense peak at the total light output equivalent for the main beam energies and additional pulses from secondary hadron interactions below the main peak.  Using the single proton band in pulse shape spectra for the four proton runs we extract the hadron component intensity as a function of the total light output by fitting a Gaussian to the intensity distribution in a series of 2 MeV bins of total light output.  The extracted $\text{N}_\text{Hadron}$ vs $\text{L}_\text{Total}$ points for the single proton bands are overlaid in Figure \ref{protonMCcompare} in Section \ref{subSectionSimValProton} where they are further discussed and are compared with simulation and numerical calculations developed in the following sections.   

To demonstrate that the hadron component model describes the proton data just as well as the neutron data presented in Section \ref{DevelopmentofModel},  we plot in Figure \ref{MonoprotonChi} the $\chi^2$ statistic for the pulses in the main energy peak of each proton data run.  From the $\chi^2$ distributions it is seen that the hadron component model performance is identical for each proton energy.

\section{Simulation Techniques for CsI(Tl) Pulse Shape Variations}
\label{simtech}

\subsection{Simulation Methodology}

As outlined by the hadron component model developed in Section \ref{DevelopmentofModel}, the total light output of \csi{}, $\text{L}_\text{Total}$, can be divided into two components, $\text{L}_{\gamma}$ and $\text{L}_\text{Hadron}$.   The shape of \csi{} scintillation pulse is then characterized by the relative intensity of the hadron component, $\text{N}_\text{Hadron}$ defined in equation \ref{nhadrons} in Section \ref{DevelopmentofModel}. The task of simulating the response of \csi{} is thus reduced to calculating the quantities $\text{L}_\text{Total}$ and $\text{L}_\text{Hadron}$.  Once these quantities are known the output response of the \csi{} dectector can be constructed from equation \ref{ThreeCompModelGeant}.

\begin{equation}
\label{ThreeCompModelGeant}
\text{L}(t) =  (\text{L}_\text{Total}-\text{L}_\text{Hadron})R_\gamma(t) + \text{L}_\text{Hadron} R_\text{Hadron}(t) 
\end{equation} 

\noindent Where $R_\gamma(t)$ and $R_\text{Hadron}(t) $ are the normalized detector response shapes for the photon and hadron scintillation emission shapes.  In the particular case of PMT readout directly by a digitizer, $R_\text{Hadron}(t) $ is an exponential with decay time equal to \tauh{} and $R_\gamma(t)$ is given by $I_\gamma(t) / \text{L}_\gamma$ where $I_\gamma(t)$ is defined in equation \ref{TwoCompModel_Current} and $\text{L}_\gamma = \text{L}_\text{Total} - \text{L}_\text{Hadron}$ by construction.  For detector systems using additional shaping electronics, $R_\gamma(t)$ and $R_\text{Hadron}(t)$ should include the response of the signal chain electronics.

In order to compute $\text{L}_\text{Total}$ and $\text{L}_\text{Hadron}$, equations \ref{BirksGEANT4} and \ref{HadonicComp} are used, respectively.  In these equations the computed light output yields are expressed in calibrated units of photon equivalent energy deposited.  Practical effects such as scintillation photon self-absorption in the crystal, absolute total light yield of the crystal and photo-detector responses are not modelled as these effects are not required for the simulation truth pulse shape spectrum results we present.

\begin{equation}
\label{BirksGEANT4}
 \text{L}_\text{Total} = \sum_i^{\text{N}_\text{Particles}}  \int\limits_{E_i^\text{Initial}}^{E_i^\text{Final}} B\Big(\tfrac{dE}{dx}_i \Big) dE_i  \mathrel{\stackrel{\makebox[0pt]{\tiny \text{MC}}}{\approx}}  \sum_i^{\text{N}_\text{Particles}}  \sum_j^{\text{N}_i^\text{Step}} \text{E}_{ij}^\text{Step} B\Big(\tfrac{dE}{dx}_{ij}^\text{avg}\Big) 
 \end{equation} 

\begin{equation}
\label{HadonicComp}
 \text{L}_\text{Hadron} = \sum_i^{\text{N}_\text{Particles}} \int\limits_{E_i^\text{Initial}}^{E_i^\text{Final}} f\Big(\tfrac{dE}{dx}_i\Big) B\Big(\tfrac{dE}{dx}_i\Big) dE_i \mathrel{\stackrel{\makebox[0pt]{\tiny \text{MC}}}{\approx}}  \sum_i^{\text{N}_\text{Particles}} \sum_j^{\text{N}_i^\text{Step}} \text{E}_{ij}^\text{Step} f\Big(\tfrac{dE}{dx}_{ij}^\text{avg}\Big) B\Big(\tfrac{dE}{dx}_{ij}^\text{avg}\Big) 
 \end{equation} 

\noindent Where\footnote{Values for ionization energy loss (\dedx{}) are computed using the GEANT4 G4EmCalculator class.  }:

\begin{align*} 
\text{N}_\text{Particles}&=  \parbox{30em}{Total number of primary and secondary particles depositing energy inside the crystal volume.}\\~\\
E_i^\text{Initial}&= \text{Initial energy of the $i^{th}$ particle.}\\~\\
E_i^\text{Final}&= \text{Final energy of the $i^{th}$ particle.}\\~\\
\tfrac{dE}{dx}_{i} &= \tfrac{dE}{dx}(P_i,E_i)= \parbox{25em}{Ionization energy loss of the particle $i$, $P_i$, with instantaneous energy $E_i$.}\\~\\
 B\Big(\tfrac{dE}{dx}\Big) &= \parbox{30em}{Birk's scintillation efficiency correction for \csi{} defined to be normalized to 1 for 662 keV photon response. See Section \ref{sectionBdedx} for additional discussion.}\\~\\
  f\Big(\tfrac{dE}{dx}\Big) &= \parbox{30em}{Hadron component emission function defined to compute the fraction of the instantaneous energy deposit which results in scintillation emission in the hadron scintillation component. See Section \ref{sectionFdedx} for additional discussion.}\\~\\
  \text{N}^\text{Step}_i&=  \parbox{30em}{Total number of simulation steps inside the crystal volume by the $i^{th}$  shower particle.}\\~\\
 \text{E}_{ij}^\text{Step} &= \parbox{29em}{Energy deposited by the $i^{th}$ shower particle in the  $j^{th}$ discrete step of a simulation.}\\~\\
 \tfrac{dE}{dx}^\text{avg}_{ij} &= \parbox{30em}{\dedx{} computed using the average kinetic energy between the post and pre-step simulation points.}
\end{align*}
 
In equations \ref{BirksGEANT4} and \ref{HadonicComp}, $\mathrel{\stackrel{\makebox[0pt]{\tiny \text{MC}}}{\approx}}$ is used to indicate the approximation made when Monte Carlo libraries such as GEANT4 \cite{geant} are used and the particles are tracked in discrete steps.  In this case, using $\tfrac{dE}{dx}^\text{avg}$ is needed to improve the accuracy of modelling the lower energy charged hadron energy deposits where the \dedx{} can vary substantially from step-to-step.  We emphasize that when simulating the response, equations \ref{BirksGEANT4} and \ref{HadonicComp} are evaluated for all primary and secondary shower particles created in an event.  In addition, the scintillation response of each particle is computed continuously as it deposits energy in the crystal in order to account for the changing \dedx{} of the particle along its track. 

\subsection{Pulse Amplitude Calculation}
\label{sectionBdedx}

The Birk's scintillation efficiency defined as, $B(\tfrac{dE}{dx}) = \frac{d L}{dE}(\tfrac{dE}{dx})$, is known to vary for \csi{} depending on the particle ionization energy loss, \dedx{} \cite{Gwin,Koba,Zeitlin,Pietras,Twenhofel,Parlog}.   Studies of the scintillation efficiency are typically conducted using measurements independent of the pulse shape variations by integrating the light emission for a long time period compared to the pulse length.  

Models of the Birk's scintillation efficiency correction for inorganic scintillators have been discussed in the literature such that the correction empirically has the form of equation \ref{modBirks} \cite{Koba}.  This correction is defined to be normalized to 1.0 relative light output for 662 keV photons \cite{Koba}.

\begin{equation}
\label{modBirks}
B(\tfrac{dE}{dx}) = \frac{a}{1 + b \frac{dE}{dx} + c (\frac{dE}{dx})^{-1}}
\end{equation} 

The applicability of two parametrizations for equation \ref{modBirks} have been evaluated for protons and alphas by references \cite{Koba,Pietras}.  These parametrizations, calculated by reference \cite{Koba}, are referred to as the Birk's and Modified Birk's parametrizations and are presented in Table \ref{BirksPara}.

\begin{table}[h]
\centering
\caption{Parametrizations for equation \ref{modBirks} studied.}
\label{BirksPara}
\begin{tabular}{|l|l|l|l|l|}
\hline
\multicolumn{1}{|c|}{Parametrization} & \multicolumn{1}{c|}{\textit{a}} & \multicolumn{1}{c|}{\textit{b}} & \multicolumn{1}{c|}{\textit{c}} & \multicolumn{1}{c|}{Reference} \\ \hline
Birk's                         & 1.08        & 1.29e-3 & 0 & \cite{Koba}   \\ \hline
Modified Birk's         & 1.26    & 1.92e-3 & 7.47e-1 & \cite{Koba}    \\ \hline
This Study         & 1.52     & 3.448e-3 & 2 & -   \\ \hline
\end{tabular}
\end{table}

We find that in order to simulate the \csi{} response for protons while maintaining linearity for electromagnetic showers from electrons with energies from 2 MeV - 250 MeV, a re-parametrization of equation \ref{modBirks} is required.  In particular, we consider three criteria for the scintillation efficiency of \csi{} that a Birk's scintillation efficiency model must follow.   These criteria are summarized by Figure 7 of reference \cite{Gwin} and are as follows. \\

\begin{enumerate}

\item As measured by reference \cite{Ikeda}, the relative light output of electromagnetic showers from electrons over the energy range of 20 MeV - 5.4 GeV must be linear.  This requirement translates to $B(\frac{dE}{dx} < 2 \text{ MeV cm}^2\text{/g} ) \approx 1$.   

 \item The relative light output of protons with kinetic energies in the approximate range of 10 - 100 MeV must be greater than 1.  
 
\item The relative light output of low energy heavy particles such as alphas and ions is less than one. This requirement roughly translates to $  B(\tfrac{dE}{dx} > 200 \text{ MeV cm}^2\text{/g} ) < 1$.\\
\end{enumerate}

In order to satisfy these requirements, we use a re-parametrized version of equation \ref{modBirks} defined by equation \ref{refitBirks} with the parameters outlined in Table \ref{BirksPara}. 

 \begin{equation}
\label{refitBirks}
B_\text{This Study}(\tfrac{dE}{dx}) = \left\{
        \begin{array}{ll}
            1 & \quad \tfrac{dE}{dx} < 10 \text{ MeV cm}^2\text{/g} \text{ and } B(\tfrac{dE}{dx})  < 1 \\
B(\tfrac{dE}{dx}) & \quad  \text{for all other values of $\tfrac{dE}{dx}$ and $B(\tfrac{dE}{dx})$}
        \end{array}
    \right.
    \newline
 \end{equation} 

\begin{figure}[h]
\centering
\includegraphics[width=0.65\textwidth]{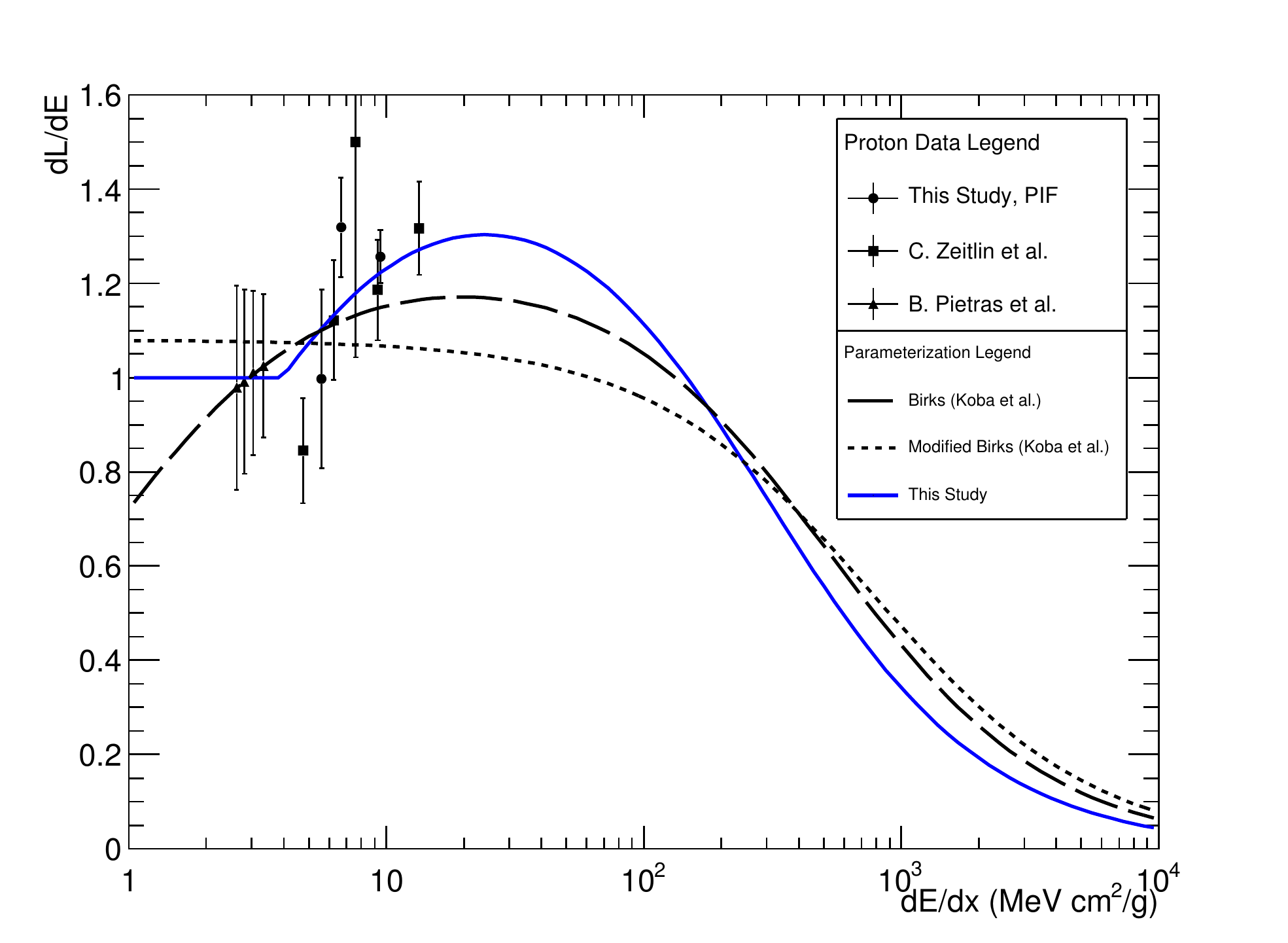}
\caption{Three parametrizations for \csi{} scintillation efficiency corrections overlaid with experimental data from PIF and references \cite{Zeitlin} and \cite{Pietras}.  Non-PIF proton data points are read from plots in references \cite{Zeitlin} and \cite{Pietras}.}
\label{BirksModels}
\end{figure}

\noindent We plot in Figure \ref{BirksModels} the three parametrizations for $B(\tfrac{dE}{dx})$.

\begin{figure}[h]
\centering
\subfloat[Proton response comparison.]{\includegraphics[width=0.5\textwidth]{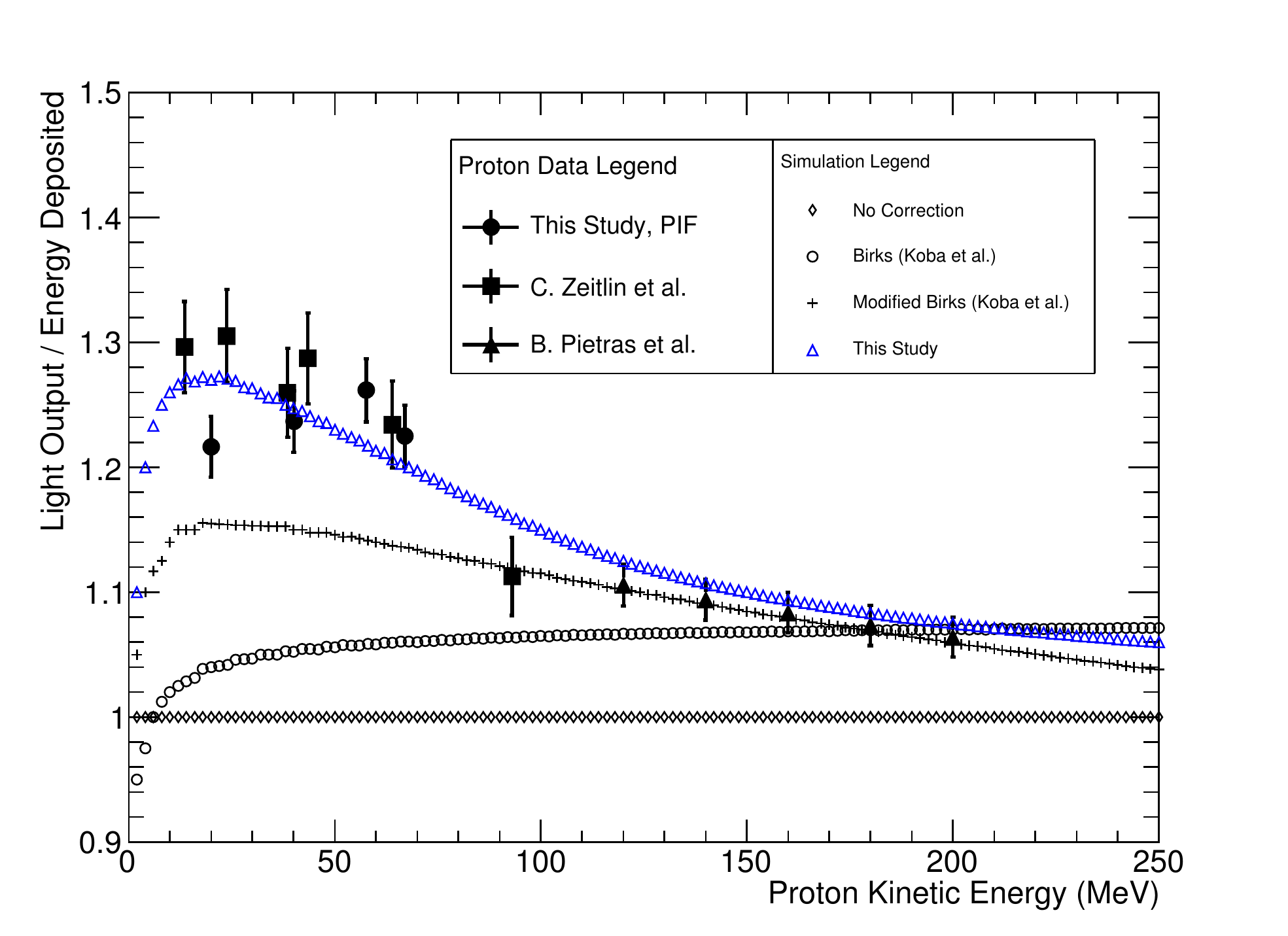}\label{SimBirksResultsa}} 
\subfloat[Electron response comparison.]{\includegraphics[width=0.5\textwidth]{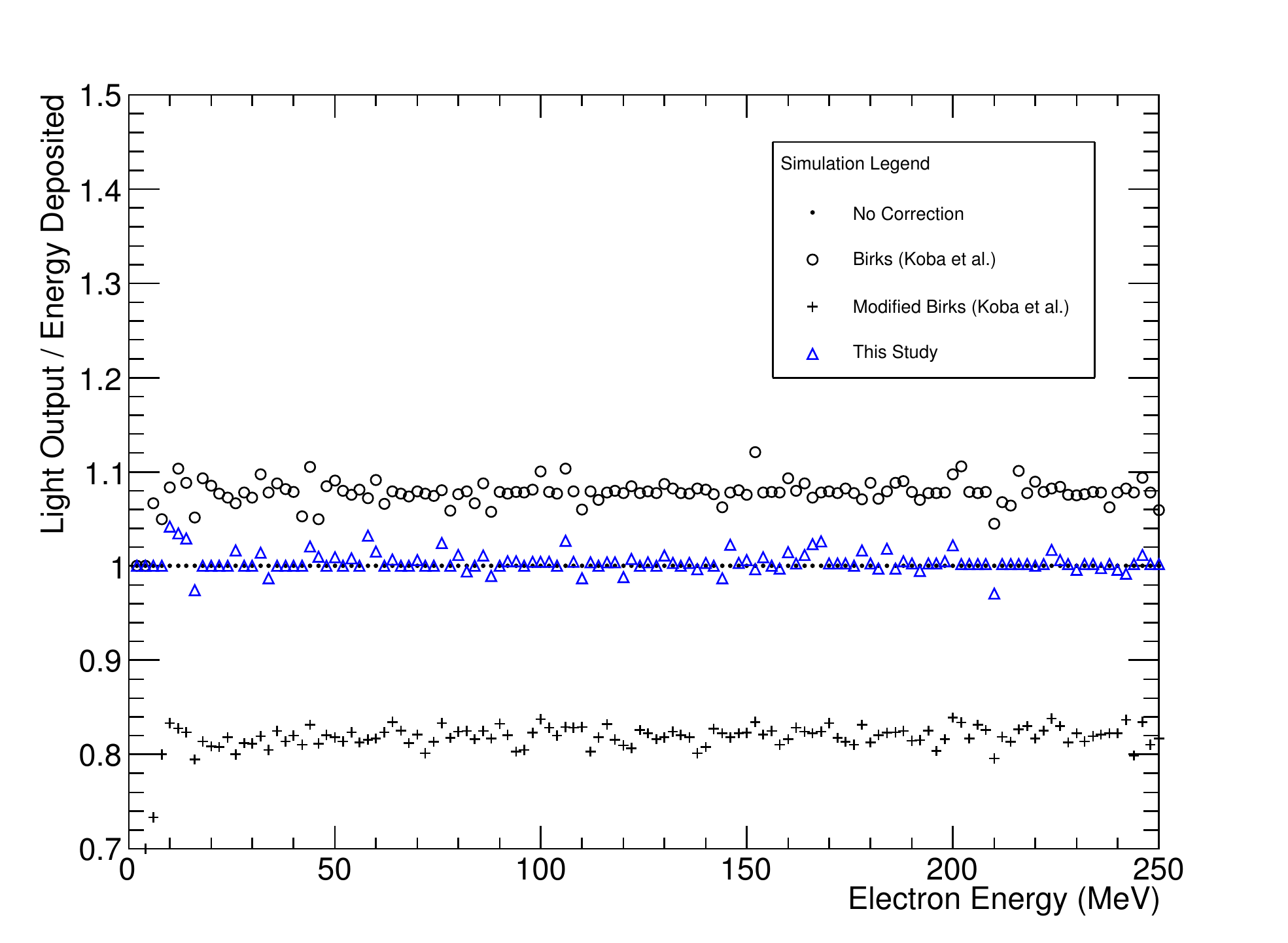}\label{SimBirksResultsb}} 
\caption{Comparisons of data and MC for simulations employing the three Birk's scintillation efficiency parametrizations outlined in Table \ref{BirksPara}.  Non-PIF proton data points are read from the plots in references \cite{Zeitlin} and \cite{Pietras}.}
\end{figure}

 The scintillation efficiency correction parametrization we propose was computed such that the simulated light output of protons at initial kinetic energies of from 2 - 250 MeV are in agreement with the \csi{} response measured with our proton testbeam data and other proton testbeam data in literature from references \cite{Zeitlin} and \cite{Pietras}. In addition we also require that the three criteria outlined above are satisfied, specifically that the scintillation response to electromagnetic showers is unchanged.  Comparisons for the simulated proton and electron response using three parametrization in Table \ref{BirksPara} with data are shown in Figures \ref{SimBirksResultsa} and \ref{SimBirksResultsb}. 

In Figure \ref{SimBirksResultsa} the simulated \csi{} response to protons is compared to proton testbeam data for the three parametrizations in Table \ref{BirksPara}.  These simulations demonstrate that, unlike the other parametrizations, the re-parametrization used in this study accurately reproduces the lower energy proton data in literature while simultaneously matching the higher energy proton response of the Modified Birk's parametrization.  In Figure \ref{SimBirksResultsa} is it also observed that the Birk's scintillation efficiency corrections can be a significant effect for protons in the low energy region as the light output can reach up to approximately 1.3 times the response with no correction.

In Figure \ref{SimBirksResultsb} the same parametrizations are evaluated for electrons.  Further support for our re-parametrization is observed in these results as both the Birk's and Modified Birk's parametrizations do not maintain a unity response for electromagnetic showers.  This indicates that the agreement of the Modified Birk's for the high energy proton response also results in a reduction in the electromagnetic response of \csi{} which is not observed in measurements reported in literature \cite{Gwin,Ikeda}.  From these results we conclude that our re-parametrization should be used to simulate the total light emission of \csi{}.  

Using the full simulation approach discussed above to evaluate the scintillation efficiency parametrizations is advantageous as all effects of secondary particle production are included in the calculation.  Past studies \cite{Gwin,Koba} evaluating  $\frac{dL}{dE}$ have used an alternate approach where data taken by various particle types and successive kinetic energies is differentiated and overlaid onto plots of $\frac{dL}{dE}$ such as Figure \ref{BirksModels}.  In the ideal case where the only energy deposited in the event is from the initial primary particle, this technique can be used to compute to the magnitude of $\frac{dL}{dE}$ at a specific value of $\frac{dE}{dx}$.  As an additional test of the models we overlay our PIF proton data combined with proton data in references \cite{Zeitlin} and \cite{Pietras} onto Figure \ref{BirksModels}.  From this analysis approach we again find agreement as expected between our scintillation efficiency re-parametrization with proton testbeam data.

\subsection{Pulse Shape Calculation}
\label{sectionFdedx}

In the literature the observed pulse shape variations of \csi{} have been correlated with the ionization energy loss of the particle as early as 1959 by Storey et al. where the \csi{} pulse shapes variations were empirically shown to be related to the ionization energy loss of the particle, \dedx{} \cite{Storey}.   As a result we assume that the hadron component emission function can be written as \fdedx{}, independent of the particle type.  

From the neutron data in Figure \ref{PMT_PSDINTENSITYvsENE} we begin by estimating the shape and bounds for \fdedx{}.  As muons and electrons have $\text{N}_\text{Hadron}=0$ we expect the lower bound $f(\frac{dE}{dx} < 2 \text{ MeV cm}^2\text{/g})\approx 0$ as this is the ionization region for these particles.  In addition, the maximum value of \fdedx{} can also be bound from Figure \ref{PMT_PSDINTENSITYvsENE} to be approximately 60-70\% emission.  This maximum emission is observed for neutron induced alpha and  $E_k<1$ MeV proton events which have  \dedx{} values beyond 100 $\text{ MeV cm}^2\text{/g}$.  From these bounds it can thus be extrapolated that in the region of approximately 2 $\text{ MeV cm}^2\text{/g} < $\dedx{} $<$ 100 $\text{ MeV cm}^2\text{/g}$, \fdedx{} will transition from 0 to 60-70\%.  Further evidence for this transition region is established by the shape of the single proton band in the proton data shown in Figure \ref{Mono67a} and Figures \ref{Monob} - \ref{Monod}.  As the kinetic energy of the primary proton decreases and approaches higher values for \dedx{}, the pulse shapes continuously approach higher values for $\text{N}_\text{Hadron}$.  From this observation we are able to extract part of the emission function from the single proton bands in the proton data we collected.

Using the single proton bands present in the proton data shown in Figures \ref{Mono67a} and  \ref{Monob}- \ref{Monod} we extract \fdedx{} by assuming the proton events to be an ideal case consisting of no secondary shower particles and only proton ionization.  Using GEANT4 we verify this assumption holds for protons with kinetic energies less than 100 MeV.  In this case, the proton of initial kinetic energy $E_k$ will ionize until it stops in the crystal volume resulting in a total energy deposit equal to the initial proton kinetic energy.  The hadron scintillation emission of the final pulse can then be written as equation \ref{protonL}.

\begin{equation}
\label{protonL}
\text{L}_\text{Hadron}  (P\text{=proton},E_k) =  \int_{0}^{E_K} B(P,k) f(P,k)  dk
 \end{equation} 

\noindent Where $k$ is the instantaneous kinetic energy of the proton. We solve equation \ref{protonL} to get an expression for the emission function as a function of the instantaneous proton kinetic energy as shown in equation \ref{protonEmission}.
 
\begin{equation}
\label{protonEmission}
f(P\text{=proton},E_k)  = \frac{1}{B(\tfrac{dE}{dx})} \frac{d}{dk}\biggr\rvert_{E_k} \text{L}_\text{Hadron} (P\text{=proton},k) 
 \end{equation}  

\begin{figure}[h]
\centering
\subfloat[Linear scale.]{\includegraphics[width=0.5\textwidth]{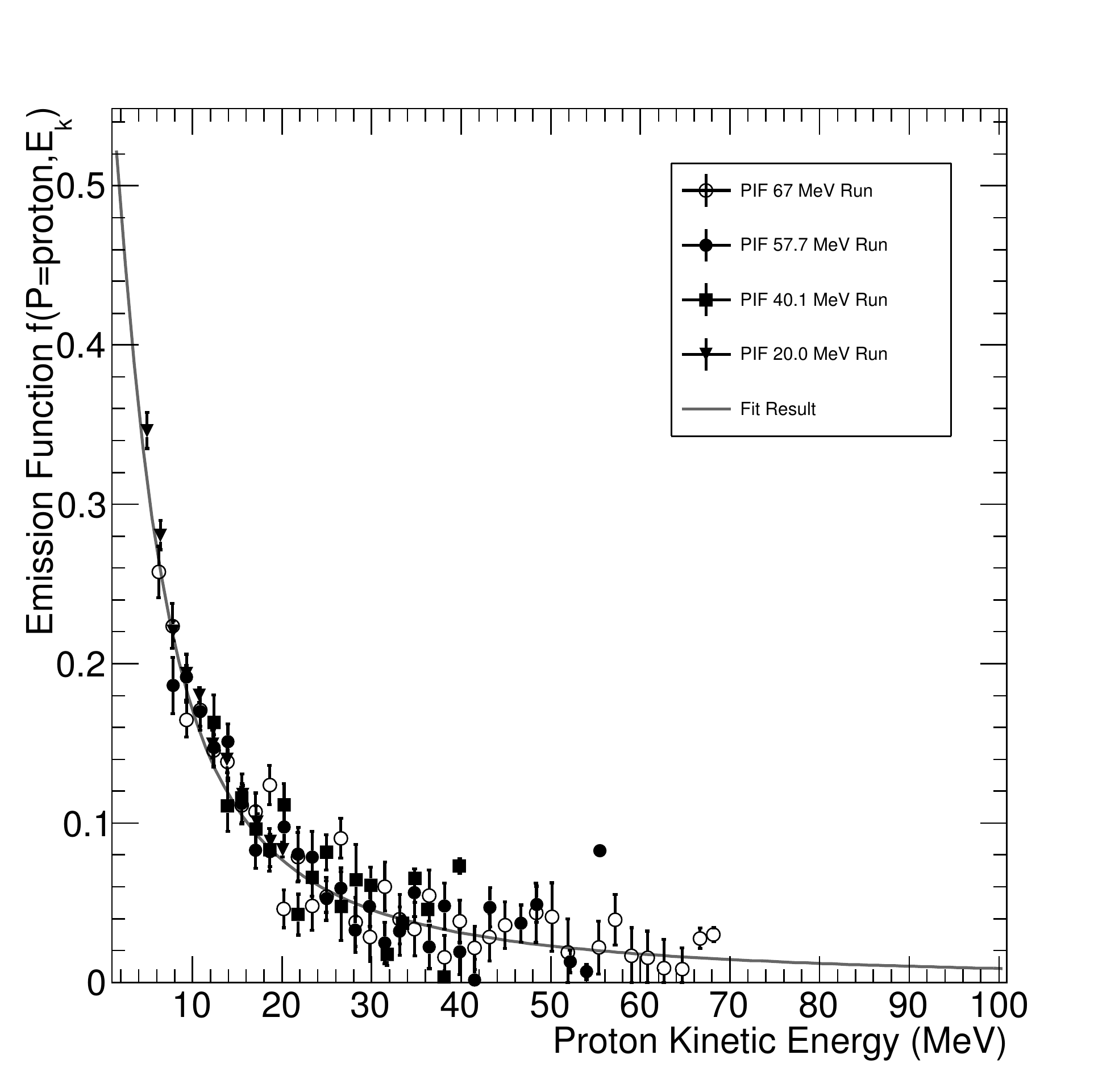}} 
\subfloat[Log scale.]{\includegraphics[width=0.5\textwidth]{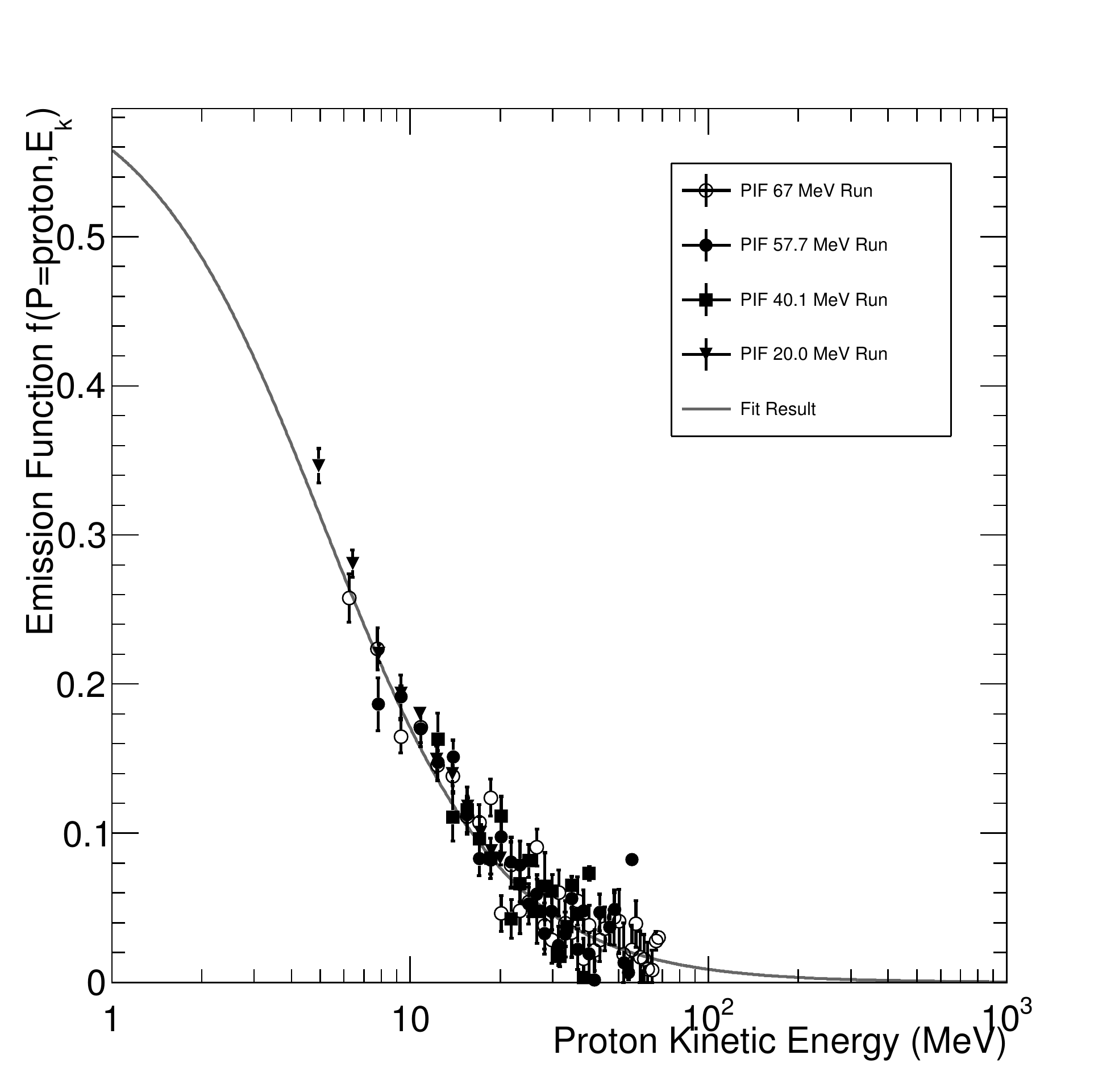}} 
\caption{Instantaneous hadron scintillation component emission extracted from single proton bands in PIF proton data using equation \ref{protonEmission}.  Analytic result for equation \ref{protonfkp} is overlaid.}
\label{fpk}
\end{figure}

Applying equation \ref{protonEmission} to the single proton bands in the PIF proton beam data in Figure \ref{Mono67a} and Figures \ref{Monob} - \ref{Monod} we plot the values of $f(P\text{=proton},E_k)$ in Figure \ref{fpk}.  

\begin{figure}[h]
\centering
\includegraphics[width=0.65\textwidth]{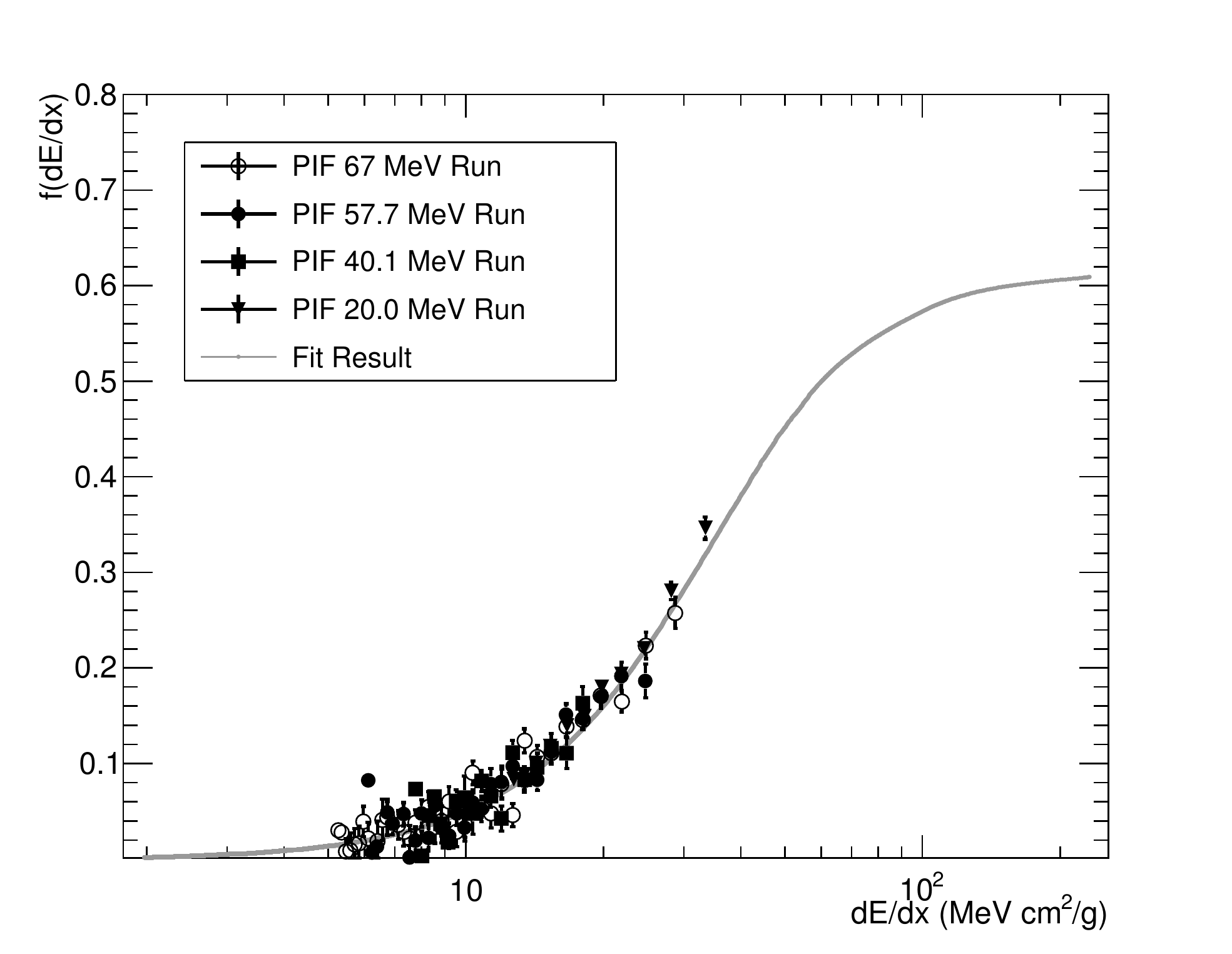}
\caption{Instantaneous hadron scintillation component emission as a function of \dedx{}.  Analytic result for equation \ref{protonfkp} is overlaid.}
\label{fdedx}
\end{figure}

As the goal is to extract the emission function as a function of \dedx{}, independent of particle type, the final step in computing \fdedx{} is to convert from proton kinetic energy to \dedx{}.  After this conversion\footnote{We perform the conversion from proton kinetic energy to ionization energy loss (\dedx{}) using the GEANT4 G4EmCalculator class.} the data values for \fdedx{} extracted from the proton bands is shown in Figure \ref{fdedx}.

From Figure \ref{fdedx} we see that using the single proton band we are able to extract the hadron emission intensity in a limited ionization region of 5-35 $\text{ MeV cm}^2\text{/g}$.  In order to extrapolate to the higher and lower \dedx{} regimes we use equation \ref{protonfkp}.
   
 \begin{equation}
\label{protonfkp}
f(P\text{=proton}, k)= \frac{A}{1 + (B  k)^{C}}
 \end{equation}

The functional form of equation \ref{protonfkp} is empirically driven as it is a simple analytic function that well describes the data in Figure \ref{fpk} and will satisfy the boundary conditions discussed above.  In order to determine the parameters for equation \ref{protonfkp} we evaluate equation \ref{protonL} numerically in 0.01 MeV step sizes and fit directly to the  $\text{N}_\text{Hadron}$ vs $\text{L}_\text{Total}$ PIF proton data points in Figure \ref{protonMCcompare} and overlay the numerical fit result in violet in this Figure.  From the fit we find the parameters for equation \ref{protonfkp} to be $A=0.612 \pm 0.003$, $B=0.194 \pm 0.001 \text{ MeV}^{-1}$ and $C=1.430 \pm 0.004$. We also overlay equation \ref{protonfkp} in Figure \ref{fpk} and in Figure \ref{fdedx} by converting from proton kinetic energy to \dedx{}.  We observe the measured data points for \fdedx{} are in agreement with the numerical fit result as expected.  In addition we confirm our expectation for the bound of \fdedx{} at high and low \dedx{}.

\subsection{Simulation Validation with Proton Data}
\label{subSectionSimValProton}

\begin{figure}[H]
\centering
\includegraphics[width=0.65\textwidth]{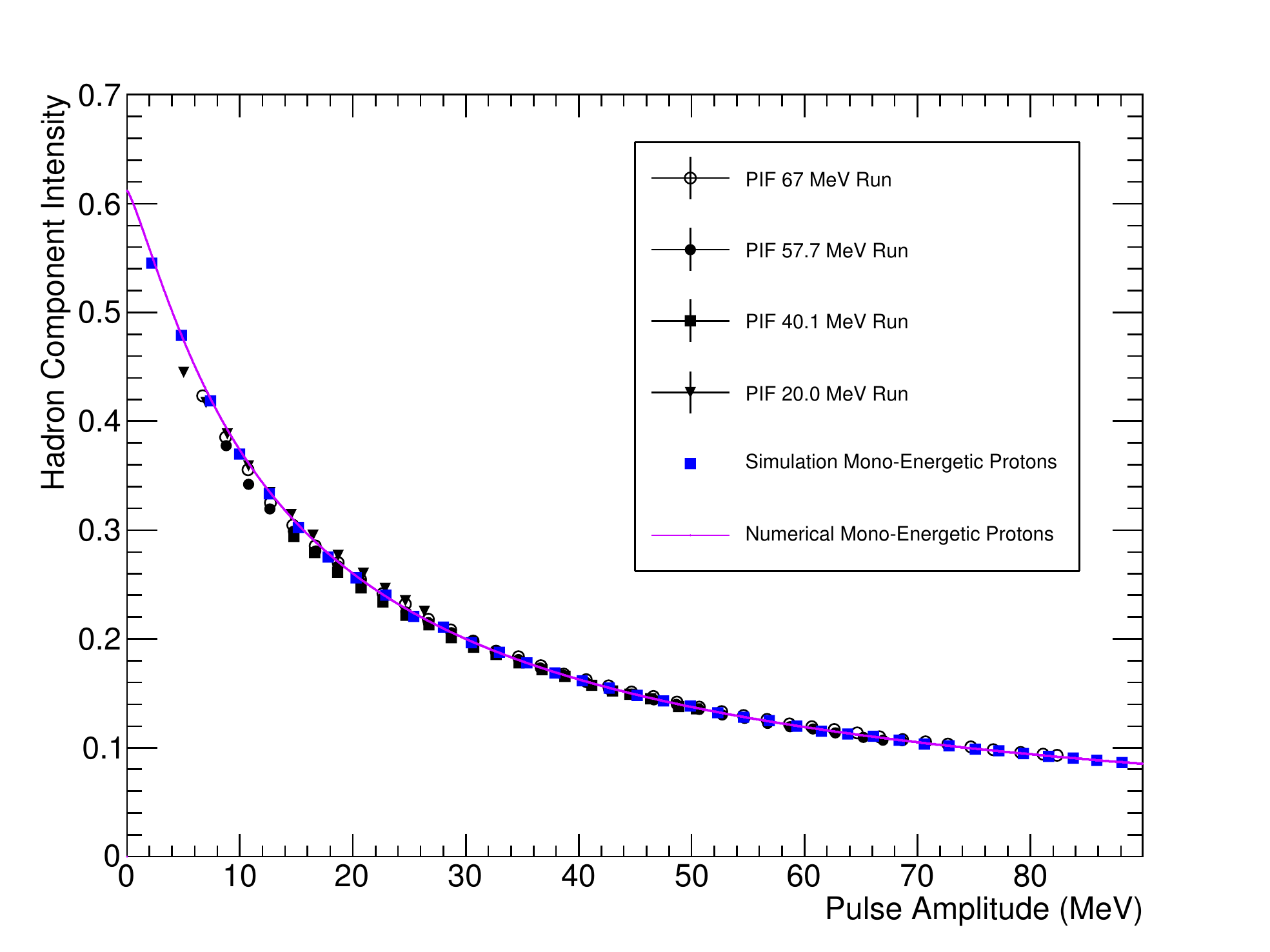}
\caption{Hadron component intensity as a function of pulse amplitude for the single proton band in the PIF proton data runs shown in Figures \ref{Mono67a} and  \ref{Monob}- \ref{Monod}. The fit result to equation \ref{protonL} is overlaid in violet.  Simulation results computed using equations \ref{BirksGEANT4} and \ref{HadonicComp} are overlaid as blue squares over the full range of 2-88 MeV. }
\label{protonMCcompare}
\end{figure}

We have now developed the tools to compute the \csi{} pulse amplitude and pulse shape based on the shower particles of the event by using equations \ref{BirksGEANT4} and \ref{HadonicComp}, respectively.  As was done with the scintillation efficiency corrections validation, we validate the proton ionization pulse shape simulations by simulating protons runs in the kinetic energy range of 2 to 70 MeV at 2 MeV intervals and extracting the $\text{L}_\text{Total}$ and $\text{L}_\text{Hadron}$ for each simulated proton energy run.  The results for the simulated proton response are overlaid in Figure \ref{protonMCcompare} with the $\text{N}_\text{Hadron}$ vs pulse amplitude values measured in the PIF proton data runs discussed in Section \ref{protondata}.  From Figure \ref{protonMCcompare} we observe reasonable agreement between data and simulation demonstrating we can accurately simulate the \csi{} scintillation response to protons.  This result also demonstrates that the instantaneous hadron scintillation component emission can be computed from the ionization energy loss of the interacting particles of the proton events.  A side-by-side data and simulation comparison for 67.0 MeV protons is shown in Figure \ref{Mono67a}.  This side-by-side comparison is discussed in Section \ref{protondata} and illustrates that in addition to proton ionization events,  the pulse shapes from the proton inelastic interactions are also reproduced in the simulation.

\subsection{Simulation Validation with PIF Neutron Data}

Using the PIF neutron run we can further validate our simulation methods.  The energy spectrum of the PIF neutrons is expected to approximately follow a 1/E energy distribution with a maximum energy of 500 MeV \cite{Ewart}.  We note that exact knowledge of the neutron energy spectrum is not critical for pulse shape validation as we expect this to only effect the relative rate of the different neutron interactions in \csi{} resulting in different relative intensities in the two dimensional pulse shape scatter plots.  As long as the energy threshold for an interaction is reached, the location of the hadron bands in the pulse shape spectra is expected to be independent of the neutron energy distribution.

\begin{figure}[H]
\centering
\includegraphics[width=0.65\textwidth]{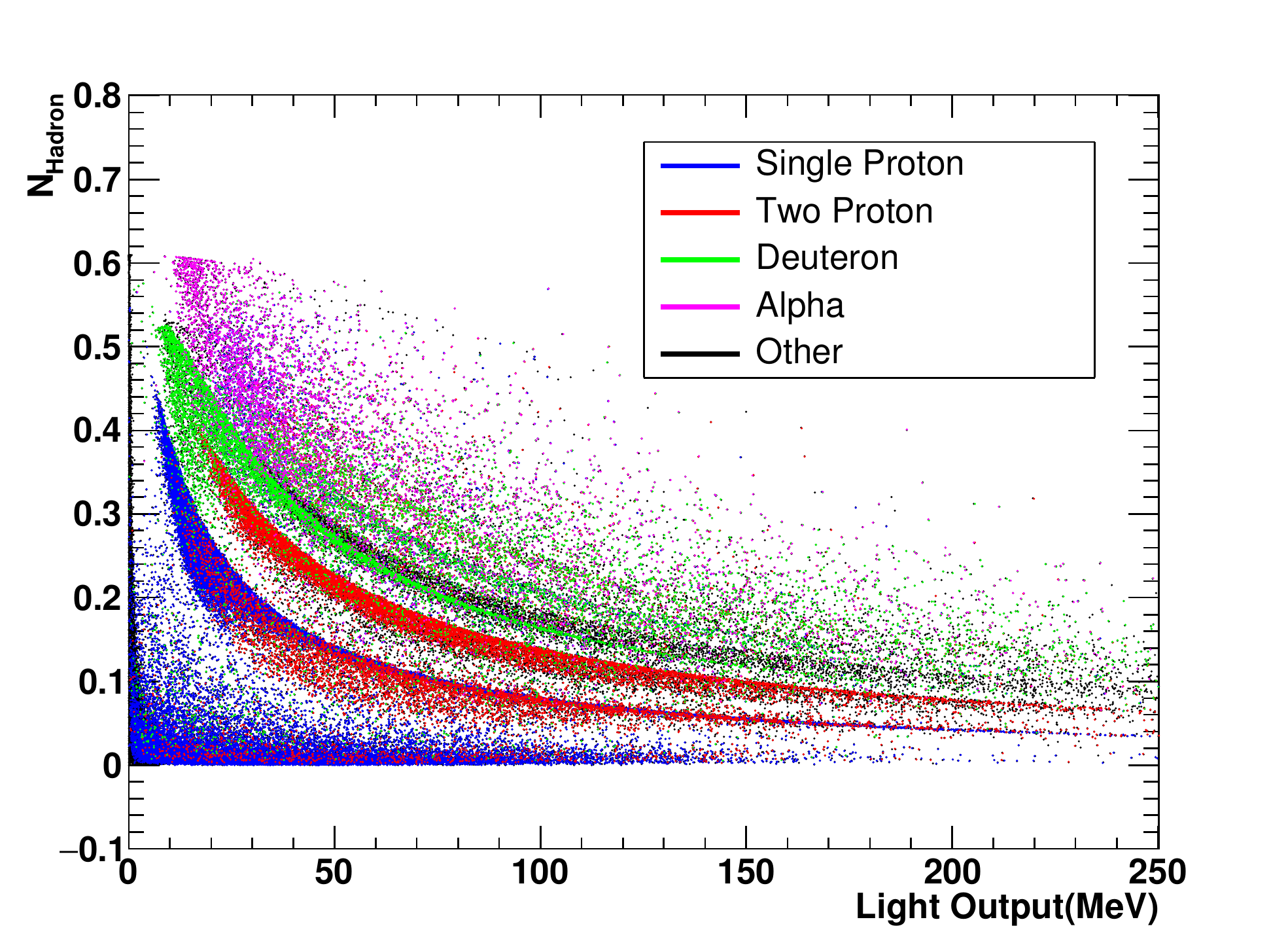}
\caption{Simulation truth pulse shape distributions for neutron events following a 1/E energy spectrum. Detector resolution is not modelled.}
\label{geant4NeutronColour}
\end{figure}

We simulate in GEANT4 neutrons following a 1/E energy distribution between kinetic energies of 1-500 MeV and compute the total light output and total hadron component light output based on the primary and secondary particles of the event.  Note detector resolution effects such as electronic noise and photo-electron statistics are not simulated.  Simulation truth results for the pulse shape spectrum of neutrons is shown in Figure \ref{geant4NeutronColour} with points colour coded based on the secondary charged hadrons generated from the neutron interaction in the event.  Note for all events in Figure \ref{geant4NeutronColour} the initial primary particle was a neutron.  This simulation result shows the same band structure as was observed in the neutron data presented in Figure \ref{PMT_PSDINTENSITYvsENE}.

As observed in Figure \ref{geant4NeutronColour}, in addition to the single proton band, the additional band structures arising from other secondary hadrons produced from inelastic neutron interactions are observed.   Detailed comparisons between data and simulation projections are plotted for different energy ranges in Figure \ref{geant4NeutronDataCompare}.  From these projections it is observed that in addition to correctly simulating the single proton response, the location of the peaks from the multi-proton and deuteron bands are in reasonable agreement between data and MC. This result confirms that although the hadron component emission function was extracted using proton data, when expressed as a function of \dedx{} the hadron scintillation component emission function has universal application to other charged hadrons. Note the excess of events at $\text{N}_\text{Hadron}=0$ in the data in the lower energy projections is due to the cosmic muon background.

\begin{figure}[H]
\centering
\subfloat[$\text{L}_\text{Total}$ = 30-35 MeV]{\includegraphics[width=0.5\textwidth]{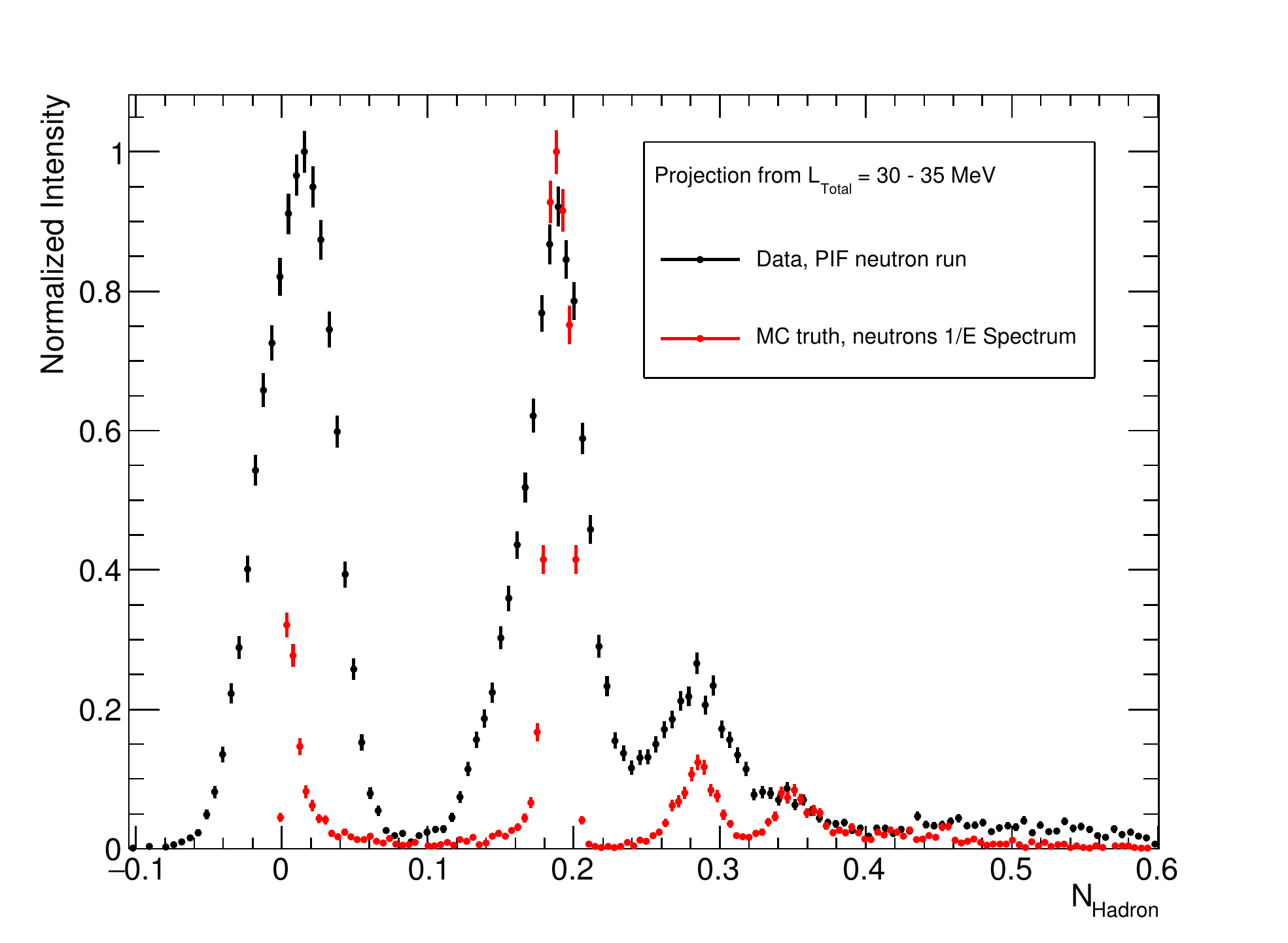}}
\subfloat[$\text{L}_\text{Total}$ = 40-45 MeV]{\includegraphics[width=0.5\textwidth]{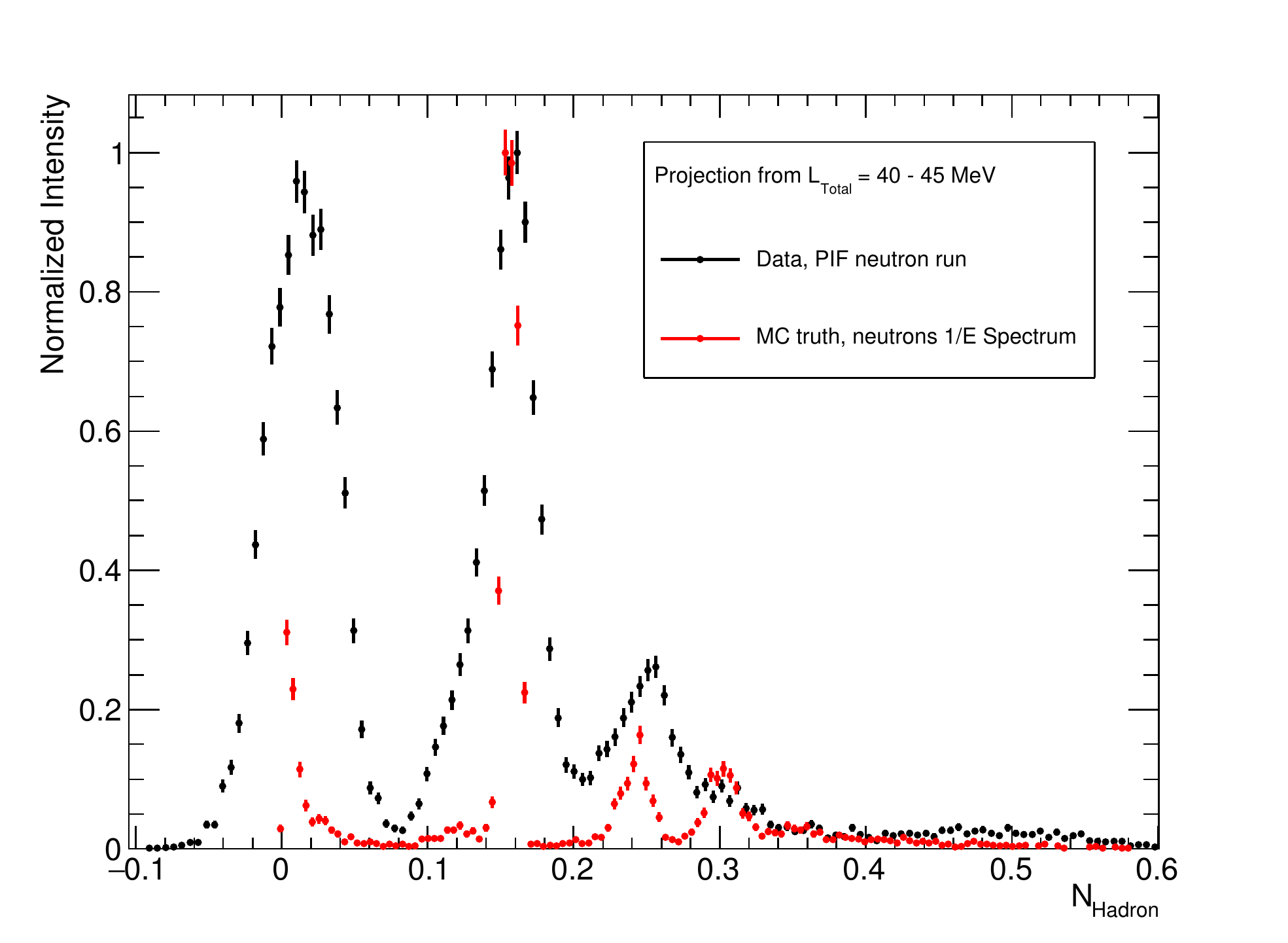}}
\end{figure}

\begin{figure}[H]
\centering
\subfloat[$\text{L}_\text{Total}$ = 50-55 MeV]{\includegraphics[width=0.5\textwidth]{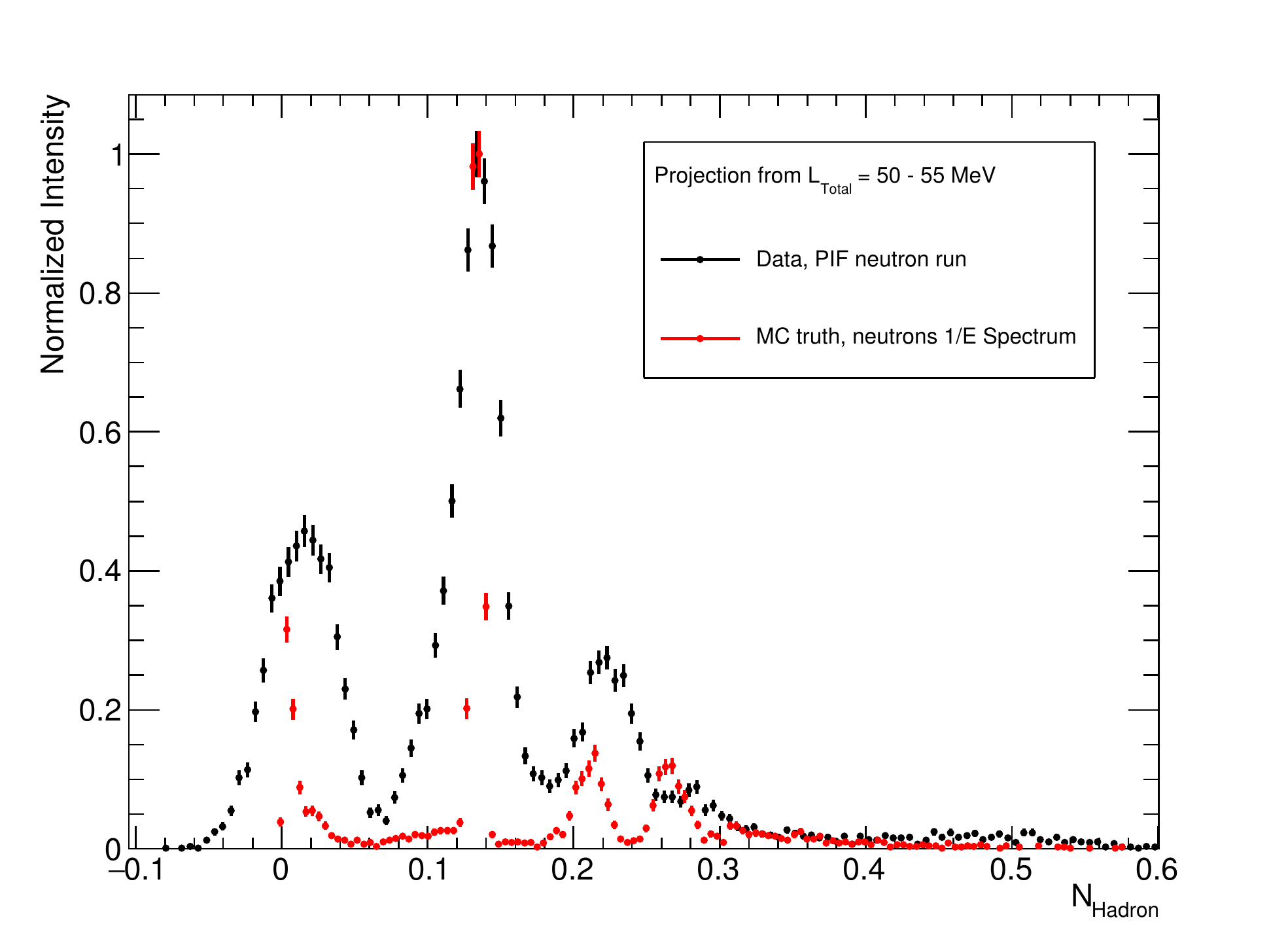}}
\subfloat[$\text{L}_\text{Total}$ = 115-120 MeV]{\includegraphics[width=0.5\textwidth]{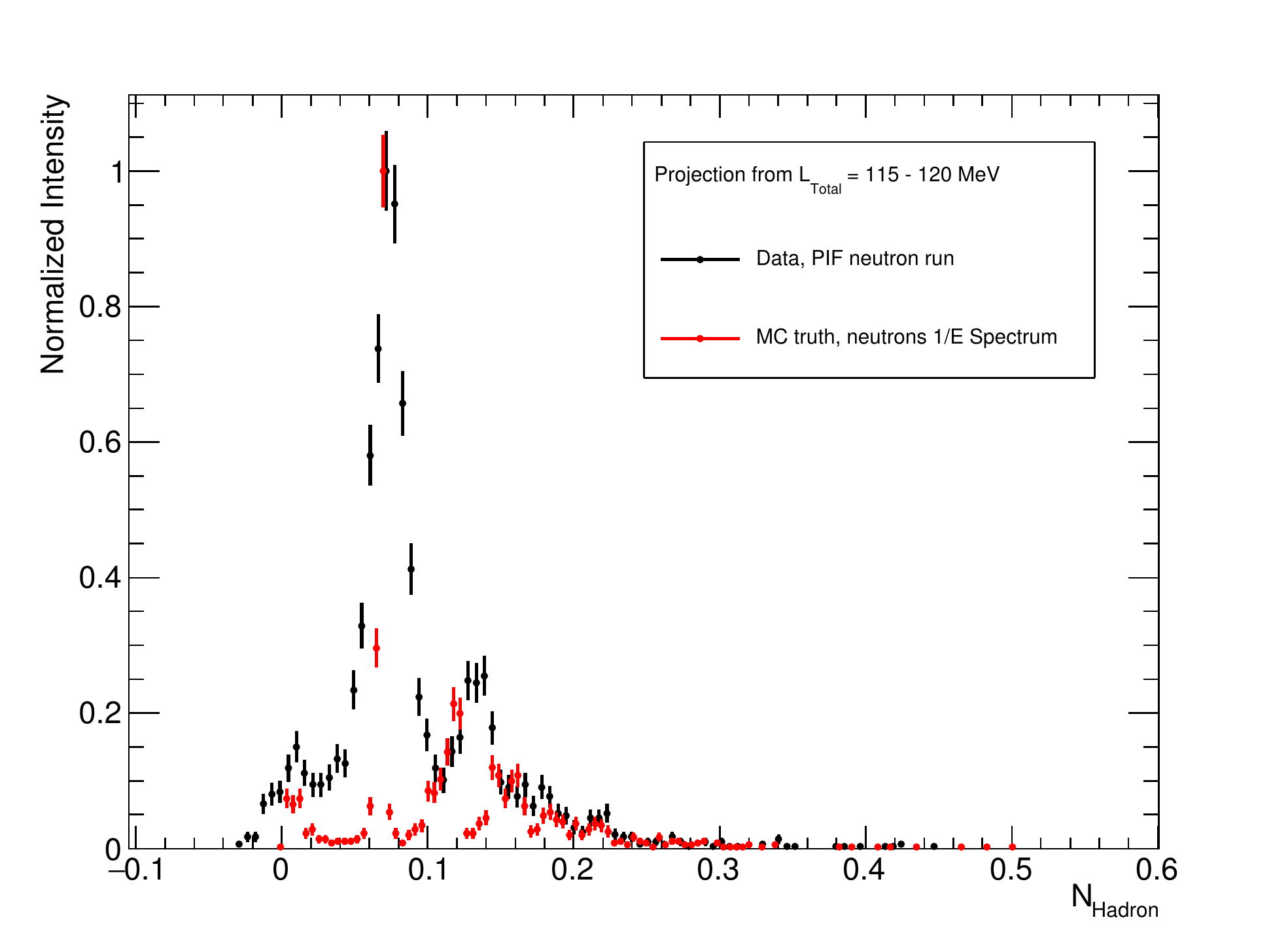}}
\caption{Data and simulation projections comparing PIF neutron run to MC truth for 1/E neutron energy distributions.  Note the cosmic muon background is not included in the simulation and as a result the data has an excess of events at $\text{N}_\text{Hadron}=0$ in the lower energy projections.  Note also that detector resolution effects are not modelled in the simulation.}
\label{geant4NeutronDataCompare}
\end{figure}

\subsection{Discussion of Simulation Validation Results}

 In order to improve the accuracy of these simulations additional studies in the high ($>35 \text{ MeV cm}^2\text{/g}$) and low ($< 5 \text{ MeV cm}^2\text{/g}$) \dedx{} regions should be pursued to verify our extrapolation of the emission function to these regions.  Such studies could potentially be completed using alpha beams for the higher \dedx{} region and high kinetic energy (> 80 MeV) protons for the low \dedx{} region. 

In addition to improving the accuracy for the simulations of the scintillation response of \csi{},  the above simulation techniques now give the ability to use \csi{} PSD for evaluation of the GEANT4 modelling for hadron material cross-sections in \csi{}.  As accurate hadronic material interaction modelling is important for minimizing the systematic errors in precision measurements made in nuclear and particle physics measurements, improving hadronic cross-section modelling could potentially have a large impact on a broad range of applications which make use of GEANT4 simulations.   Specifically by performing two dimensional cuts on the pulse shape distributions one could isolate the energy deposited spectra for specific secondary interactions and in principle measure the cross-sections for the specific inelastic interactions in CsI.  This type of analysis for example has been explored, using data only, by reference \cite{Bendel} to separate photon and proton interactions in \csi{}.  The simulation techniques we developed now allow for data vs simulation comparisons for these studies.  

\section{Pulse Shape Discrimination for Neutral Hadron vs Photon Separation in $e^+ e^-$ Collider Experiments}
\label{ClusterStudy}

One of the central questions to determine if PSD will be viable for neutral hadron detection in $e^+ e^-$ collider experiments deploying CsI(Tl) crystals (such as Belle II and BESIII) is:  will the energy deposited from the secondary charged hadrons of a hadronic shower be significant compared to the electromagnetic component of the shower, so that the \csi{} pulse shape variations can be measured in the crystals associated with a hadronic shower candidate?  In this section we address this question using the simulation methods developed and demonstrate the potential for neutral hadron identification using \csi{} PSD.  To study cluster effects we simulate in GEANT4 a $5 \times 5$ \csi{} crystal cluster constructed from crystals with rectangular prism geometry and dimensions of $5 \times 5 \times 30 \text{ cm}^3$.  These crystal dimensions are similar to the electromagnetic calorimeters used in the Belle II and BESIII experiments \cite{Belle2TDR,BESIIITDR}.  $K_L^0$ and neutrons of fixed momenta of 0.5 and 1 GeV/c are sent into the centre of the cluster and the quantities $\text{L}_\text{Total}$ and $\text{L}_\text{Hadron}$ are computed for all crystals in the cluster.  Photons with energy uniformly distributed between 0.2-1 GeV are also generated to serve as the electromagnetic shower control sample.  To ensure the particle interacted with the cluster, only events with total cluster light output ($\text{L}^\text{Cluster}_\text{Total}$) greater than 10 MeV are analysed. $\text{L}^\text{Cluster}_\text{Total}$ is defined as the sum of the light output from all the crystals in the $5 \times 5$ matrix of crystals.  For the 0.5 and 1 GeV/c \kl{} samples 44\% and 56\% of generated events pass the 10 MeV energy threshold, respectively.  For the 0.5 and 1 GeV/c neutron samples 41\% and 49\% of generated events pass the 10 MeV energy threshold, respectively. 

\subsection{Crystal Level Analysis}

\begin{figure}[h]
\centering
\includegraphics[width=0.65\textwidth]{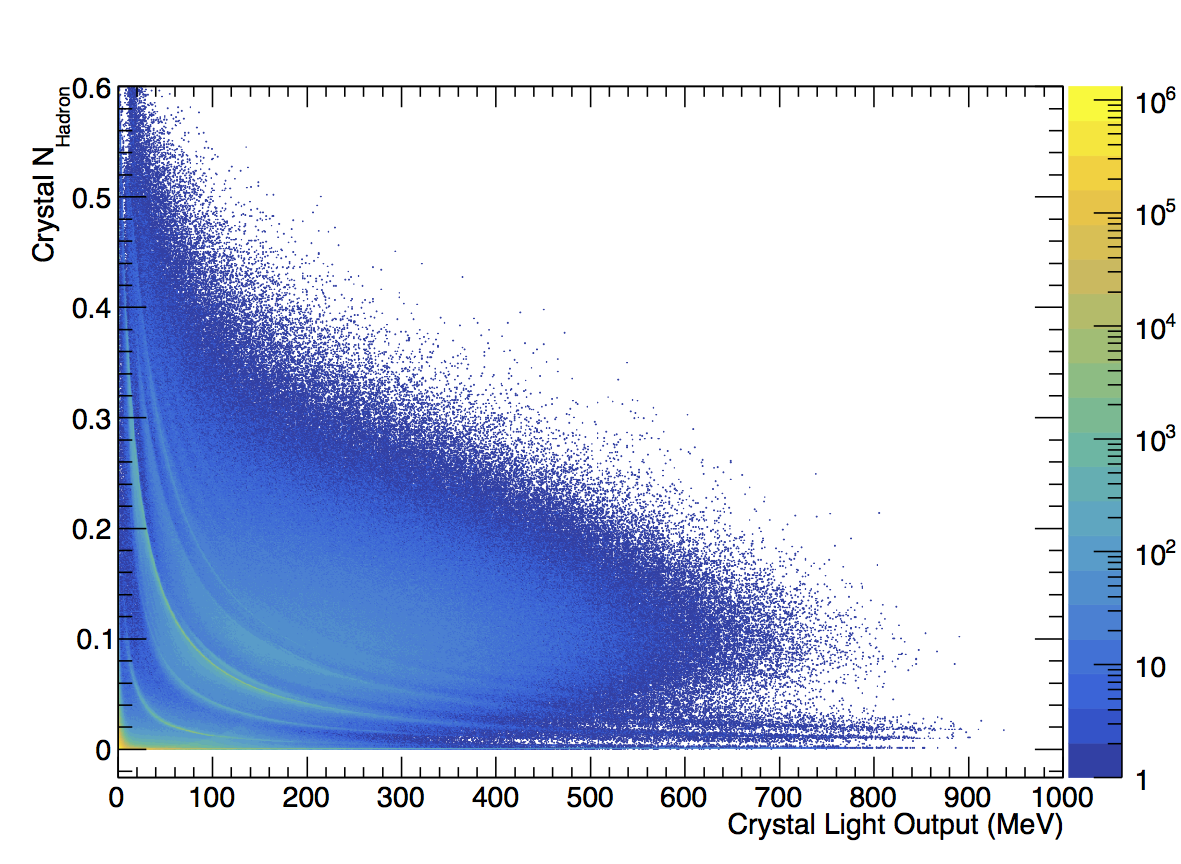} 
\caption{Calculated pulse shapes of cluster crystals contained in hadronic showers from simulated 1 GeV/c (620 MeV kinetic energy) $K_L^0$. }\label{SimKaonResultsa}
\end{figure}

In Figure \ref{SimKaonResultsa} we plot the computed hadron intensity distribution vs total light output for the cluster crystals in showers from 1 GeV/c \kl{}'s.  From this we see that a large variety of hadron pulse shapes are produced in the cluster crystals.  Compared to the distributions in the PIF neutron data, it is observed that the crystals in the high momentum \kl{} showers produce similar band structures.  From simulation truth we find that in addition to single and double proton bands, there are pulse shape bands also originating from higher multiples such as triple and quadruple proton events and also multiples of secondary deuterons and tritons.  In the region of high pulse amplitude and high hadron intensity, a smoother distribution is observed.  Pulses in this region are produced by combinations of different secondary hadrons.

\begin{figure}[h]
\centering
\includegraphics[width=0.495\textwidth]{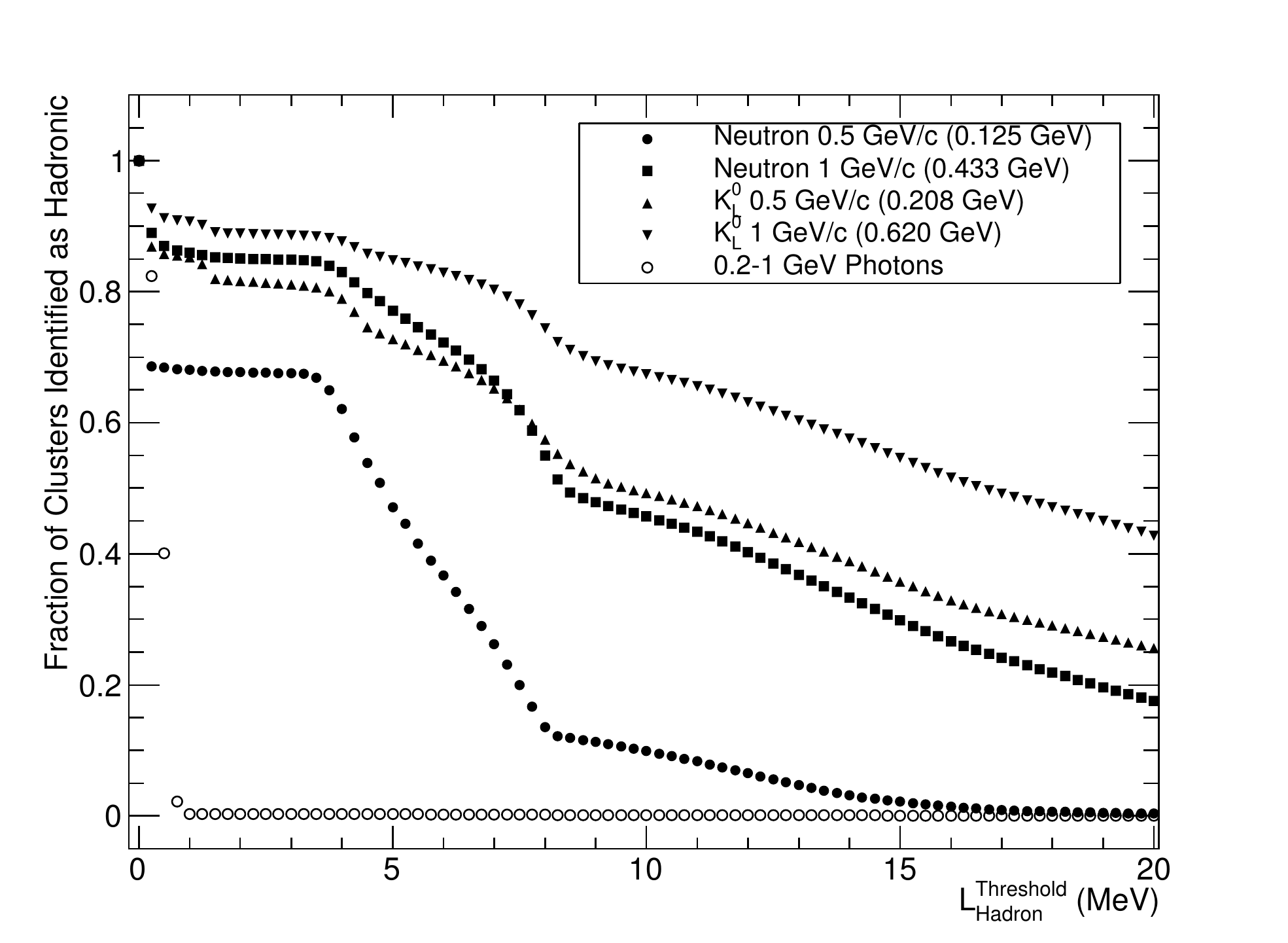}
\caption{Fraction of clusters with $\text{L}^\text{Cluster}_\text{Total}>10$  MeV that contained a minimum of one cluster with $\text{L}_\text{Hadron}>\text{L}_\text{Hadron}^\text{Threshold}$.}
\label{CrystalEffResults_0}
\end{figure}

 In addition to the hadron pulse shapes, a number of crystals in the 1 GeV/c \kl{} clusters result in photon pulse shapes due to the electromagnetic component of the shower.   In principle electromagnetic vs hadronic cluster separation however would only require a minimum of one crystal in the cluster to contain large contribution from the hadron scintillation component to allow for that cluster to be classified as hadronic.  This is because electromagnetic showers from photons are expected to consistently have zero hadron component intensity, independent of photon energy.  To estimate the performance of PSD for neutral hadron ID we compute the fraction of clusters which contained at least one crystal with a significant contribution from the hadronic scintillation component.   For a given detector system the pulse shape resolution of the system will determine the minimum amount of hadron component light output which can be resolved.  This resolution will be the limiting factor in the effectiveness of PSD.  Factors which are expected to degrade the pulse shape resolution such as electronic noise and pulse pile-up from background sources will vary between different detector systems.  To illustrate the potential for hadron identification, we define the quantity $\text{L}_\text{Hadron}^\text{Threshold}$ to be the minimum hadron component light output which a detector system can resolve in the pulse shapes.  In order to classify a cluster as hadronic we then require that the hadron component light output of at least one crystal in the cluster is greater than $\text{L}_\text{Hadron}^\text{Threshold}$.  We show in Figure \ref{CrystalEffResults_0} the efficiency for identifying hadronic clusters as a function of $\text{L}_\text{Hadron}^\text{Threshold}$.   
 
From Figure \ref{CrystalEffResults_0} it is predicted that if a detector is able to resolve 3 MeV or less of hadron component light output then PSD alone has the potential to provide excellent performance for neutral hadron vs photon separation as we see that a very small fraction of photon clusters are mis-identified as hadrons.  Quantitatively, at $\text{L}_\text{Hadron}^\text{Threshold}= 3 \text{ MeV}$ the percentage of photons mis-identified is $<0.35$\% and decreases to $<0.14$ \% at $\text{L}_\text{Hadron}^\text{Threshold}= 10 \text{ MeV}$.  The majority of photon clusters containing crystals with $\text{L}_\text{Hadron}>3$ MeV are found to be from events where the photon underwent a photo-nuclear interaction resulting in a secondary proton being produced in the cluster. 

Examining the change in hadron identification efficiencies with increasing $\text{L}_\text{Hadron}^\text{Threshold}$, three $\text{L}_\text{Hadron}^\text{Threshold}$ regions are identified where the efficiency has a distinct trend.  In the initial region corresponding to $\text{L}_\text{Hadron}^\text{Threshold} < 3 \text{ MeV}$ it is observed that the hadron identification efficiency for all hadron samples studied is approximately constant demonstrating that there is not a significant gain in setting $\text{L}_\text{Hadron}^\text{Threshold} < 3 \text{ MeV}$.  This is related to the minimum energy threshold for secondary proton production by a hadronic interaction in \csi{}.  Specifically for neutrons, the kinetic energy threshold for secondary proton production in CsI is 8 MeV \cite{Bartle}.  As a result the minimum secondary proton kinetic energy generated in a single proton neutron scatter event is expected to be approximately 8 MeV.  This 8 MeV kinetic energy threshold can be seen in our neutron data presented in Figure \ref{PMT_PSDINTENSITYvsENE} where it is observed that 8 MeV is the total light output threshold for pulses with hadron intensity greater than 10\%.  Now considering the event where a single 8 MeV proton is produced from a hadronic scatter in a crystal, we expect from the proton data shown in Figure \ref{protonMCcompare} that this will result in $0.38 \times 8$ MeV = 3.0 MeV of hadron component light output.  Thus due to the \csi{} energy threshold for secondary charged hadron production, $\sim$3 MeV is the minimum magnitude of hadron component light output expected in a hadronic scatter.
 
In the region of $3 \text{ MeV} < \text{L}_\text{Hadron}^\text{Threshold} < 8 \text{ MeV}$ these simulations predict that as $\text{L}_\text{Hadron}^\text{Threshold}$ increases there is a noticeable drop in identification efficiency for all particles.  This drop is a result of the $\text{L}_\text{Hadron}^\text{Threshold}$ surpassing the hadron component light output of the single proton band in the pulse shape spectrum.  As single proton production is dominant for lower energy hadrons, the impact of not resolving the single proton band is largest for the 0.5 GeV/c (0.125 GeV kinetic energy) neutrons.  In the final region of $\text{L}_\text{Hadron}^\text{Threshold}>$8 MeV, the efficiency does not drop as quickly as the previous region.  This is observed for all particles studied and is because in this region of high hadron component light out, multiple different charged hadrons are produced in the hadronic shower creating large amounts of hadron component light output.
 
\begin{figure}[h]
\centering
\includegraphics[width=0.495\textwidth]{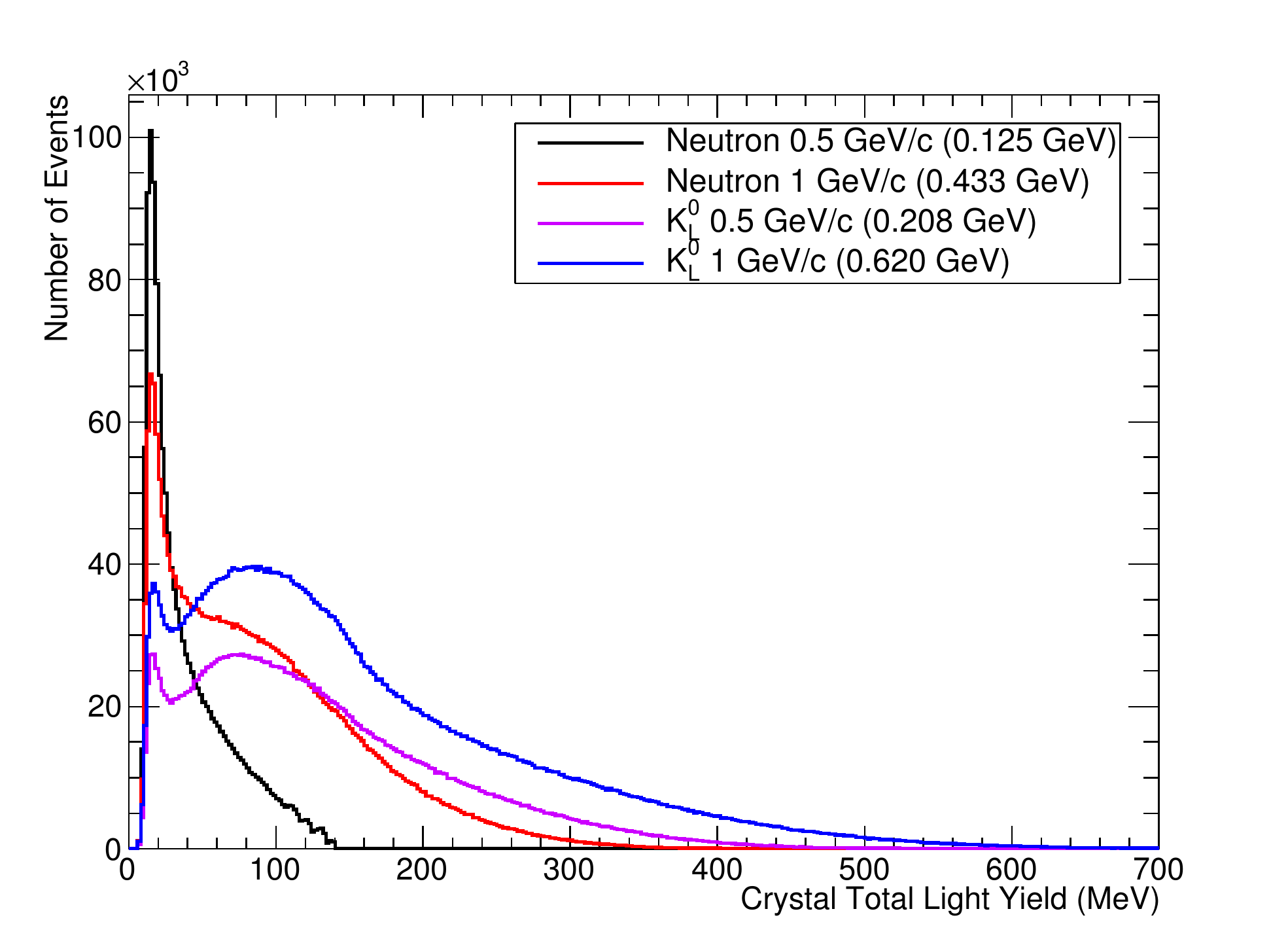} \label{CrystalLY_gt3MeV}\caption{Total energy distribution for cluster crystals with $\text{L}_\text{Hadron}>3$ MeV.}
\label{SimcrystalEnergy}
\end{figure}

\begin{figure}[h]
\centering
\includegraphics[width=0.495\textwidth]{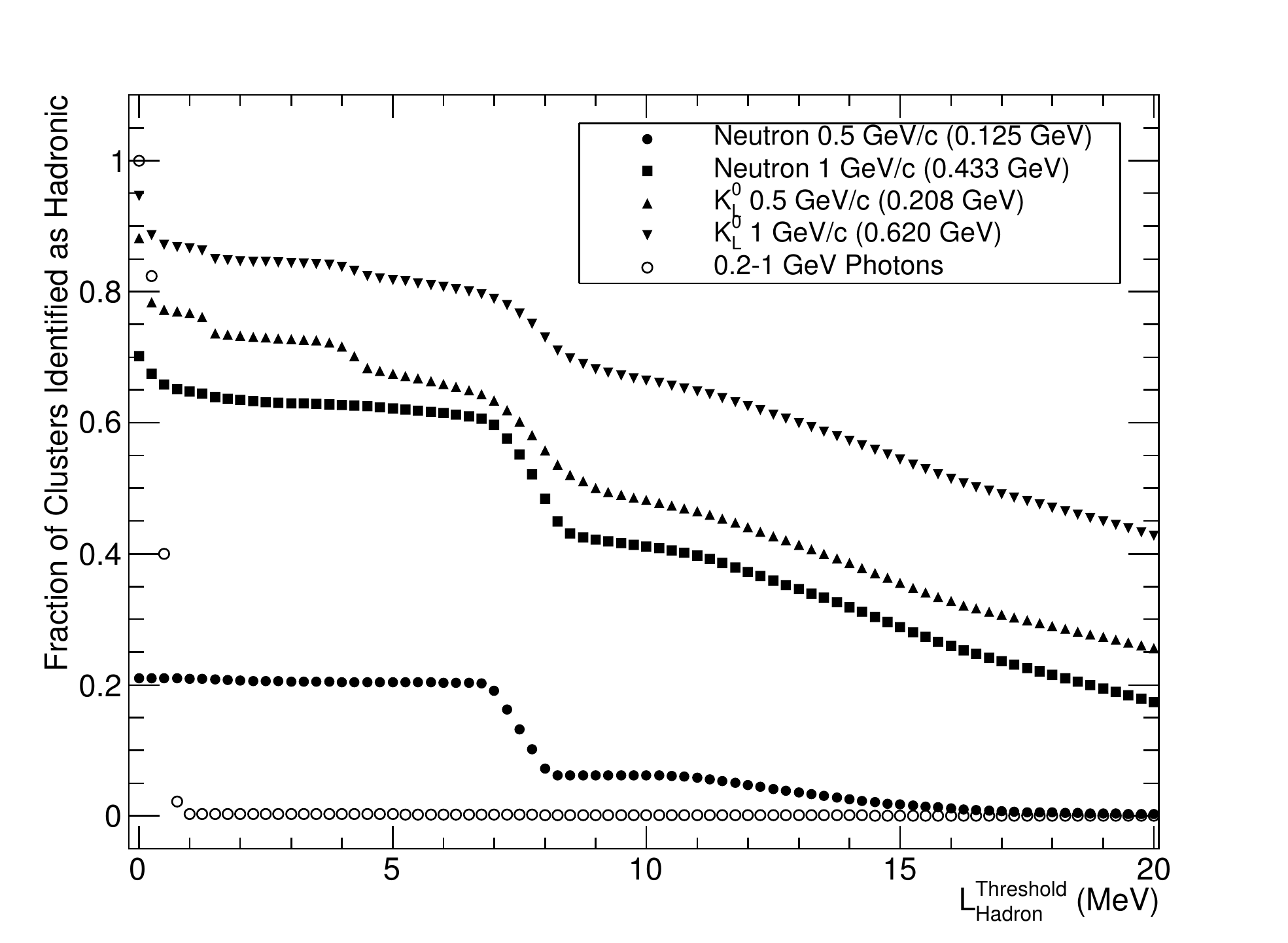} 
\caption{Fraction of clusters  studied that contain one crystal with both $\text{L}_\text{Hadron}>\text{L}_\text{Hadron}^\text{Threshold}$ and $\text{L}_\text{Total} > 50$ MeV.}
\label{CrystalEffResults_50}
\label{SimEffs_crystal}
\end{figure}

To study the energy distribution of hadron identified crystals we plot in Figure \ref{SimcrystalEnergy} the distribution of the total light output for all cluster crystals with $\text{L}_\text{Hadron}^\text{Threshold}>$3 MeV.  From Figure \ref{SimcrystalEnergy} we see for the higher kinetic energy samples studied, that a significant number of hadron identified cluster crystals are expected to also have relatively high energy deposits. We now consider a more constrained case where an experiment might only be able to characterize higher energy deposits in the calorimeter. We study this by applying a tighter criteria for hadron identification where in addition to a cluster containing a crystal with $\text{L}_\text{Hadron}>\text{L}_\text{Hadron}^\text{Threshold}$, we also require that the same crystal has a total light output greater than 50 MeV in order to be identified as hadronic.  We show in Figure \ref{CrystalEffResults_50}, the expected hadron identification efficiencies after applying this tighter criteria.  Compared to the results shown in Figure \ref{CrystalEffResults_0} where no crystal energy threshold was applied, we see that applying a 50 MeV crystal total light output threshold has the greatest impact for all hadron samples on the hadron detection efficiencies for low $\text{L}_\text{Hadron}^\text{Threshold}$ values.  This means that detectors with good pulse shape resolution will have the largest impact.  This result is expected as events with low hadron component light output are typically from lower energy single proton energy deposits.  The drop in efficiency in Figure \ref{CrystalEffResults_50} thus is a result of now requiring the threshold hadronic interaction to be a single secondary proton with total light output above 50 MeV.  For a single proton event with total light output of 50 MeV the equivalent hadron component light output is  $\sim$7 MeV as shown in our proton data in Figure \ref{protonMCcompare}.  As a result the 50 MeV total light output threshold results in the efficiency in the low $\text{L}_\text{Hadron}^\text{Threshold}$ region to be reduced to approximately the value of the efficiency curve in Figure \ref{CrystalEffResults_0} evaluated at $\text{L}_\text{Hadron}^\text{Threshold} = 7$ MeV.  For the region of $\text{L}_\text{Hadron}^\text{Threshold}>8$ MeV in Figure \ref{CrystalEffResults_50} the detection efficiency is not significantly affected by the 50 MeV total light output requirement as hadron events in this region of high hadron component light output will typically also have high total light output and pass the 50 MeV cut.

\subsection{Cluster Level Analysis}
 
\begin{figure}[h]
\centering
\includegraphics[width=0.55\textwidth]{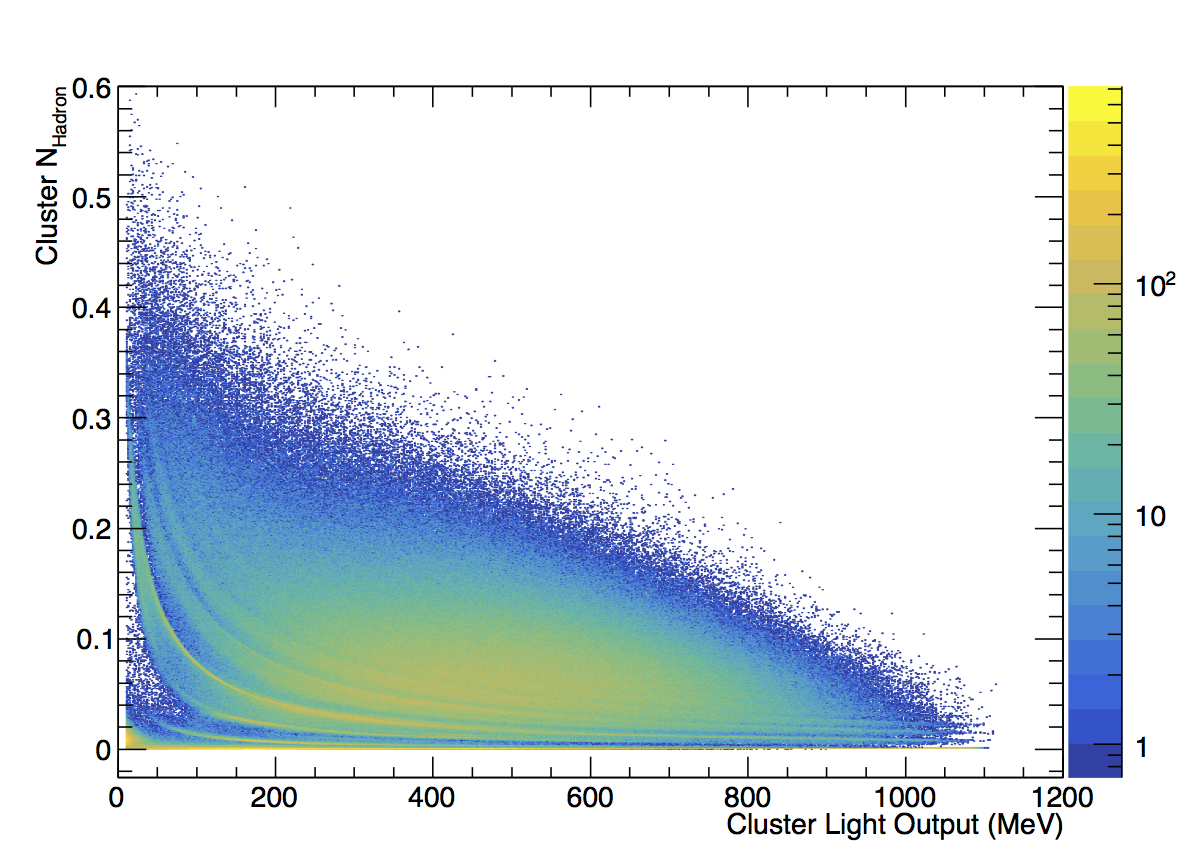}
\caption{Calculated cluster hadron intensity vs total light output distribution for 1 GeV/c (620 MeV) $K_L^0$.}\label{clustera}
\end{figure}

The analysis's presented in Figures \ref{CrystalEffResults_0} and \ref{CrystalEffResults_50} are completed using the cluster crystal information individually.  Considering an alternative algorithm for hadron shower identification, we now combine the crystal information of a $5 \times 5$ cluster and compute the cluster hadron component light output, $\text{L}_\text{Hadron}^\text{Cluster}$, by summing the hadron component light output of all cluster crystals. Analogous to the single crystal, the cluster hadron intensity defined by $\text{L}_\text{Hadron}^\text{Cluster} / \text{L}_\text{Total}^\text{Cluster}$ is also computed.  In Figure \ref{clustera} we plot the distribution of cluster hadron intensity vs $\text{L}_\text{Total}^\text{Cluster}$, for the same 1 GeV/c \kl{} events as Figure \ref{SimKaonResultsa}.  Using the cluster variables it is predicted that the individual band structures are now less prominent and the intensity of the events in the smooth high energy and high hadron intensity region increases compared to the individual crystal distribution in Figure \ref{SimKaonResultsa}.  This is expected to improve the PSD performance as pulses in this region have the largest magnitude of hadron component light output.

\begin{figure}[h]
\centering
\subfloat[]{\includegraphics[width=0.5\textwidth]{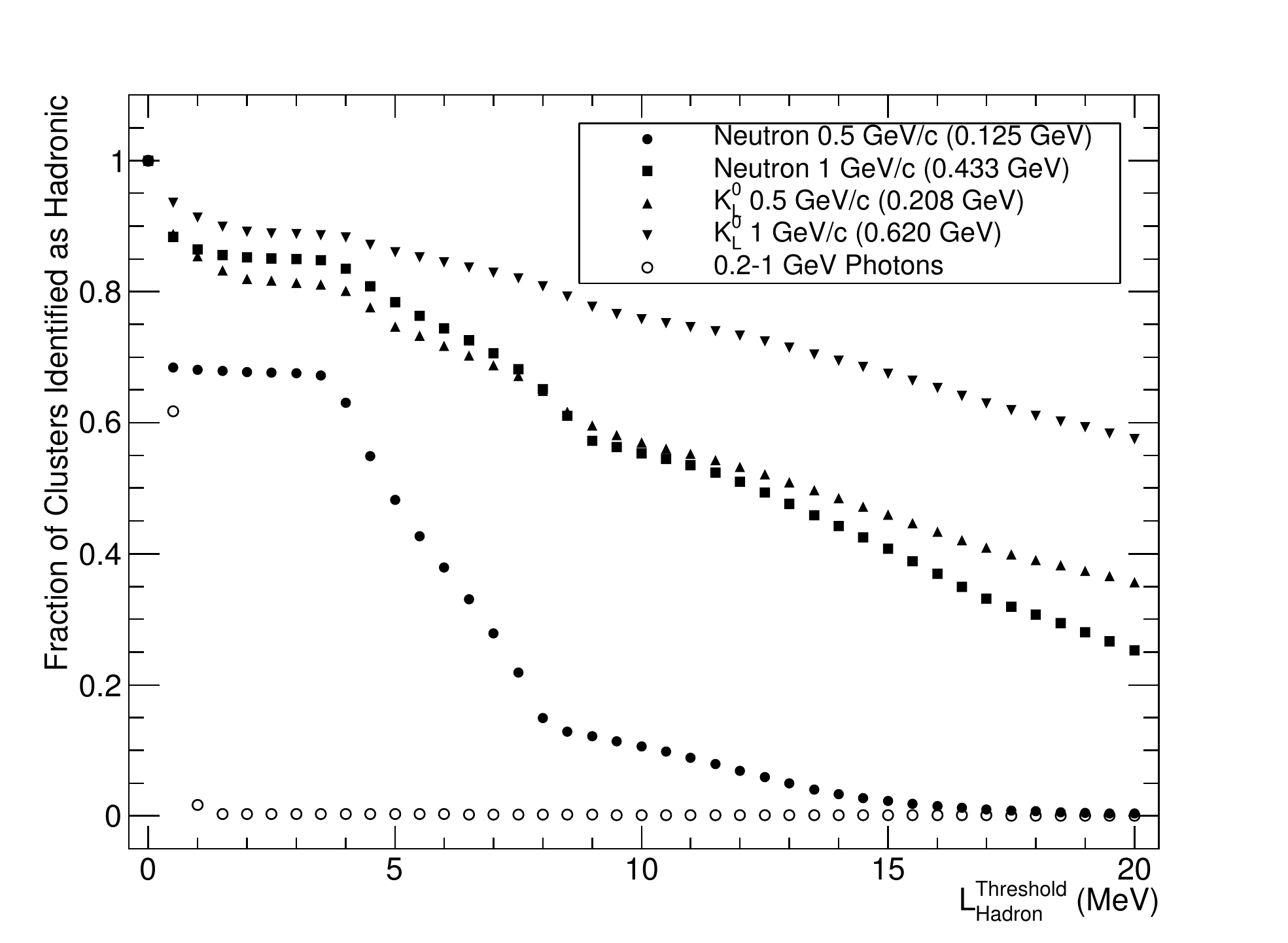} \label{clusterEff0}}
\subfloat[]{\includegraphics[width=0.5\textwidth]{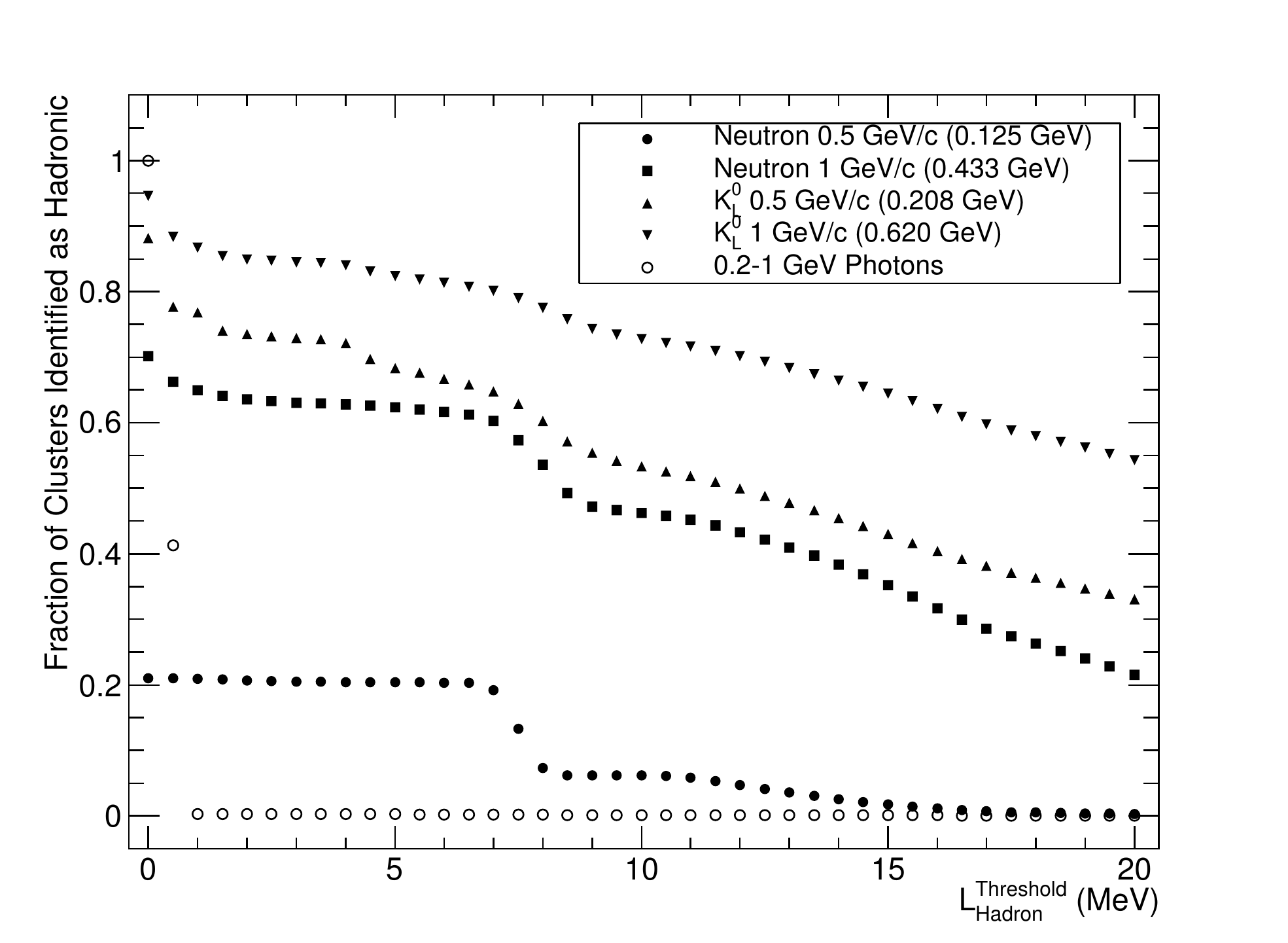} \label{clusterEff50}}

\caption{a)  Fraction of clusters with $\text{L}_\text{Hadron}^\text{Cluster} > \text{L}_\text{Hadron}^\text{Threshold}$. b) Fraction of clusters with $\text{L}_\text{Hadron}^\text{Cluster} > \text{L}_\text{Hadron}^\text{Threshold}$ where $\text{L}_\text{Hadron}^\text{Cluster}$ is computed using only cluster crystals with total light output greater than 50 MeV.}
\end{figure}

We evaluate the hadron identification efficiency using the cluster variables by plotting in Figure \ref{clusterEff0} the fraction of clusters with $\text{L}_\text{Hadron}^\text{Cluster}>\text{L}_\text{Hadron}^\text{Threshold}$ as a function of $\text{L}_\text{Hadron}^\text{Threshold}$.   Comparing with the crystal level analysis in Figure \ref{CrystalEffResults_0}, it is predicted for the 0.5 and 1 GeV/c \kl{} that combining the crystal information in the cluster will have significant impact in improving the $\text{L}_\text{Hadron}^\text{Threshold} > 10 \text{ MeV}$ region of the efficiency plot.  This is because the high energy particles produce many secondary charged hadrons throughout the cluster.   This result demonstrates that detectors with higher noise resulting in high $\text{L}_\text{Hadron}^\text{Threshold}$ values can have significant benefit from combining the crystal information in a cluster.  In the 0.5 GeV/c neutron case the increase in performance is not as significant compared to the higher energy neutron and \kl{} cases.  This is because the dominant interaction for the low energy neutrons is single proton production which is expected to be contained in a single crystal volume thus the other cluster crystals will likely not contain additional hadron component light output. 

In Figure \ref{clusterEff50} we consider a second scenario of the cluster algorithm where only cluster crystals with greater than 50 MeV of total light output are used for computing $\text{L}_\text{Hadron}^\text{Cluster}$.  In this case the hadron identification efficiencies are shown in Figure \ref{clusterEff50}.  In Figure \ref{clusterEff50} we see a similar trend as observed in the crystal level analysis when only high energy deposits were used for hadron identification, that is the hadron identification efficiently for low values of $\text{L}_\text{Hadron}^\text{Threshold}$ are the most affected by the energy threshold. 
  
\subsection{Discussion  of Neutral Hadron PSD Results}

In the above section we demonstrate that the hadron component light output from charged secondary particles in hadronic showers will be large enough for \csi{} PSD to provide discrimination between electromagnetic and hadronic showers.  The analysis presented in the previous sections used only the magnitude of the hadron component light output as a discrimination variable and predicts the ability to cleanly discriminate between neutral hadron clusters with high efficiency using PSD alone.  In general we observe the trend that the PSD performance is predicted to improve with increasing hadron energy as multi-hadron production becomes more likely.  It is important to recognize that this PSD observable uses information that is independent of information used in existing techniques to discriminate between hadronic and electromagnetic showers, such as differences in the spatial distribution of energy deposited by hadronic and electromagnetic showers as captured in e.g. lateral shower shape variables and longitudinal shower properties.  Combining the PSD information with these existing shower spatial discriminators in multivariate analyses will lead to significantly improved performance in distinguishing between hadronic and electromagnetic showers.  Moreover, the differences in PSD characteristics for $K_L^0$ and neutrons suggest that there is potential for using CsI(Tl) PSD as a tool for identifying different types of hadrons. 

\section{Conclusions}

The results in this paper show that PSD can be an effective tool for neutral hadron vs photon separation at current and upcoming high energy physics experiments using \csi{} calorimeters.  With experiments such as Belle II and BESIII applying FPGA waveform analysis in the front-end electronics, online pulse shape characterization is now feasible for these experiments.  In addition we focus in this article on using \csi{} scintillators for PSD however we note that the same principles we use may be applied to other inorganic scintillators which are known to have analogous pulse shape discrimination properties, such as NaI(Tl) \cite{Bartle}. 

To demonstrate the potential for \csi{} PSD we began by using neutron data and proton beam data collected at the TRIUMF PIF to analyse the pulse shape differences observed in \csi{} for photon and charged hadron energy deposits.  From the pulse shape differences, we demonstrated that the pulse shape variations observed in \csi{} for the charged hadron energy deposits can be characterized using a third scintillation component with decay time of $630\pm10$ ns, referred to as the hadron scintillation component.  This defined a new hadron scintillation component model for \csi{} where the pulses shapes are characterized by the relative intensity of the hadron scintillation light output to the total light output.  This new method for pulse shape description reduces the number of parameters required to describe the pulse shape variations in \csi{} compared to present techniques.

Techniques for computing the total and hadron scintillation component light output as a function of the shower particle's instantaneous ionization energy loss were developed in order to simulate these pulse shape variations.   By incorporating these methods in the GEANT4 simulation libraries we were able to reproduce the pulse shape distributions observed in the neutron and proton data.  

Using the pulse modelling and simulation techniques, the predicted pulse shape vs pulse amplitude spectra for $5 \times 5 \times 30 \text{ cm}^3$ \csi{} crystals in a $5 \times 5$ array was computed for hadronic showers generated by 0.5 and 1 GeV/c samples of simulated neutron and \kl{} mesons entering the $5 \times 5$ array.  Using a couple of simple identification algorithms we demonstrate that if the detector system can resolve the single proton band, excellent separation efficiency can be achieved for \kl{} vs photon clusters using PSD alone.  

\acknowledgments

The authors would like to thank TRIUMF for provision of the PIF facility as well its support and kind hospitality. We particularly thank E.W. Blackmore and  M. Trinczek  for their operation of the PIF facility and technical assistance provided for this experiment. The assistance of C. Hearty, J. Coffey and Z. Li during data-taking is also gratefully acknowledged. This work is supported by the Natural Sciences and Engineering Research Council (Canada).

\end{document}